\documentclass[aps,floatfix,superscriptaddress,twocolumn,longbibliography]{revtex4-1}

\usepackage{amssymb}
\usepackage{graphicx}
\usepackage{graphics}
\usepackage{amsmath}
\usepackage{amsthm}
\usepackage{color}

\usepackage{mathtools}
\usepackage{relsize}
\usepackage{amsfonts}
\usepackage{url}
\usepackage{booktabs}
\usepackage{braket}
\usepackage{bbm}
\usepackage{bbold}
\usepackage{float}
\usepackage{enumerate}
\usepackage{color}
\usepackage{hyperref}

\theoremstyle{plain}
\newtheorem{theorem}{Theorem}

\newtheorem{lemma}[theorem]{Lemma}
\newtheorem{corollary}[theorem]{Corollary}

\newtheorem{observation}[theorem]{Observation}
\newtheorem{definition}[theorem]{Definition}

\theoremstyle{remark}

\newcommand{\bea}{\begin{eqnarray}}
\newcommand{\eea}{\end{eqnarray}}

\def\bi{\begin{itemize}}
\def\ei{\end{itemize}}
\def\bc{\begin{center}}
\def\ec{\end{center}}

\newcommand{\C}{\mathbb{C}}
\def\R{\hbox{$\mit I$\kern-.6em$\mit R$}}
\def\N{\hbox{$\mit I$\kern-.6em$\mit N$}}

\def\be{\begin{equation}}
\def\ee{\end{equation}}
\def\ba{\begin{align}}
\def\ea{\end{align}}

\newcommand{\mH}{\mathcal{H}}

\newcommand{\eqdef}{\equiv}

\def\ket#1{|#1\rangle}
\newcommand{\one}{\mbox{$1 \hspace{-1.0mm}  {\bf l}$}}

\def\ket#1{\left| #1\right>}
\def\bra#1{\left< #1\right|}

\newcommand{\proj}[1]{\ket{#1}\bra{#1}}

\newcommand{\abs}[1]{\left| #1 \right|}

\definecolor{old}{rgb}{.0,.5,.2}

\begin{document}
\title{Discrete And Differentiable Entanglement Transformations}
\author{David Sauerwein}
\affiliation{Institute for Theoretical Physics, University of Innsbruck, 6020 Innsbruck, Austria}
\affiliation{Max-Planck-Institut f{\"u}r Quantenoptik, 85748 Garching, Germany}
\author{Katharina Schwaiger }
\affiliation{Institute for Theoretical Physics, University of Innsbruck, 6020 Innsbruck, Austria}
\author{Barbara Kraus}
\affiliation{Institute for Theoretical Physics, University of Innsbruck, 6020 Innsbruck, Austria}

\begin{abstract}

The study of transformations among pure states via Local Operations assisted by Classical Communication (LOCC) plays a central role in entanglement theory. The main emphasis of these investigations is on the deterministic, or probabilistic transformations between two states and mainly tools from linear algebra are employed. Here, we go one step beyond that and analyze all optimal protocols. We show that for all bipartite and almost all multipartite (of arbitrarily many $d$--level systems) pairs of states, there exist infinitely many optimal intermediate states to which the initial state can first be transformed (locally) before it is transformed to the final state. The success probability of this transformation is nevertheless optimal. We provide a simple characterization of all intermediate states. We generalize this concept to differentiable paths in the Hilbert space that connect the two states of interest. With the help of survival analysis we determine the success probability for the (probabilistic) continuous transformation along such a path and derive necessary and sufficient conditions on the optimality of it. We show, in strong contrast to previous results on state transformations, that optimal paths are characterized as solutions of a differential equation. Whereas, for almost all pairs of states, there exist infinitely many optimal paths, we present examples of pairs of states for which not even a single optimal intermediate state exists. Furthermore, we introduce a physically motivated distance measure, the interconversion metric, and show that (generically) any minimal geodesic with respect to the interconversion metric is an optimal path.  Moreover, we identify infinitely many easily computable entanglement monotones for generic multipartite pure states. We show that a given finite set of these entanglement monotones can be used to completely characterize the entanglement contained in a generic state.
\end{abstract}

\maketitle

\section{Introduction}

The investigation and better understanding of multipartite entanglement is essential to advance several fields in physics, such as quantum information theory \cite{NielsenChuang,Zyczkowski2006,Horodecki2009}, and, much more generally, quantum many-body physics \cite{PerezGarcia2006,Amico2008,Verstraete2008,Zeng2015}. Hence, a huge effort is performed to uncover some of the intriguing and useful properties of multipartite entanglement.

Entanglement is the resource to overcome the natural limitations of possible operations that can be used by spatially separated parties. These operations consist of Local Operations assisted by Classical Communication (LOCC). Whereas the mathematical description of LOCC is rather complicated (see e.g. \cite{Donald2002, Chitambar2014}), physically it simply comprises all possible operations which spatially separated parties can realize. That is, the first party can perform a measurement on his/her system and send the information about the outcome, $i$, to all all other parties. One of them performs a measurement on their share of the system that depends on $i$ and obtains outcome $j$, which he/she communicates to the other parties, etc. The difficulty in analyzing LOCC protocols stems, on the one hand, from the fact that in each round of the protocol the used local operations must correspond to completely positive trace-preserving maps and on the other hand from the fact that it has been shown that there exist instances where infinitely many rounds of LOCC are required \cite{Chitambar2011}.

Despite the difficulties encountered here, the investigation of LOCC transformations among pure states is particularly useful, as it (i) is of practical relevance in settings where the parties are spatially separated, such as in quantum networks; and (ii) allows to partially order the amount of entanglement contained in states and to identify new entanglement measures, i.e. functions which are nonincreasing under LOCC \footnote{This results form the fact that entanglement is a resource theory, where the free operations are LOCC}. That is, if an initial state, $\ket{\Psi}$, can be deterministically transformed into a final state, $\ket{\Phi}$, via LOCC, it holds that $E(\Psi)\geq E(\Phi)$ for any entanglement measure $E$. In this case the state $\ket{\Psi}$ is clearly at least as useful as $\ket{\Phi}$ for any desired task.
In case a deterministic transformation is not possible, a state might still be probabilistically transformable into the desired state. Such a probabilistic transformation is called optimal if the success probability for transforming the initial to the final state is maximal. Apart from the viewpoint of practical applications, investigations of probabilistic transformations are needed to extend the studies to transformations from pure states to ensembles of states \cite{Vidal1999}.

We focus here on deterministic as well as probabilistic transformations among single copies of fully entangled pure states, i.e. states whose single party reduced states all have maximal rank; which we consider to be identical for all subsystems. In the bipartite case it is well-known that any entangled pure state can be transformed into another entangled pure state via LOCC. The necessary and sufficient conditions for a state transformation via LOCC are given by the so-called majorization condition \cite{Nielsen1999}. In case a deterministic transformation between a pair of bipartite states is not possible, the optimal success probability and a corresponding  optimal protocol has been presented in Ref. \cite{Vidal1999}. The maximal success probability, $P(\Psi,\Phi)$, to transform the initial state, $\ket{\Psi}$, into the final state, $\ket{\Phi}$, is given by the minimal ratio of all entanglement monotones \cite{Vidal2000mono}, i.e. $P(\Psi,\Phi)= \min_\mu \mu(\Psi)/\mu(\Phi)$. Whereas this expression also holds in the multipartite setting, it has been shown that, in the bipartite setting, only a small, known set of entanglement monotones has to be considered \cite{Vidal1999}. Hence, the maximal success probability can be easily computed. The optimal protocol, i.e. the protocol which achieves this success probability, consists of a LOCC transformation from $\ket{\Psi}$ to some specific state $\ket{\chi}$ succeeded by a so-called One Successful Branch Protocol (OSBP) \cite{Vidal1999}. An OSBP has the property that one of the outcomes leads (with largest possible probability) to the desired state, whereas all other outcomes are no longer fully entangled and, hence, cannot be transformed to $\ket{\Phi}$ (not even probabilistically) \cite{Acin2000}.

Recently, it has been shown that LOCC transformations among multipartite pure states, with constant local dimensions, almost never exists \cite{Gour2017,Sauerwein2018}. That is, a pure state $\ket{\Psi} \in {\C^d}^{\otimes n}$ (for $n>4$ and arbitrary local dimension $d$) can almost never be (nontrivially) transformed into any other fully entangled state. Here, nontrivially means via different operations than Local Unitary (LU) operations, which do not alter the entanglement and can always be applied \footnote{Recall that LU-operations are invertible and hence do not alter the entanglement properties of a state.}. In retrospect, one can say that the reason why LOCC transformations among multipartite states were only considered for very special cases (such as GHZ states \cite{Turgut2010}, W states \cite{Kintas2010} and small system sizes \cite{DeVicente2013, Sauerwein2015, Spee2016, Hebenstreit2016}) is because, generically, there are no transformations possible in the single copy regime. Hence, in the generic multipartite case, probabilistic transformation are even more relevant than in the bipartite case. In the following we call a state generic, if it belongs to a specific full-measure set of states, which will be defined in Section \ref{sec:mainresults}. In Ref. \cite{Gour2017} a simple expression for the maximum success probability has been derived (for generic multipartite states) and it has been shown that the optimal protocol is a OSBP. Note, however, that this formula was derived without using entanglement monotones. In particular, the set of entanglement monotones which are required to compute the success probability via the general expression, $P(\Psi,\Phi)= \min_\mu \mu(\Psi)/\mu(\Phi)$, were unknown.

Here, we go one step beyond the investigations of transformations between pairs of states. In particular, we study {\it how} one can transform one state into the other. Considering an initial state, $\ket{\Psi}$, and a final state, $\ket{\Phi}$, we investigate whether there exists an \emph{intermediate state}, $\ket{\chi}$, to which one can transform the initial state first and then transform $\ket{\chi}$ to the final state $\ket{\Phi}$ (see also Fig. \ref{fig:IntStates}). The main requirement is, of course, that the transformation is still optimal. After presenting a simple characterization of all (optimal) intermediate states, we show that an optimal protocol can (generically) also be divided into finitely many steps, i.e. a sequence of optimal intermediate states. This is true in both cases, deterministic LOCC transformations as well as probabilistic ones. The question which presents itself is whether it is even possible to transfrom the initial state into the final one along a path in the Hilbert space. That is, is it possible to divide the optimal protocol into infinitesimal steps? We present here a simple characterization of \emph{optimal paths}, along which an initial state can be optimally transformed into a given final state. The standard tools used to study state transformations, which are based on linear algebra, allow to determine the success probability along a path only in a rather cumbersome way (see Appendix \ref{sec:altder}). We show that a more elegant way is provided by the so-called \emph{survival model}, an often used model in data analysis. This model can be used to determine the expected time of occurrence of an event, such as the death of a biological organism, or the failure of a device. The application of this model in the context considered here is particularly interesting, as it can be easily generalized to other quantum resource theories. Interestingly, we show that optimal paths can be characterized as solutions of a (very simple) "differential equation". Whereas, we show that, in the bipartite as well as the generic multipartite case, intermediate states and even (infinitely many) optimal paths always exist, we present examples of pairs of multipartite (nongeneric) states, for which no intermediate state exists. Hence, in this case, the optimal protocol cannot be divided and one really needs to "jump" from the initial to the final state. Let us mention that, in the context investigated here, it turns out that it is often easier to handle generic multipartite states than bipartite states. The reason for this is that the optimal protocol is always an OSBP in the generic multipartite case. However, as the example mentioned above clearly shows, the nongeneric multipartite setting is, as expected, much more involved than the bipartite setting.

Besides their practical relevance, the main motivation for these investigations stems from their potential insight into entanglement theory. Apart from a simple characterization of optimal intermediate states and optimal paths (via a differential equation), we will see that this research leads naturally to a distance measure between multipartite pure states, which is a metric on local-unitary-equivalence (LU-equivalence) classes. Hence, the metric is not only invariant under the physically irrelevant LU operations, but it vanishes iff two states are LU-equivalent. We show that, in the bipartite as well as in the generic multipartite case, any \emph{minimal geodesic} (i.e. any shortest path between two states) is also an optimal path. This establishes a connection between geometry and pure state transformations. Furthermore, we introduce a family of new entanglement monotones for generic multipartite pure states (see also Ref.\ \cite{Gour2011}, where entanglement monotones where introduced, that are included in this new family of entanglement monotones) that can be regarded as the analog of the well-known and often used bipartite entanglement monotones mentioned above and introduced in Ref. \cite{Vidal1999}. In contrast to many known entanglement monotones, these monotones are not invariant under stochastic LOCC (SLOCC) transformations. Note that, whereas entanglement monotones that are invariant under stochastic local operations are very useful to differentiate between SLOCC classes \cite{Gour2013, Eltschka2012}, the here introduced monotones can be used to compare the entanglement of states within the same SLOCC class. The latter is highly relevant as states in different SLOCC classes correspond to {\it different types of entanglement} \cite{Dur2000}. In fact, we show that a given finite set of these monotones, together with the knowledge of the SLOCC class to which a given state belongs to, completely characterizes the entanglement of the state. We use these monotones to characterize optimal intermediate states (and paths). Moreover, we show that, as in the case of bipartite states, these monotones allow to determine the maximal success probability of pure state transformations among generic multipartite states.

The outline of the remainder of the paper is the following.
In Section \ref{sec:mainresults}, we present the main results of this paper and explain their physical relevance. In particular, we present the characterization of optimal intermediate states and optimal paths connecting two states. Moreover, we introduce a metric among LU-equivalence classes and show that, in the bipartite and in the generic multipartite case, any minimal geodesic (with respect to this metric) is an optimal path. Furthermore, we introduce an infinite set of easily computable entanglement monotones for multipartite pure states and show their relevance in entanglement theory. The analysis performed here shows many more similarities between bipartite and multipartite states than the results about direct state transformations (see e.g. \cite{Turgut2010, Kintas2010, DeVicente2013, Gour2017,Sauerwein2018} and references therein) would suggest. In Section \ref{sec:preli} we first recall some known results related to bipartite and multipartite (S)LOCC transformations and introduce the survival model.
In Section \ref{sec:defintstates} we consider the possibility of transforming one state into the other by transforming the initial state first into an intermediate state (possibly with some probability different from unity) and then transforming the intermediate state to the final state. For a given pair of states we determine necessary and sufficient conditions for the intermediate states to be optimal. In Section \ref{sec:contLOCC} we use the survival model to investigate the success probability for a transformation along a path in Hilbert space. We consider piecewise differentiable paths that connect the initial to the final state and use tools from survival analysis to characterize optimal paths. In Section \ref{sec:metric} we introduce the {\it interconversion metric}, which is a metric on LU-equivalence classes, and prove a strong relation between the minimal geodesics with respect to this metric and optimal paths. In Section \ref{sec:bipex} we then consider bipartite states and derive new optimal protocols. First, we consider deterministic transformations and determine all LOCC paths (see also Ref. \cite{Schwaiger2018}). Along these paths, the transformation is deterministic. We characterize all optimal intermediate states with the so-called majorization lattice \cite{Cicalese2002}. The survival model is then employed to study optimal probabilistic transformations. In Section \ref{sec:multiex} we turn to the multipartite scenario. As LOCC transformations are generically not possible in this setting, we immediately consider probabilistic transformations. We derive a differential equation which characterizes optimal paths connecting a pair of generic states and show a strong relation between minimal geodesics with respect to the interconversion metric and optimal paths. Furthermore, we demonstrate that there are pairs of nongeneric multipartite states for which not even one intermediate state exists. In Section \ref{sec:monotones}, we introduce an infinite set of entanglement monotones for multipartite pure states and show that they can be regarded as a generalization of the well-known bipartite entanglement monotones. In particular, a known finite set of these monotones (together with a description of the SLOCC class the state belongs to) completely characterizes the entanglement contained in the state. In Section \ref{sec:conclusion} we present our conclusions.

\section{Main results and implications}
\label{sec:mainresults}
We summarize here the main results of this paper and discuss their consequences within entanglement theory.

\subsection{Notation}
Let us first introduce our notation and recall some results in the context of state transformations. We consider pure states belonging to the Hilbert space $\mH_{n,d}\eqdef\otimes^{n}\mathbb{C}^{d}$, i.e. the Hilbert space of $n$ qudits \footnote{Note that we consider here only systems where the local dimension of the Hilbert spaces all coincide. The reason for that is that only in these cases the results presented in Refs.  \cite{Gour2017,Sauerwein2018} ensure that the set of states for which there exists no nontrivial (i.e. different than the identity operator) local symmetry is of full measure.}. Whenever we do not need to be specific about the local dimensions, we simply write ${\cal H}_n$ instead of $\mH_{n,d}$. Unless stated otherwise, we consider normalized fully entangled pure states. Throughout this paper, the set of local invertible operators (local unitary operators) on $\mH_n$ is denoted by $\tilde{G}$ (${\cal U}$) resp. Two states, $\ket{\Psi}, \ket{\Phi}$ are said to be SLOCC (LU) equivalent if there exists a $g \in G$ ($u \in {\cal U}$) such that $\ket{\Psi}= g \ket{\Phi}$ ($\ket{\Psi}= u \ket{\Phi}$). Note that $\ket{\Psi},\ket{\Phi}$ are SLOCC (LU) equivalent iff $\ket{\Psi}$ can be transformed locally into $\ket{\Phi}$ and back to $\ket{\Psi}$ with some finite probability of success (with probability one) respectively \cite{Dur2000, Gingrich2002}.  As mentioned above, local unitary operators do not alter the entanglement. Hence, in what follows we always consider a LU--equivalence class, instead of considering a single state. For a bipartite state $\ket{\Psi}$ we therefore always consider the Schmidt decomposition, which we denote by $\ket{\Psi}=\sum_i \sqrt{\Psi_i} \ket{i,i}$, with $\Psi_i \geq \Psi_{i+1} \geq 0$ and $\sum_i \Psi_i=1$.

We call a multipartite state a generic state if it belongs to the full-measure set of states that do not possess any nontrivial (i.e. different than the identity operator) local symmetry \cite{Gour2017,Sauerwein2018}. For homogeneous systems, i.e. n $d$-level systems with $n>4$ and $d$ arbitrary, such a full-measure set of states always exists \cite{Gour2017,Sauerwein2018}. We consider transformations among fully entangled $n$-partite generic states. Unless stated differently, we denote by $\ket{\Psi} = g\ket{\Psi_s}$ ($\ket{\Phi} = h\ket{\Psi_s}$) the initial (final) state respectively. Both of them belong to the same SLOCC class as the generic state $\ket{\Psi_s}$. Here and in the following, $g$ and $h$ always denote local invertible operators, i.e. $g=g_1 \otimes \ldots \otimes g_n$ with $g_i \in GL(d,\C)$ for $i\in \{1,\ldots ,n\}$ and similarly for $h$ and $G=g^\dagger g$, $H=h^\dagger h$.

As we study here deterministic and probabilistic transformations, we recall that the optimal success probability to transform $\ket{\Psi} \in \mH_n$ into $\ket{\Phi} \in \mH_n$, which we denote by $P(\Psi,\Phi)$ throughout the paper, can be expressed as \cite{Vidal2000mono},
\bea
P(\Psi, \Phi) = \min_\mu \frac{\mu(\Psi)}{\mu(\Phi)}. \label{eq:opttrans1Intro}
\eea
Here, the minimization is to be taken over all entanglement monotones $\mu$ \footnote{The monotones are rescaled so that they vanish on separable states. Moreover, recall that we consider here only fully entangled states.}. In the case of bipartite states, it has been shown that it suffices to perform the optimization only over the entanglement monotones
\begin{align}
 \label{eq:EnMonBip} E_l(\Psi)=\sum_{i=l}^{d} \Psi_i, \mbox{ for } l\in\{1,\ldots d\},
\end{align}
That is, for bipartite states the following holds \cite{Vidal1999},
\begin{align}
 P(\Psi,\Phi) = \min_{k \in \{1,\ldots,d\}} \frac{E_k(\Psi)}{E_k(\Phi)}. \label{eq:fewmonotones}
\end{align}

Note that the set of entanglement monotones $\{E_l\}_{l=2}^d$ determines the Schmidt coefficients and therefore the entanglement contained in the state. Note further that the success probability is equal to one, i.e. the transformation can be done deterministically via LOCC, iff $E_l(\Psi)\geq E_l(\Phi)$ for $l\in\{1,\ldots d\}$ (majorization condition) \cite{Nielsen1999}.

In Ref. \cite{Gour2017} it was shown that the maximum probability to convert a generic multipartite state $\ket{\Psi} = g\ket{\Psi_s}$ to $\ket{\Phi} = h\ket{\Psi_s}$ via $LOCC$ is given by

\begin{align}
 P(\Psi,\Phi) =  \frac{\|\Phi\|^2}{\|\Psi\|^2}
\frac{1}{\lambda_{max}(G^{-1} H)}, \label{eq:optprob}
\end{align}
where $\lambda_{max}(X)$ denotes, here and in the following, the maximal eigenvalue of $X$ and $G=g^\dagger g$ and $H=h^\dagger h$. Note that the maximal eigenvalue can be easily computed as $G^{-1} H$ is a local operator. Note further that (nontrivial) LOCC transformations are impossible in the generic case \cite{Gour2017,Sauerwein2018}; that is, $P(\Psi,\Phi)=1$ only holds if the states are LU-equivalent.

\subsection{Complete set of entanglement monotones}

By scrutinizing the equivalence of the optimal success probability given in Eq. (\ref{eq:opttrans1Intro}) and Eq. (\ref{eq:optprob}), we identify here an infinite set of entanglement monotones for pure generic multipartite states. This leads to a natural generalization of the simple formula given in  Eq. (\ref{eq:fewmonotones}) to the generic multipartite case, as stated in the following lemma (see Section \ref{sec:monotones} for the derivation).\\

\noindent \textit{{\bf Lemma.}
Let $\ket{\Psi}$ be a normalized state in the SLOCC class of a generic state $\ket{\Psi_s} \in \mH_{n,d}$, i.e. $\ket{\Psi} = g\ket{\Psi_s}$ normalized.
The functions \bea \label{eq:EnMons} E^{\Psi_s}_{\vec{x}}(\ket{\Psi})=\bra{\vec{x}}G\ket{\vec{x}}\eea
are entanglement monotones for generic pure multipartite states. Here, $\ket{\vec{x}}$
denotes a product state, i.e. $\ket{\vec{x}}=\ket{x_1}\otimes
\ldots \otimes \ket{x_n}$, with $\ket{x_i}\in \C^d$ and $G=g^\dagger g=\otimes_{i=1}^d G_i$.  Moreover, in the case of generic states, it holds that
\bea
P(\Psi,\Phi) = \min_{\vec{x}} \frac{E^{\Psi_s}_{\vec{x}}(\Psi)}{E^{\Psi_s}_{\vec{x}}(\Phi)} \label{eq:opttransmultiIntro}
\eea
and in the nongeneric case the right hand side of Eq. (\ref{eq:opttransmultiIntro}) gives a lower bound on the success probability.\\
}

In the following we call the entanglement monotones for generic pure multipartite states simply entanglement monotones \footnote{ It should be noted, however, that (as it is commonly the case), these functions are shown to be monotonic only in case the initial state is pure.}.

In Eq. (\ref{eq:EnMons}), infinitely many entanglement monotones are defined. However, as can be easily shown, a given finite set of them, $\{E^{\Psi_s}_{\vec{x}}\}_{i\in I}$, with $I$ a finite index set (together with $\ket{\Psi_s}$) uniquely defines the state up to LUs (see Sec \ref{sec:monotones}). Stated differently, given the SLOCC class to which a state belongs to, a known finite set of entanglement monotones (as defined in Eq. (\ref{eq:EnMons})) completely characterizes the entanglement contained in the state. This is in complete analogy to the bipartite entanglement monotones (see Eq. (\ref{eq:EnMonBip})). Moreover, in the multipartite case the minimization in Eq. (\ref{eq:opttransmultiIntro}) does not even need to be computed, as it is given by Eq. (\ref{eq:optprob}).

Let us mention here that, generically, a so-called critical state $\ket{\Psi_s}$ can be easily derived from $\ket{\Psi}$ \cite{Verstraete2003}. However, any other representative $\ket{\Psi_s}$ of the generic SLOCC class can also be used to define the measures. Hence, as $G$ in Eq. (\ref{eq:EnMons}) is local and $\ket{\vec{x}}$ is a product state, the entanglement monotones can be very easily computed.

Particularly important, and also necessary for a complete characterization of entanglement, is that these monotones are not SLOCC invariant. In fact, they are determined by the local invertible operator, $g$, which relates the state of interest to the representative of the SLOCC class, $\ket{\Psi_s}$ in Eq. (\ref{eq:EnMons}). This is why these monotones are very useful for the comparison of entanglement contained in states belonging to the same SLOCC class. To highlight this relevance, let us note here that there exist methods to construct entanglement monotones from so-called SL-invariant polynomials \cite{Eltschka2012,Gour2013} (i.e. polynomials that are invariant under the action of local invertible matrices with determinant one). However, these monotones coincide for all states in one SLOCC class, per construction. Hence, they are very useful to differentiate SLOCC classes, but they cannot be employed to compare the amount of entanglement contained in states belonging to the same SLOCC class. Note, however, that this is a particularly interesting comparison, as states in different SLOCC classes possess very distinct entanglement properties. The last statement is due to the fact that states in different SLOCC classes cannot even probabilistically be transformed into each other. Prominent examples of different SLOCC classes are the three-qubit $GHZ$ class and the $W$ class \cite{Dur2000}.

Combined with previous results, the following picture for generic multipartite states emerges. Given a generic state, $\ket{\Psi}$, first its SLOCC class is determined using e.g. the algorithm presented in Ref. \cite{Verstraete2003} \footnote{Note that one could also use the SL-invariant polynomials \cite{Eltschka2012,Gour2013,Osterloh2012} to identify the SLOCC class.}. This algorithm gives as an output a so-called critical state of the corresponding SLOCC class, $\ket{\Psi_s}$ (see Refs. \cite{Verstraete2003,Gour2010}). Then, a given finite set of entanglement monotones defined in Eq. (\ref{eq:EnMons}), $\{E^{\Psi_s}_{\vec{x}}\}_{i\in I}$, uniquely characterizes the entanglement contained in the state.

Using these entanglement monotones for the multipartite and the bipartite setting, we show that optimal intermediate states and paths can be easily characterized (Section \ref{sec:bipex} and Section \ref{sec:multiex}).

\subsection{Characterization of optimal intermediate states}

In the following we introduce the concept of intermediate states of a transformation from an initial state $\ket{\Psi} \in \mH_n$ to a final state $\ket{\Phi} \in \mH_n$. We define an optimal intermediate state of this transformation as a state $\ket{\chi} \in \mH_n$, which is neither LU-equivalent to $\ket{\Psi}$, nor to $\ket{\Phi}$, and for which it holds that $\ket{\Psi}$ can be transformed into $\ket{\chi}$ and then $\ket{\chi}$ can be transformed into $\ket{\Phi}$ and yet the resulting transformation from $\ket{\Psi}$ to $\ket{\Phi}$ is optimal (for the precise definition see Def. \ref{def:intstate} in Section \ref{sec:defintstates}). That is, $\ket{\chi} $ is an optimal intermediate state if
\bea
P(\Psi,\Phi) = P(\Psi,\chi)P(\chi,\Phi).
\eea
The set of all such states $\ket{\chi}$ is denoted by $\mathcal{I}(\Psi,\Phi)$. Note that we include here also deterministic transformations, for which $P(\Psi,\Phi) = 1$.\\

\begin{figure}
   \centering
   \includegraphics[width=0.5\textwidth]{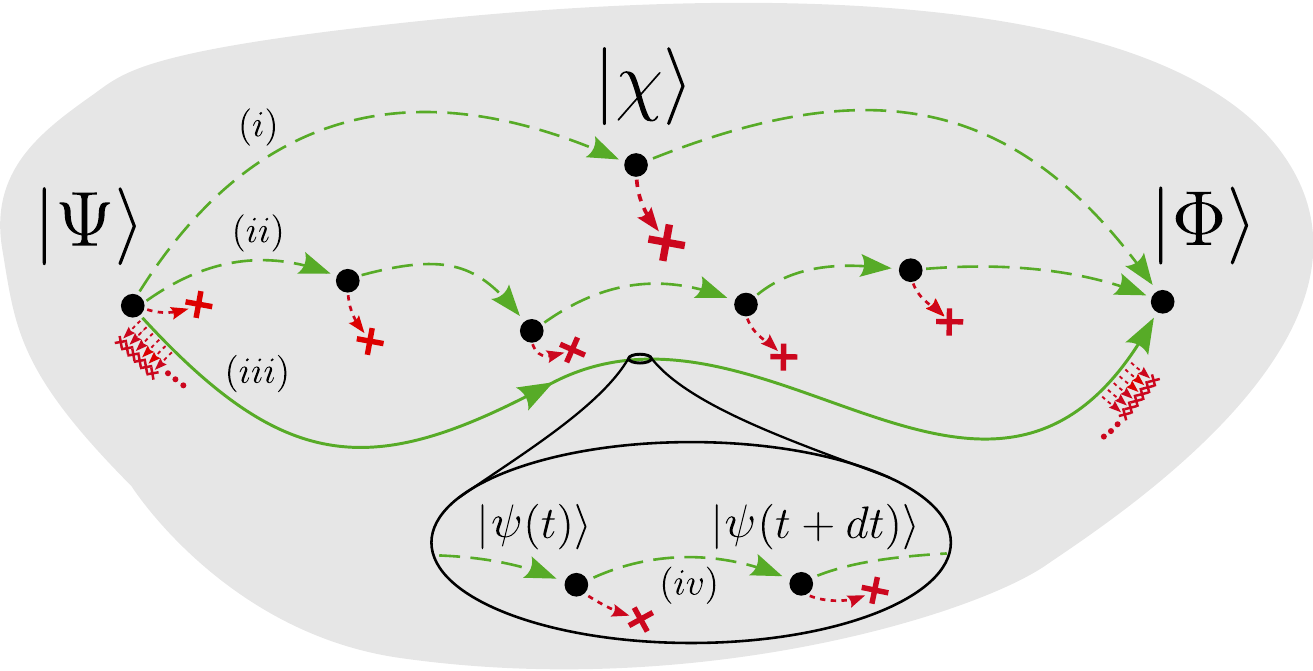}
   \caption{(color online). In this figure, black dots represent fully entangled states, while red crosses correspond to (not fully entangled) states outside of the SLOCC class. The optimal SLOCC transformation from $\ket{\Psi}$ to $\ket{\Phi}$ is performed via (i) an optimal intermediate state $\ket{\chi} \in \mathcal{I}(\Psi,\Phi)$, (ii) a sequence of optimal intermediate states, (iii) a differentiable transformation along an optimal SLOCC path $\{\ket{\psi(t)}\}_{0\leq t \leq 1}$ from $\ket{\Psi}$ to $\ket{\Phi}$. The green dashed arrows (which lead from one fully entangled state to the other) correspond to the successful branch(es) of a transformation, while the red arrows correspond to the unsuccessful branches that lead to a state that is no longer fully entangled. In the differentiable transformation along the path $\{\ket{\psi(t)}\}_{0\leq t \leq 1}$ (see (iii)) the state $\ket{\psi(t)}$ is optimally transformed into the infinitesimally close state $\ket{\psi(t+dt)}$ with probability $P[\psi(t),\psi(t+dt)]$ for $0 \leq t < 1$ (see (iv)). See Section \ref{sec:theory} of the main text for details.}
   \label{fig:IntStates}
   \end{figure}

Let us mention here that there always exists a state to which the initial state $\ket{\Psi}$ can be transformed locally (possibly probabilistically) and which is then transformed (locally) to the final state $\ket{\Phi}$. However, it is, \emph{a priori}, not clear if an optimal intermediate state exists for a given pair of states ($\ket{\Psi}$, $\ket{\Phi}$); that is, a state such that "passing by" that state does not diminish the success probability of the whole transformation from $\ket{\Psi}$ to $\ket{\Phi}$ (see Figure \ref{fig:IntStates}, (i)).

We show in Section \ref{sec:bipex} that there always exist infinitely many optimal intermediate states for transformations among bipartite and generic multipartite states \footnote{Note that in this case deterministic transformations are not possible \cite{Gour2017, Sauerwein2018}}. However, in Section \ref{sec:noint} we also provide examples of multipartite LOCC transformations for which the set of intermediate states is empty. In this case it is, therefore, impossible to split the transformation into several steps.
\\

Using the bipartite entanglement monotones defined in Eq. (\ref{eq:EnMonBip}) we derive, from a general characterization of intermediate states (see Theorem \ref{thm:monotones1} in Section \ref{sec:defintstates}), the following simple characterizations of optimal intermediate states for bipartite transformations.\\

\noindent \textit{{\bf Corollary.}
Let $\ket{\Psi}, \ket{\Phi} \in \C^d \otimes \C^d$ be bipartite states with $P(\Psi ,\Phi) = \min_i \frac{E_i(\Psi)}{E_i(\Phi)} = \frac{E_l(\Psi)}{E_l(\Phi)}$. A state $\ket{\chi} \in \C^d \otimes \C^d$  is an optimal intermediate state, i.e. $\ket{\chi} \in \mathcal{I}(\psi, \phi)$ iff
\begin{equation}
P (\Psi, \chi) = \frac{E_l(\Psi)}{E_l(\chi)} \ \text{and} \ P (\chi , \Phi) = \frac{E_l(\chi)}{E_l(\Phi)}.
\end{equation}
}\\

Using this characterization we show that, for both, probabilistic and deterministic transformations, optimal intermediate states exist in the bipartite case. For instance, for transformations that can be performed via LOCC, one can always identify intermediate states that can be reached from the initial state via LOCC and that can then be further transformed into the final state via LOCC. We characterize in Section \ref{sec:bipex} all these states using the notion of majorization lattices \cite{Cicalese2002}. In particular, we show that a state is an optimal intermediate state iff it is contained in a subset (an interval) defined by the initial and final state.

In case the transformation is not deterministic, an optimal protocol has been presented in Ref. \cite{Vidal1999}. This protocol consists of an LOCC transformation to a particular state $\ket{\xi}$, followed by an OSBP. As the transformation is optimal, the state $\ket{\xi}$ is an  optimal intermediate state. Furthermore, it has been shown that $\ket{\xi}$ also optimizes the fidelity to the final state \cite{Vidal2000}. We show here that this is not necessary and that there exist infinitely many other optimal intermediate states. Furthermore, we identify extreme intermediate states which have the property that they are the most (least) entangled states to which $\ket{\Psi}$ can be transformed via LOCC (OSBP) such that the intermediate state can still be transformed to the final state via an OSBP (LOCC) in an optimal way, respectively.

In the generic multipartite case, we prove that the entanglement monotones, $E^{\Psi_s}_{\vec{x}}$, lead to a similar characterization of optimal intermediate states as in the bipartite case.
Using the fact that the optimal protocol is always given by an OSBP \cite{Gour2017, Sauerwein2018} in the case of generic multipartite states, the following characterization of intermediate states can be shown (see Theorem \ref{thm:neccsuff0} in  Section \ref{sec:multiex}).\\

\noindent \textit{{\bf Theorem.}
Let $\ket{\Psi} \in \mH_n$ be a generic state and let $\ket{\Phi} = h\ket{\Psi}$ be a state in the SLOCC class of $\ket{\Psi}$. A state $\ket{\chi} = g\ket{\psi}$ is an optimal intermediate state, i.e. $\ket{\chi} \in \mathcal{I}(\Psi,\Phi)$, iff
\bea  \label{Eq:InterMulti}
  \lambda_{max}(G^{-1} H) = \frac{\lambda_{max}(H)}{\lambda_{max}(G)}.
\eea
}\\

Recall that both, $G=\otimes_i G_i$ and $H=\otimes_i H_i$, are positive local operators. Hence, the computation of the eigenvalues and the inspection of the necessary and sufficient condition is very easily performed. In fact, as for any two operators $G,H$ the right hand side of Eq. (\ref{Eq:InterMulti}) is an upper bound to the left hand side, it is easy to show that $\ket{\chi} \in \mathcal{I}(\Psi,\Phi)$, if the following conditions hold for all $j \in \{1,\ldots,n\}$

\begin{itemize}
\item[(i)] $[G_j,H_j] = 0$, and
\item[(ii)] $\frac{G_j}{\lambda_{max}(G_j)} \leq \frac{H_j}{\lambda_{max}(H_j)}$.
\end{itemize}

For generic multiqubit states we show that these conditions are necessary and sufficient. Thus, not only can the optimal intermediate states be easily characterized, but using the conditions above, one can also easily identify a subset of optimal intermediate states.

Note that the characterization within the generic multipartite case is actually simpler than in the bipartite case. Here, the characterization is simply given by an equivalence of eigenvalues of local operators. The reason for the simplicity originates from the fact that the optimal protocol in the generic multipartite case is always an OSBP, whereas in the bipartite case either LOCC, OSBP, or combinations of them are optimal.

In the bipartite as well as the generic multipartite case, this shows that optimal intermediate states always exist. Interestingly, this is not always the case. In fact, we present in Section \ref{sec:noint} an example of a pair of (nongeneric) states for which there exists no intermediate state. The results presented here then show that one really needs to "jump" to the final state and is prevented to first transform the initial state to some intermediate state (deterministically) and then transform the intermediate state to the final state (again deterministically).

The concept of optimal intermediate states can be generalized straightforwardly to sequences of optimal intermediate states. Any state in the sequence is then an optimal intermediate state of the proceeding and the subsequent state in the sequence (see Section \ref{sec:defintstates} for the precise definition). An optimal sequence allows to divide the optimal transformation from $\ket{\Psi} \in \mH_n$ to $\ket{\Phi} \in \mH_n$ into a finite number of optimal steps (see Fig. \ref{fig:IntStates}, (ii)). Moreover, it is easy to see that the concept of optimal intermediate states can also be generalized to other quantum resource theories. In the following we call optimal intermediate states simply intermediate states. \\

\subsection{Characterization of optimal paths}
\label{sec:CharOptPath}
The question which presents itself after this characterization of intermediate states, which allow to split the transformation into finitely many steps, is, whether it is also possible to identify a continuous path in the Hilbert space along which an optimal transformation is possible. That is, we are interested in a path in the Hilbert space, denoted by $\{\ket{\psi(t)}\}_{0\leq t \leq 1}$, for which $\ket{\Psi(0)}=\ket{\Psi}$, $\ket{\Psi(1)}=\ket{\Phi}$ and that consists of intermediate states. We show how, for certain paths of this kind, $\ket{\Psi}$ can be optimally transformed into $\ket{\Phi}$ in, what we call, an optimal transformation along this path (see also Fig. \ref{fig:IntStates}, (iii)).

\begin{figure}
   \centering
   \includegraphics[width=0.53\textwidth]{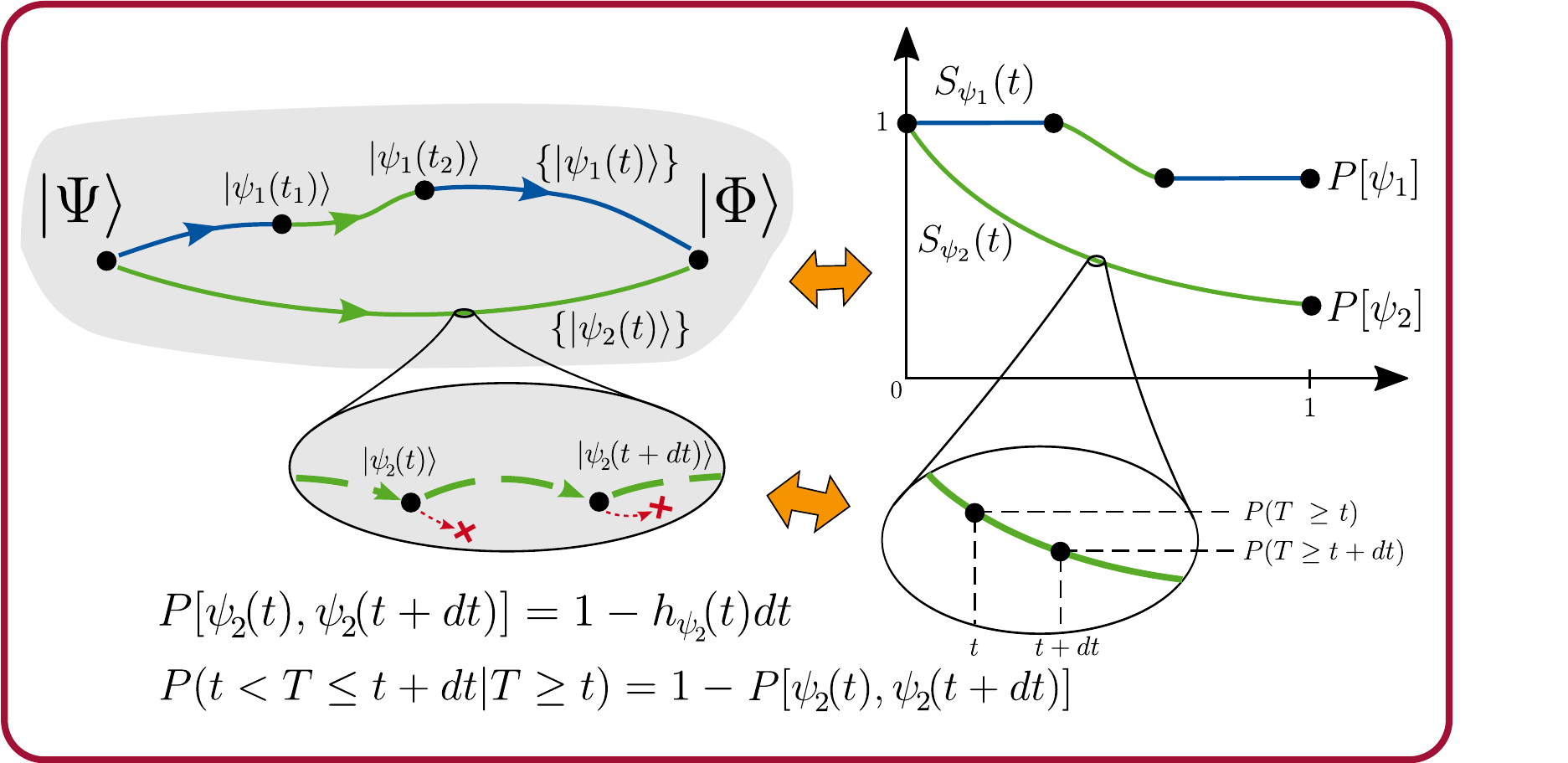}
   \caption{(color online). Transformations along an SLOCC path, viewed as a survival process. For the paths $\{\ket{\psi_i(t)}\}_{0\leq t \leq 1}$, where $i \in \{1,2\}$, on the left, the survival functions $S_{\psi_i}(t)$ are plottet on the right. For the path $\{\ket{\psi_1(t)}\}$ the transformations in the time intervals $[0,t_1]$ and $[t_2,1]$ are deterministic and hence $S_{\psi_1}(t)$ is constant for these times. The connection between the infinitesimal optimal transformation from $\ket{\psi_2(t)}$ to $\ket{\psi_2(t+dt)}$ and its representation in the survival model are depicted. See Section \ref{sec:difftrans} of the main text for details.}
   \label{fig:survival}
   \end{figure}

\begin{figure}
   \centering
   \includegraphics[width=0.4\textwidth]{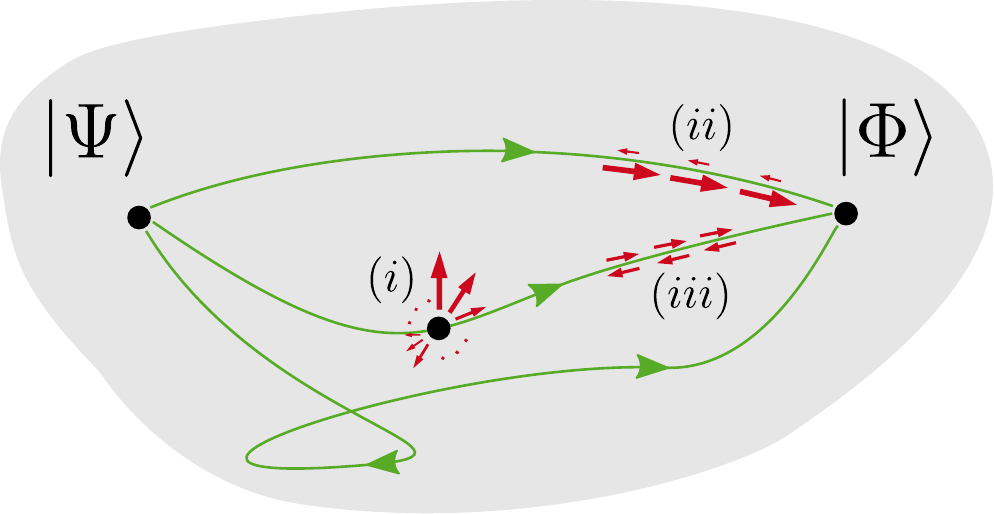}
   \caption{(color online). Different SLOCC paths from $\ket{\Psi}$ to $\ket{\Phi}$. The arrows in (i) correspond to the magnitude of the hazard rate if one goes through this point along an SLOCC path that is coming from the indicated direction. This illustrates that the hazard rate is not only position dependent, but also direction dependent. This is in analogy to the refractive index in an anisotropic medium (see Section  \ref{sec:fermat}). (ii) The hazard rates experienced when going along a path from $\ket{\Psi}$ to $\ket{\Phi}$ can be very small, while the hazard rate corresponding to the transformation from $\ket{\Phi}$ back to $\ket{\Psi}$ (along the same path) can be very high. (iii)  An SLOCC path is a distance minimizing geodesic in the interconversion metric if the transformation along this path is in both directions optimal. For this it is necessary that the hazard rates in both directions are small enough (see Section \ref{sec:geodesic}).}
   \label{fig:geodesics}
   \end{figure}

The question addressed here leads to a different viewpoint on state transformations. On the one hand, because pure state transformations via local operations have so far only been considered between a pair of states, while we consider here transformations along a whole set of states. On the other hand, because this approach allows us to use tools from calculus to investigate state transformations, as we see below. This is in contrast to the usual approach to study state transformations, which is mainly based on tools from linear algebra.\\
We are not interested in any path connecting the initial and the final state, but in optimal paths; that is, in paths along which the initial state can be optimally transformed into the final state. To this end, we first determine the success probability $P[\psi]$ for the transformation along a general (maybe nonoptimal) path $\{\ket{\psi(t)}\}_{0\leq t \leq 1}$ from $\ket{\Psi} = \ket{\psi(0)}$ to $\ket{\Phi} = \ket{\psi(1)}$. One way of computing $P[\psi]$ would be to consider the infinite product of the success probabilities for the infinitesimal transformations along the path (see Appendix \ref{sec:altder}). Here, we use an approach based on the theory of survival analysis. Survival analysis is commonly used to determine the expected duration until an event happens, for example the death of a biological organism. Here, we use it to determine the expected duration until the entanglement contained in the state is lost in a (S)LOCC transformation along a path in Hilbert space (see Section \ref{sec:contLOCC} and Fig. \ref{fig:survival}).  In this way, we find the success probability $P[\psi]$ for the transformation along a given path $\{\ket{\psi(t)}\}$ (see Fig. \ref{fig:survival} and Fig. \ref{fig:geodesics}).  We can use a variational approach to find optimal paths as those paths that maximize $P[\psi]$ with the value $P[\psi] = P(\Psi,\Phi)$ and find necessary and sufficient conditions for optimality (see Section \ref{sec:contLOCC} for the details). To state the resulting simple characterization of optimal paths for transformations of bipartite states requires notions of survival analysis. It can be found in Corollary \ref{cor:optbippath} in Section \ref{sec:bipex}. Here, we present the even simpler characterization of optimal paths for the generic multipartite case (see Theorem \ref{thm:charpath1} in Section \ref{sec:genintcont}).\\

\noindent \textit{{\bf Theorem.}
Let $\ket{\Phi} = g\ket{\Psi}$ be a state in the SLOCC class of a generic state $\ket{\Psi}$. A differentiable path $\{\ket{\psi(t)} = g(t)\ket{\Psi}\}$, with $\lambda_{max}[G(t)] = 1$, from $\ket{\Psi} = \ket{\psi(0)}$ to $\ket{\Phi} = \ket{\psi(1)}$ is an optimal path of the transformation from $\ket{\Psi}$ to $\ket{\Phi}$ iff
 \begin{align}
  \lambda_{max}\left[G(t)^{-1} G'(t)\right] = 0 \ \text{for all} \ t \in (0,1). \label{eq:localbndIntro}
 \end{align}
}\\
Here, $G'(t) = \frac{d}{dt}G(t)$ denotes the derivative of $G(t)$.
Recall that $G(t)=g(t)^\dagger g(t)> 0$ is a local operator. Note further that the condition $\lambda_{max}[G(t)] = 1$ is not a restriction as any path can be normalized in this way.
This theorem allows one to find all optimal paths, and hence all optimal protocols to transform $\ket{\Psi}$ into $\ket{\Phi}$ via generic pure states, as the solution of a "differential equation". This is in contrast to previous results on state transformations, which were mainly based on tools from linear algebra and focused on one protocol that achieves the transformation.\\

In order to illustrate the usefulness of this characterization, we present some simple examples of optimal paths. Consider a path, $\{\ket{\psi(t)}\}_{0 \leq t \leq 1}$, where $\ket{\psi(t)}=g(t)\ket{\psi(0)}$, with $G(t)=g(t)^\dagger g(t)=\bigotimes_{i=1}^n U^{(i)} \text{diag}[1,r_2^{(i)}(t),\ldots,r_d^{(i)}(t)]{U^{(i)}}^\dagger$.
It follows from the theorem above that this path is optimal iff $\frac{d}{dt}[r_k^{(i)}(t)] \leq 0$ for all $k$ and $i$ (see Section \ref{sec:genintcont}, Theorem \ref{thm:charpath1}).

These examples show that there always exist infinitely many optimal paths in the case of generic multipartite states. Moreover, in the multipartite qubit case all optimal paths are of the form presented above, as we show in Section \ref{sec:genintcont}.

In order to analyze the optimal paths further and to establish a connection to geometry, we introduce the following {\it interconversion metric} on LU-equivalence classes:
\begin{align}
d_I:& \mH_n \times \mH_n \rightarrow [0,\infty]: \nonumber\\
&(\Psi,\Phi) \mapsto d_I(\Psi,\Phi) =  -\log[P(\Psi,\Phi)P(\Phi,\Psi)].
\end{align}
The interconversion metric quantifies the difficulty of interconverting the states $\ket{\Psi}$ and $\ket{\Phi}$ via SLOCC. Moreover, $d_I$ is a metric on the set of all LU-equivalence classes. In particular, it vanishes iff the states are LU--equivalent. Furthermore, $d_I(\Psi,\Phi) = \infty$ iff $\ket{\Psi}$ and $\ket{\Phi}$ are not SLOCC equivalent, reflecting the fact that these two states then contain different kinds of entanglement.

We demonstrate that the minimal geodesics with respect to the interconversion metric (i.e. all paths of minimal length) are strongly related to paths that are \emph{twofold optimal}. We call an optimal path
$\{\ket{\psi(t)}\}_{0 \leq t \leq 1}$ twofold optimal if its reverse path,  $\{\ket{\psi_R(t)} = \ket{\psi(1-t)}\}_{0 \leq t \leq 1}$, is optimal too. We show that, for bipartite and generic multipartite states, any minimal geodesics $\{\ket{\psi(t)}\}_{0 \leq t \leq 1}$ is twofold optimal. In fact, for generic multiqubit states, the sets of optimal paths, twofold optimal paths and minimal geodesics all coincide. This establishes a strong connection between geometry and optimal state transformations.

\section{Preliminaries and Survival Analysis}
\label{sec:preli}
In this section we review some concepts of entanglement theory that were not mentioned before and that we are going to use in the subsequent sections. Moreover, we review methods used in survival analysis.

\subsection{LOCC and SLOCC transformations}
\label{Prel_ent}
We start by reviewing two important sets regarding LOCC transformation, the source and the accessible set. Then, we recall the majorization lattice for bipartite states and some results regarding transformations among multipartite states.  \par
In Ref. \cite{Schwaiger2015} the following two sets associated to a state $\ket{\Psi} \in \mH_n$ were introduced,
\begin{align}
 &M_s(\Psi) = \{\ket{\Phi}\in \mH_n \ \mbox{s.t.} \ \ket{\Phi} \xrightarrow{LOCC} \ket{\Psi}\}, \nonumber\\
 &M_a(\Psi) = \{\ket{\Phi}\in \mH_n \ \mbox{s.t.} \ \ket{\Psi} \xrightarrow{LOCC} \ket{\Phi}\}. \label{eq:MaMs}
\end{align}
Note that we consider here, as throughout this paper, only nontrivial LOCC transformations (i.e. we exclude LU-transformations).
The set $M_s(\Psi)$ is referred to as the \emph{source set} of $\ket{\Psi}$ and contains all the states in $\mH_n$ (up to LUs) from which $\ket{\Psi}$ can be reached via LOCC. The set $M_a(\Psi)$ is referred to as the accessible set and   contains all the states in $\mH_n$ that can be reached from $\ket{\Psi}$ via LOCC. The volume of these sets can be used to define operational entanglement measures \cite{Schwaiger2015}. The source and the accessible set, and the corresponding measures, were calculated for bipartite and other few-body states \cite{Schwaiger2015, Sauerwein2015}. For generic multipartite states $\ket{\Psi}$ it holds that $M_s(\Psi) = M_a(\Psi) = \{\}$ since nontrivial LOCC tranformations are not possible \cite{Gour2017, Sauerwein2018}. Here, we use $M_s(\Psi)$ and $M_a(\Psi)$ to define new optimal protocols for bipartite pure state transformations in Section \ref{sec:bipex}.\\

Let us now review the concept of the majorization lattice, which we use in Section \ref{sec:bipex} to investigate LOCC transformations among bipartite pure states.
Recall first that we express a bipartite state $\ket{\Psi} \in \C^d \otimes \C^d$ in the Schmidt decomposition as $\ket{\Psi} = \sum_{i=1}^d \sqrt{\Psi_i} \ket{i,i}$, where $\Psi_i \geq \Psi_{i+1}$ and $\sum_i \Psi_i = 1$. The vector $\lambda(\Psi) = (\Psi_1,\ldots,\Psi_d)$ is then referred to as the Schmidt vector of $\ket{\Psi}$. For states $\ket{\Psi},\ket{\Phi} \in \C^d \otimes \C^d$, it holds that $\ket{\Psi} \xrightarrow{LOCC} \ket{\Phi}$ iff the Schmidt vector of $\ket{\Psi}$ is majorized by the Schmidt vector of $\ket{\Phi}$, i.e. iff $\lambda(\Psi) \prec \lambda(\Phi)$ \cite{Nielsen1999} (see also Eq. (\ref{eq:fewmonotones})). This majorization order induces a lattice structure, which was first investigated in Ref. \cite{Cicalese2002} by introducing the so-called \textit{majorization-lattice}, $\mathcal{L}_M$. This lattice is induced by the partial order relation defined by majorization. A lattice is a partially ordered set with the property that, for any two elements $\Psi, \Phi$ in the lattice, there exists a unique \textit{infinum} $\Psi \wedge \Phi$ and a unique \textit{supremum} $\Psi \vee \Phi$. In case of the majorization lattice, $\mathcal{L}_M$, the infimum fulfills $\Psi \wedge \Phi\prec \Psi$ and $\Psi \wedge \Phi \prec \Phi$ and for all $\xi \prec \Psi, \Phi$ it holds that $\xi  \prec  \Psi \wedge \Phi$. The supremum fulfills $\Psi \vee \Phi \succ \Psi $, $\Psi \vee\Phi\succ \Phi $ and for all $\chi$ with $ \chi \succ \Psi, \Phi$ it holds that $\chi \succ \Psi \vee \Phi$. The supremum and infinum for all bipartite pure states $\ket{\Psi}, \ket{\Phi} \in \C^d \otimes \C^d$ were explicitly constructed in Ref. \cite{Cicalese2002}, showing that majorization indeed induces a lattice structure on all states.
Note that the infinum is equivalent to the optimal common resource state for bipartite pure states introduced in Ref. \cite{Guo2016}. The supremum is the most entangled state that can be obtained from $\ket{\Psi}$ and $\ket{\Phi}$ deterministically. Moreover, the induced metric on $\mathcal{L}_M$ given by $d(\Psi,\Phi) = H(\Psi)+H(\Phi) - 2 H(\Psi \vee \Phi)$, with the Shannon entropy $H(\Psi) = -  \sum_{i=1}^d \Psi_i \ln(\Psi_i)$, fulfills for $\lambda(\Psi) \prec \lambda(\xi) \prec \lambda(\Phi)$ that $d(\Psi, \Phi) = d(\Psi, \xi) + d( \xi, \Phi)$.\\
Whenever a deterministic transformation from the initial state, $\ket{\Psi} \in \C^d \otimes \C^d$ to the final state, $\ket{\Phi} \in \C^d \otimes \C^d$, is not possible, the optimal success probability is given by the minimal ratio of the entanglement monotones $E_l$ (see Section \ref{sec:mainresults}). As already mentioned before, in the generic multipartite case, nontrivial LOCC transformations are not possible \cite{Gour2017,Sauerwein2018}. Moreover, the optimal (probabilistic) protocol has been shown to be an OSBP and the maximal success probability is given by Eq. (\ref{eq:optprob}). In the nongeneric case, possible LOCC transformations were characterized in e.g. Refs. \cite{Turgut2010, Kintas2010, DeVicente2013, Gour2011} and references therein.

\subsection{Survival analysis}
\label{sec:survival}
In this section we briefly review methods used in survival analysis (see e.g. \cite{Cox1984, Rodriguez2007}), which we employ in subsequent sections to study entanglement transformations.
In survival analysis, one is interested in the following scenario that has applications in many different fields of science. Suppose $T$ is a non-negative continuous random variable; for example the lifespan of a person, the lifetime of a component or of an isotope. Let $F(t) = P(T < t)$ denote the probability that $T$ is smaller than $t$, i.e. that the event (e.g. of death) has already occurred at time $t$, and let $f(t) = F'(t)$ denote the probability density of the random variable, $T$.
For our purposes it will be more convenient to consider the \emph{survival function} which is defined as \footnote{Note that we use here that $F(\infty) = 1$ (and hence $S(\infty) = 0$), i.e. that the event happens at some point. We comment on this at the end of the section.}
\bea
S(t) = P( T \geq t) = 1 - F(t) = \int_{t}^\infty f(s) ds. \label{eq:surv}
\eea
$S(t)$ gives the probability that the event has not yet occurred at time $t$; that is, it gives the probability of still being alive after time $t$.
The probability that the event occurs in the interval $[t,t+\Delta t]$, given that it did not happen up to time $t$, can then be expressed as
\begin{align}
P(t \leq T < t + \Delta t | T \geq t) &= \frac{P(t \leq T < t + \Delta t)}{P(T \geq t)} \nonumber\\
&= \frac{F(t + \Delta t) - F(t)}{S(t)}, \label{eq:interv1}
\end{align}
where we used Eq. (\ref{eq:surv}). The hazard rate associated to the random variable $T$ is defined as
\bea
h(t) = \lim_{\Delta t \rightarrow 0} \frac{P(t \leq T < t + \Delta t | T \geq t)}{\Delta t}. \label{eq:hazard0}
\eea
That is, the hazard rate, $h(t)$, is the rate at which the event happens at time $t$, given that it has not happend until time $t$.
Using Eq. (\ref{eq:interv1}) and the fact that $f(t) = F'(t)$ we see that $h(t)$ can be expressed as,
\bea
h(t) = \frac{f(t)}{S(t)}. \label{eq:rateequation}
\eea
Hence, the rate of occurrence of the event at time $t$ can be expressed as the density of events at $t$, divided by the probability of surviving to time $t$. Note that $f(t) = -S'(t)$ and thus, we can rewrite Eq.(\ref{eq:rateequation}) to obtain a differential equation for $S(t)$, namely
\bea
h(t) = -\frac{d}{dt} \log[S(t)].
\eea
Integrating this equation from $0$ to $t$ (note that this requires that $h(t)$ is integrable) we obtain
\bea
S(t) = \exp\left(-\int_{0}^t h(s) ds\right). \label{eq:Sint}
\eea
Hence, the probability to survive until time $t$, $S(t)$, can be expressed in terms of the hazard rate, and vice versa. That is, $S(t)$ and $h(t)$ are different, but equivalent, ways to describe the distribution of the random variable $T$. Note that the probability that the event occurs in the infinitesimal time interval $[t,t+dt]$ is $P(t \leq T < t + dt) = h(t) dt$. The \emph{cumulative hazard} at time $t$ is defined as
\bea
\Lambda(t) = \int_0^t h(s) ds, \label{eq:cumhazard}
\eea
and can be used to express the survival function as,
\begin{align*}
 S(t) = e^{-\Lambda(t)}.
\end{align*}

Note that in this section we assumed that the event described by the random variable $T$ has to occur at some point and hence $S(\infty) = 0$. However, the model described above can also be applied to more general scenarios where this is not the case (see e.g. \cite{Cox1984}).
In Section \ref{sec:contLOCC} we use the concepts and quantities defined above to study pure state (S)LOCC transformations.

\section{Optimal intermediate states and differentiable transformations}
\label{sec:theory}

In this section we introduce the concepts of intermediate states and use survival analysis to define differentiable transformations along a path in Hilbert space. We show how optimal intermediate states and optimal differentiable (S)LOCC transformations can be characterized via entanglement monotones. Furthermore, we show how optimal differentiable transformations can be found variationally.

\subsection{Optimal intermediate states}
\label{sec:defintstates}

Let us consider an (S)LOCC transformation from a state $\ket{\Psi} \in \mH_n$ to a state $\ket{\Phi} \in \mH_n$. We define an optimal intermediate state of this transformation as follows.
\begin{definition}
\label{def:intstate}
Let $\ket{\Psi}, \ket{\Phi} \in \mH_n$ be $n$-partite states and let $P(\Psi,\Phi)$ be the optimal probability to transform $\ket{\Psi}$ into $\ket{\Phi}$ via SLOCC.
A state $\ket{\chi} \in \mH_n$ is called an optimal intermediate state of the transformation from $\ket{\Psi}$ to $\ket{\Phi}$ if $\ket{\chi}$ is neither LU equivalent to $\ket{\Psi}$ nor to $\ket{\Phi}$ and
\bea
P(\Psi,\Phi) = P(\Psi,\chi)P(\chi,\Phi).
\eea
The set of all such states $\ket{\chi}$ is denoted by $\mathcal{I}(\Psi,\Phi)$.
\end{definition}
That is, $\ket{\chi} \in \mathcal{I}(\Psi,\Phi)$ iff one can first (nontrivially) transform $\ket{\Psi}$ into $\ket{\chi}$ and then $\ket{\chi}$ into $\ket{\Phi}$ and yet the resulting transformation from $\ket{\Psi}$ to $\ket{\Phi}$ is optimal.
If it is clear from the context which initial state $\ket{\Psi}$ and final state $\ket{\Phi}$ we consider, we often refer to a state $\ket{\chi} \in \mathcal{I}(\Psi,\Phi)$ as an optimal intermediate state (without mentioning $\ket{\Psi},\ket{\Phi}$). Note that there always exist non-optimal intermediate states of the transformation from $\ket{\Psi}$ to $\ket{\Phi}$; that is, states $\ket{\xi}$ s.t. $P(\Psi,\xi)P(\xi,\Phi) < P(\Psi,\Phi)$. In order to increase readability we refer to optimal intermediate states as intermediate states (without mentioning optimality) and explicitly mention it if we consider intermediate states that are not optimal.
After these definitions, some remarks are in order.\\
Firstly, note that we include here also deterministic transformations, for which $P(\Psi,\Phi) = 1$.
Secondly, it is, \emph{a priori}, not clear if an intermediate state exists for a given intial state $\ket{\Psi}$ and final state $\ket{\Phi}$. We show in Section \ref{sec:bipex} and Section \ref{sec:genmulti} that transformations among bipartite and generic multipartite states always have infinitely many intermediate states. However, in Section \ref{sec:noint} we also provide examples of nongeneric multipartite LOCC transformations for which the set of intermediate states is empty. This highlights once again the complexity of the entanglement of nongeneric multipartite states compared to the entanglement of bipartite states. Finally, note that the concept of intermediate states can be easily applied to other scenarios, e.g. transformations between states of different Hilbert spaces, non-entangling transformations \cite{Chitambar2017}, mixed state transformations or even other quantum resource theories \cite{Chitambar2018}.\\

Intermediate states can be characterized with the help of entanglement monotones. Recall that the optimal success probability to transform $\ket{\Psi} \in \mH_n$ into $\ket{\Phi} \in \mH_n$ can be expressed as \cite{Vidal2000mono},
\bea
P(\Psi,\Phi) = \min_\mu \frac{\mu(\Psi)}{\mu(\Phi)}, \label{eq:opttrans1}
\eea
where the minimization is to be taken over all entanglement monotones $\mu$. Using Eq. (\ref{eq:opttrans1}) we prove the following theorem.
\begin{theorem}
\label{thm:monotones1}
Let $\ket{\Psi}, \ket{\Phi} \in \mH_n$ be $n$-partite states and let $\mu$ be an entanglement monotone for which $P(\Psi,\Phi) = \frac{\mu(\Psi)}{\mu(\Phi)}$ holds.
A state $\ket{\chi}$ is an optimal intermediate state of the transformation from $\ket{\Psi}$ to $\ket{\Phi}$, i.e. $\ket{\chi} \in \mathcal{I}(\Psi,\Phi)$, iff $P(\Psi,\chi) = \frac{\mu(\Psi)}{\mu(\chi)}$ and $P(\chi,\Phi) = \frac{\mu(\chi)}{\mu(\Phi)}$.
\end{theorem}
\proof{The "if-part`` of the theorem is obvious. We show the "only if"-part indirectly. Let $\ket{\chi}$ be an intermediate state such that the success probability of one of the intermediate transformations cannot be expressed via the monotone $\mu$, i.e. w.l.o.g. $P(\Psi,\chi) < \frac{\mu(\Psi)}{\mu(\chi)}$. Using $P(\chi,\Phi) \leq \frac{\mu(\chi)}{\mu(\Phi)}$ (see Eq. (\ref{eq:opttrans1})) we then see that the following holds,
\begin{align*}
P(\Psi,\Phi) = \frac{\mu(\Psi)}{\mu(\Phi)} = \frac{\mu(\Psi)}{\mu(\chi)}\frac{\mu(\chi)}{\mu(\Phi)} > P(\Psi,\chi) P(\chi,\Phi).
\end{align*}
This is a contradiction to the definition of optimal intermediate states and hence $P(\Psi,\chi) = \frac{\mu(\chi)}{\mu(\Psi)}$ holds. \qed
}\\

Note that, for bipartite states and, as we will show here, generic multipartite states, the success probability in Eq. (\ref{eq:opttrans1}) can be expressed as a minimization over a subset of all entanglement monotones. Hence, in these cases, only this subset is needed for the characterization of intermediate states (see Section \ref{sec:IntBip} and Section \ref{sec:monotones}).\\

The concept of optimal intermediate states can be generalized straightforwardly to sequences of optimal intermediate states. A sequence $\chi^{(N)} = (\ket{\chi_i})_{i=0}^{N+1}$ of $N$ intermediate states in $\mH_n$ with $\ket{\chi_0} = \ket{\Psi}, \ket{\chi_{N+1}} = \ket{\Phi}$ is said to be a sequence of $N$ optimal intermediate states (of the optimal SLOCC transformation from $\ket{\Psi}$ to $\ket{\Phi}$) if $\ket{\chi_i} \in \mathcal{I}(\chi_{i-1},\chi_{i+1})$ for $i \in \{1, \ldots, N\}$. That is, $\chi^{(N)}$ provides one possible way to divide the transformation from $\ket{\Psi} \in \mH_n$ to $\ket{\Phi} \in \mH_n$ into $N+1$ optimal transformations via $N$ intermediate states.\\

In Secs. \ref{sec:bipex} and \ref{sec:multiex} we characterize optimal intermediate states for all bipartite and generic multiqudit transformations and provide applications of these results. In the following section we generalize the concept of a sequence of optimal intermediate states to the concept of differentiable paths of optimal intermediate states.

\subsection{Optimal paths and differentiable transformations}
\label{sec:contLOCC}

We introduce here the concepts of (S)LOCC paths and use tools from survival analysis to determine the success probability of the differentiable (S)LOCC transformation along these paths. We then define optimal SLOCC paths and show how they can be characterized.\\

\subsubsection{SLOCC paths}
\label{sec:pathdef}
We call a path $\{\ket{\psi(t)} = g(t)\ket{\Psi}\}_{0\leq t \leq 1}$ in the SLOCC class of a state $\ket{\Psi} \in \mH_n$ an \emph{SLOCC path} if it is (i) piecewise continuously differentiable and (ii) there is an integrable function $r_{\psi}(t)$ such that
\begin{align}
P[\psi(t),\psi(t+dt)] = 1 - r_{\psi}(t) dt \label{eq:rate}
\end{align}
for all $t \in [0,1)$.\\
The physical interpretation of the function $r_{\psi}(t)$ and the relevance of condition (ii) will become clear below, when we define SLOCC transformations along an SLOCC path. However, let us note here that condition (ii) is generically not a strong restriction on $\{\ket{\psi(t)}\}$. That is, all paths $\{\ket{\psi(t)}\}$ of bipartite or generic multiqudit states that fulfill condition (i) also fulfill condition (ii), as we show in Section \ref{sec:bipex} and Section \ref{sec:multiex}.\\

\subsubsection{Differentiable transformations}
\label{sec:difftrans}
For an SLOCC path $\{\ket{\psi(t)}\}_{0\leq t \leq 1}$ we define the \emph{SLOCC transformation along this path} as the transformation in which, starting at $\ket{\Psi} = \ket{\psi(t=0)}$, the state $\ket{\psi(t)}$ is optimally converted with probability $P[\psi(t),\psi(t+dt)]$ into the infinitesimally close state $\ket{\psi(t+dt)}$ for all $t \in [0,1)$ (see Fig. \ref{fig:survival}). The transformation from $\ket{\Psi}$ to $\ket{\Phi}$ via these infinitesimal steps along $\{\ket{\psi(t)}\}$ succeeds with a (in general non-optimal) probability $P[\psi]$. In the following we bridge the theory of SLOCC transformations with concepts used in survival analysis (see Section \ref{sec:survival}) to determine $P[\psi]$.\\

The central idea is the following. After the SLOCC path $\{\ket{\psi(t)}\}_{0\leq t \leq 1}$ has been fixed, the parameter $t \in [0,1]$ can be interpreted as a time variable (see Fig. \ref{fig:survival}). The transformation from $\ket{\Psi} = \ket{\psi(0)}$ to the state $\ket{\psi(t)}$ along $\{\ket{\psi(t)}\}$ can then be viewed as a continuous survival process in which the random variable is the lifetime $T$ of the entanglement. If the system is at time $t \in [0,1)$ of the transformation in the state $\ket{\psi(t)}$, i.e. the entanglement is still alive, one of the following two things can happen in the subsequent infinitesimal time interval $[t,t+dt]$: either the state $\ket{\psi(t)}$ is successfully transformed via the optimal protocol into the intermediate state $\ket{\psi(t+dt)}$, or the infinitesimal transformation fails and leads to a state that is no longer fully entangled \footnote{Recall that the failure branch of an optimal SLOCC protocol that transforms a fully entangled state $\ket{\Psi}$ into an SLOCC equivalent state $\ket{\Phi} = g\ket{\Psi}$ has to lead to a state that is no longer fully entangled, as otherwise the failure branch could still be transformed to $\ket{\Phi}$ with a non-vanishing probability. Note also that we include here the possibility of having several (un)successful branches.}. Put differently, the entanglement either survives the period $[t,t+dt]$ with probability $P[\psi(t),\psi(t+dt)]$, or it dies with probability $1-P[\psi(t),\psi(t+dt)]$. If we denote by $P(T\geq t)$ the probability that the entanglement is still alive at time $t$, the probability that the entanglement dies in the time interval $[t,t+dt]$, given that it has survived until time $t$ thus reads,
\begin{align}
 P(t < T \leq t+dt| T \geq t) &= 1-P[\psi(t),\psi(t+dt)] \nonumber \\
 &= r_{\psi}(t) dt. \label{eq:condprob}
\end{align}
Note that we used here Eq. (\ref{eq:rate}).\\

The hazard rate of this survival process (as defined in Eq. (\ref{eq:hazard0})) can then be expressed as
\begin{align}
 h_{\psi}(t) &= \lim_{\Delta t \rightarrow 0} \frac{P(t < T \leq t + \Delta t | T \geq t)}{\Delta t} , \nonumber\\
 &= \lim_{\Delta t \rightarrow 0} \frac{1 - P[\psi(t),\psi(t+\Delta t)]}{\Delta t} = r_{\psi}(t) \label{eq:hazard}
\end{align}
From this equation we can draw several conclusions.
Firstly, it shows that the function $r_{\psi}(t)$ defined in condition (ii) of the definition of an SLOCC path is equal to the hazard rate and thus has a clear operational meaning; namely as the rate at which the entanglement dies at time $t$ in the transformation along $\{\ket{\psi(t)}\}$, given that it has survived until time $t$. Hence, we have 
\begin{align}
 P[\psi(t),\psi(t+dt)] = 1 -h_{\psi}(t)dt. \label{eq:hazardinft}
\end{align}
Secondly, Eq.  (\ref{eq:hazard}) shows that $h_{\psi}(t) = r_{\psi}(t)$ is, due to the definition of an SLOCC path (see Section \ref{sec:pathdef}), an integrable function and thus the survival function $S_{\psi}(t) \equiv P(T\geq t)$ of the survival process can be expressed as (see Section \ref{sec:survival}),
\bea
S_{\psi}(t) = \exp\left(-\int_0^t h_{\psi}(s)ds\right). \label{eq:PfuncInt}
\eea
Since $S_{\psi}(t) \equiv P(T \geq t)$ is the probability that the transformation from $\ket{\Psi}$ to $\ket{\psi(t)}$ along the path suceeds, this provides a formula for $P[\psi] = S_{\psi}(1)$, namely
\begin{align}
 P[\psi] = e^{-\Lambda[\Psi]}, \label{eq:probfunc}
\end{align}
where $\Lambda[\Psi]$ is the cumulative hazard rate (see Eq. \ref{eq:cumhazard}), i.e. the functional
\begin{align}
 \Lambda[\psi] = \int_0^1 h_{\psi}(s)ds.
\end{align}

In the following we refer to an SLOCC path $\{\ket{\psi(t)}\}_{0\leq t \leq 1}$ as \emph{LOCC path} if $P[\psi] = 1$ (see also Ref.\ \cite{Schwaiger2018}).
Note that the function $S_{\psi}(t)$ is differentiable for all $t \in (0,1)$. Because of this, we refer to (S)LOCC transformations along (S)LOCC paths simply as \emph{differentiable (S)LOCC transformations} whenever the specific path is not relevant. Note further that $P[\psi] > 0$ means that the lifetime $T$ of the entanglement in the transformation along $\{\ket{\psi(t)}\}$ is not described by a properly normalized probability density. However, one can always artificially extend the path $\{\ket{\psi(t)} = g(t)\ket{\Psi}\}$ to times $t \in [0,\infty]$ in such a way that the probability to reach the state $\ket{\psi(t)}$ via a transformation along this path converges to zero for $t \rightarrow \infty$, i.e. $S_{\psi}(t) \rightarrow 0$ for $t \rightarrow \infty$ \footnote{Consider, for example, the transformation from $\ket{\Psi}$ into $\ket{\Phi}$ via $\{\ket{\psi(t)}\}$ and then back to $\ket{\Psi}$ along the same path (but now in reverse direction). The path that one obtains by repeating this ''loop`` infinitely many times has the property that $S_{\psi}(t) \rightarrow 0$ for $t \rightarrow \infty$.}. Then, the probability density describing $T$ is normalized and $S_{\psi}(t)$ is unchanged for $t \in [0,1]$. Furthermore, note that $P[\psi]$ can also be derived as a so-called product integral. We briefly discuss this alternative derivation in Appendix \ref{sec:altder}.
Finally, note that, like the concept of intermediate states, the concept of SLOCC transformations can be generalized straightforwardly to more general scenarios, e.g. mixed state transformations or other quantum resource theories.

\subsubsection{Transformations via optimal paths}
\label{sec:optpaths}
The probability to transform $\ket{\psi(0)}$ along an SLOCC path $\{\ket{\psi(t)}\}_{0 \leq t \leq 1}$ into $\ket{\psi(1)}$ clearly cannot exceed the optimal transformation probability $P[\psi(0),\psi(1)]$, i.e.
\begin{align}
P[\psi] \leq P[\psi(0),\psi(1)]. \label{eq:nonopt}
\end{align}
While Eq. (\ref{eq:nonopt}) is a strict inequality for most SLOCC paths, we show in Section \ref{sec:bipex} and Section \ref{sec:multiex} that there are, in many cases, SLOCC paths for which equality holds. We call an SLOCC path with this property an \emph{optimal SLOCC path} (or simply an \emph{optimal path}). In the following we discuss some further properties of these special paths and show how they can be characterized.\\

In analogy to the criterion in Theorem \ref{thm:monotones1} for the existance of optimal intermediate states, optimal SLOCC paths can be characterized in terms of entanglement monotones. In order to obtain this characterization, we first show how the hazard rate of a general SLOCC path can be expressed in terms of entanglement monotones. To this end, we use the formula $P(\Psi,\Phi) = \min_{\mu}\frac{\mu(\Psi)}{\mu(\Phi)}$ of Ref. \cite{Vidal2000mono} for the optimal success probability, where the minimization is performed over all entanglement monotones $\mu$. The hazard rate of an SLOCC path $\{\ket{\psi(t)}\}$ (see Eq. (\ref{eq:hazard})) can then be expressed as,
\begin{align}
 h_{\psi}(t) &= \lim_{\Delta t \rightarrow 0} \frac{1 - P[\psi(t),\psi(t+\Delta t)]}{\Delta t} \nonumber\\
 &= \lim_{\Delta t \rightarrow 0} \frac{1 - \min_{\mu} \frac{\mu[\psi(t)]}{\mu[\psi(t+\Delta t)]}}{\Delta t}. \label{eq:minim}
\end{align}
Since $\{\ket{\psi(t)}\}$ is an SLOCC path (see Section \ref{sec:pathdef}), the limit of $\Delta t \rightarrow 0$ is well-defined. This is the case iff the minimum in Eq. (\ref{eq:minim}) is attained by a monotone $\mu$ for which $\mu[\psi(t)]$ is differentiable at $t$ (in the following a monotone with this property is simply referred to as a differentiable monotone). Hence, we can restrict the minimization in Eq. (\ref{eq:minim}) to differentiable monotones and we obtain the following equation.
\begin{align}
 &h_{\psi}(t) = \lim_{\Delta t \rightarrow 0} \frac{1 - \min_{\mu, \text{diff.}} \frac{\mu[\psi(t)]}{\mu[\psi(t+\Delta t)]}}{\Delta t}, \label{eq:minim2}\\
 &= \lim_{\Delta t \rightarrow 0} \frac{1 - \min_{\mu, \text{diff.}}\left[1 - \Delta t \frac{\frac{d}{dt}\mu[\psi(t)]}{\mu[\psi(t)]} + O(\Delta t^2) \right]}{\Delta t}, \nonumber\\
 &= \max_{\mu, \text{diff.}} \frac{1}{\mu[\psi(t)]} \frac{d}{dt}\mu[\psi(t)]. \label{eq:maximization}
\end{align}
Equation (\ref{eq:maximization}) shows how the hazard rate of an SLOCC path can be expressed in terms of a maximization over all differentiable entanglement monotones \footnote{Note that the hazard rate in Eq. (\ref{eq:maximization}) can be zero. In this case the transformation from $\ket{\psi(t)}$ to $\ket{\psi(t+dt)}$ is deterministic.}. Using this result we obtain the following characterization of optimal paths in terms of entanglement monotones.
\begin{theorem}
\label{thm:pathcar}
 Let $\{\ket{\psi(t)}\}_{0 \leq t \leq 1}$ be an SLOCC path and let $\mu$ be an entanglement monotone for which $P[\psi(0),\psi(1)] = \frac{\mu[\psi(0)]}{\mu[\psi(1)]}$ holds. Then the following conditions are equivalent:
 \begin{itemize}
  \item[(i)] $\{\ket{\psi(t)}\}$ is an optimal SLOCC path.
  \item[(ii)] For all $t \in (0,1)$ the success probability
  \begin{align}
   P[\psi(t),\psi(t+dt)] = 1 - h_{\psi}(t)dt, \label{eq:inftprob}
  \end{align}
  is given by
  \begin{align}
   h_{\psi}(t) = \frac{1}{\mu[\psi(t)]}\frac{d}{dt}\mu[\psi(t)]. \label{eq:hazardthm}
  \end{align}
  \item[(iii)] For all $t \in (0,1)$ the following holds,
\begin{align}
 \frac{1}{\mu[\psi(t)]}\frac{d}{dt}\mu[\psi(t)] = \max_{\tilde{\mu}, \text{diff.}} \frac{1}{\tilde{\mu}[\psi(t)]}\frac{d}{dt}\tilde{\mu}[\psi(t)], \label{eq:stronggrow}
\end{align}
where the maximization is performed over all differentiable monotones.
  \item[(iv)] $P[\psi(t_1),\psi(t_2)] = \frac{\mu[\psi(t_1)]}{\mu[\psi(t_2)]}$ for all $0 \leq t_1 < t_2 \leq 1$.
  \item[(v)] $\ket{\psi(t_1)} \in \mathcal{I}[\psi(0),\psi(t_2)]$ for all $0 \leq t_1 < t_2 \leq 1$.
 \end{itemize}
\end{theorem}
Note that this shows, in particular, that an optimal SLOCC path from $\ket{\Psi}$ to $\ket{\Phi}$ can only exist if the success probability $P(\Psi,\Phi) = \frac{\mu(\Psi)}{\mu(\Phi)}$ is given by a
monotone $\mu$ that is differentiable along a path connecting these two states. Moreover, (iii) shows that a path is optimal iff there is one monotone $\mu$ that has for all times, $t$, the largest relative growth along this path (among all entanglement monotones).

\proof{
(ii) $\Leftrightarrow$ (iii): This follows directly from Eq. (\ref{eq:maximization}).\\
(iv) $\Leftrightarrow$ (v): This follows directly from the characterization of optimal intermediate states given in Theorem \ref{thm:monotones1}.\\
(i) $\Rightarrow$ (v): Clearly, an SLOCC path $\{\ket{\psi(t)}\}_{0 \leq t \leq 1}$ can only be optimal if, for any $0 \leq t_1 < t_2 \leq 1$, the transformation along this path from $\ket{\psi(0)}$ to $\ket{\psi(t_1)}$ and then to $\ket{\psi(t_2)}$ is still optimal, i.e. if (iv) holds.\\
(iv) $\Leftrightarrow$ (ii): Equation (\ref{eq:inftprob}) holds because $\{\ket{\psi(t)}\}$ is an SLOCC path (see Eq. (\ref{eq:hazardinft})). If we insert $P[\psi(t_1),\psi(t_2)]$ given in (iv) into the definition of the hazard rate in Eq. (\ref{eq:hazard}), we find that it is, indeed, given by Eq. (\ref{eq:hazardthm}).\\
(ii) $\Rightarrow$ (i): If we insert the expression for the hazard rate in Eq. (\ref{eq:hazardthm}) into the formula for $P[\psi]$ (see Eq. (\ref{eq:probfunc})) we find the following.
\begin{align*}
&P[\psi] = \exp\left(-\int_{0}^1 \frac{1}{\mu[\psi(t)]}\frac{d}{dt}\mu[\psi(t)] dt\right) \\
&= \frac{\mu[\psi(0)]}{\mu[\psi(1)]} = P[\psi(0),\psi(1)].
\end{align*}
Hence, $\{\ket{\psi(t)}\}$ is an optimal path. \qed \\
}

Optimal paths can also be found using a variational approach. To see this, recall that a path $\{\ket{\psi(t)}\}_{0 \leq t \leq 1}$ connecting $\ket{\Psi} = \ket{\psi(0)}$ with $\ket{\Phi} = \ket{\psi(1)}$ is optimal iff it attains the maximum possible value of the probability functional $P[\psi]$, i.e. iff $P[\psi] = P(\Psi,\Phi)$. Equivalently, the path $\{\ket{\psi(t)}\}$ is optimal iff it minimizes the cumulative hazard rate (see Eq. (\ref{eq:cumhazard})) of the survival process corresponding to the differentiable transformation along $\{\ket{\psi(t)}\}$, i.e. the functional
\bea
\Lambda[\psi] = \int_0^1 h_{\psi}(s)ds, \label{eq:hazard1}
\eea
such that $\Lambda[\psi] = -\log[P(\Psi,\Phi)]$ holds.
Hence, if we denote by $\mathcal{Q}(\Psi,\Phi)$ all SLOCC paths connecting $\ket{\Psi}$ and $\ket{\Phi}$, we get the following characterization of all optimal paths.
\begin{observation}
\label{obs:var}
An SLOCC path $\{\ket{\psi(t)} = g(t)\ket{\Psi}\}_{0\leq t \leq 1}$ from $\ket{\Psi} = \ket{\psi(0)}$ to $\ket{\Phi} = \ket{\psi(1)}$ is optimal iff it minimizes the cumulative hazard rate s.t.
\bea
\Lambda[\psi] = -\log[P(\Psi,\Phi)] = \min_{\gamma \in \mathcal{Q}(\Psi,\Phi)} \Lambda[\gamma]. \label{eq:functional3}
\eea
\end{observation}
In Section \ref{sec:multiex} we use Observation \ref{obs:var} to characterize all optimal paths for transformations of generic multipartite states. In particular, we show how the minimization of the functional in Eq. (\ref{eq:functional3}) leads to a differential equation that is satisfied by a differentiable path iff it is an optimal path. This variational approach is reminiscent of variational problems in other fields of physics and mathematics. In the following subsection we discuss a strong analogy with \emph{Fermat's principle} in classical optics.\\

\subsubsection{Analogy with Fermat's principle}
\label{sec:fermat}
Consider an anisotropic optical medium whose refractive index $n_{\vec{e}}(\vec{x})$ depends on the position $\vec{x}$ and on the direction $\vec{e}$ at which a light ray is incident. \emph{Fermat's principle} then states that, among all paths in this medium connecting an inital point $\vec{\gamma}(0)$ with a final point $\vec{\gamma}(1)$, a light ray follows a path that can be traversed in the least time (see e.g. \cite{Lakshminarayanan2002}). That is, it follows the path $\{\vec{\gamma}(t)\}_{0\leq t \leq 1}$ of least optical length $L_o[\vec{\gamma}]$, which is defined as,
\begin{align}
 L_o[\vec{\gamma}] = \int_{0}^{1} n_{\frac{d}{dt}\vec{\gamma}(t)}[\vec{\gamma}(t)] dt.
\end{align}
Comparing this with Observation \ref{obs:var} we see that differentiable SLOCC paths (optimal and non-optimal) can be viewed as paths in an anisotropic optical medium whose refractive index is given by the hazard rate. In this picture the cumulative hazard rate associated to a path is its optical length (see Eq. (\ref{eq:hazard1})),  and optimal paths are those paths that minimize this optical length (see Observation \ref{obs:var}).\\

The variational approach to finding optimal SLOCC paths is also in analogy to the problem of finding geodesics in curved spaces, which we discuss in the following section.

\section{An operational metric on entangled quantum states}\hfill\\
\label{sec:metric}

In this section we define an operational metric on quantum states that is induced by their interconversion properties. As we explain below, this metric fulfills many properties that are natural to demand of a function that measures the distance of quantum states in terms of how difficult it is to interconvert them. We further show that, in many cases, optimal SLOCC paths can in fact be viewed as geodesics with respect to this metric. This provides an operational connection between geometry and state transformations and opens the possibility to use tools from geometry to study optimal SLOCC transformations.\\

The metric is defined as,
\begin{align}
d_I:& \mH_n \times \mH_n \rightarrow [0,\infty]: \nonumber\\
&(\Psi,\Phi) \mapsto d_I(\Psi,\Phi) =  -\log[P(\Psi,\Phi)P(\Phi,\Psi)]. \label{eq:defdist}
\end{align}
We call $d_I$ the \emph{interconversion metric} as it quantifies the difficulty of interconverting the states $\ket{\Psi}$ and $\ket{\Phi}$ via SLOCC. The generalization to mixed states or other resource theories is straightforward. Note that $d_I$ is a metric on the set of all LU-equivalence classes, i.e. it fulfills the following conditions for all  $\ket{\Psi},\ket{\Phi},\ket{\Omega} \in \mH_n$:
\begin{itemize}
 \item[(i)] $d_I(\Psi,\Phi) \geq 0$
 \item[(ii)] $d_I(\Psi,\Phi) = 0$ iff $\ket{\Psi}$ and $\ket{\Phi}$ are LU-equivalent.
 \item[(iii)] $\ d_I(\Psi,\Phi) = d_I(\Phi,\Psi$),
 \item[(iv)] $d_I(\Psi,\Phi) \geq d_I(\Psi,\Omega) + d_I(\Omega,\Phi)$.
\end{itemize}
It is easy to see that (i) and (iii) hold. In order to see (ii) note that it was shown in Ref.  \cite{Gingrich2002} that $P(\Psi,\Phi)P(\Phi,\Psi) = 1$ iff $\ket{\Psi}$ is LU equivalent to $\ket{\Phi}$. Note further that (iv) is equivalent to the inequality $P(\Psi,\Phi)P(\Phi,\Psi) \geq P(\Psi,\Omega)P(\Omega,\Phi)P(\Phi,\Omega)P(\Omega,\Psi)$, which is obviously fulfilled.\\

The interconversion metric has several interesting properties. Firstly, two states are at zero distance to each other iff they are LU equivalent (see (ii)), i.e. iff they are indiscernible in terms of their entanglement properties. Secondly, $d_I(\Psi,\Phi) = \infty$ iff $\ket{\Psi}$ and $\ket{\Phi}$ are not SLOCC equivalent, reflecting the fact that these two states then contain different kinds of entanglement (see Section \ref{Prel_ent}). Note that, e.g., the trace distance does not fulfill these properties.\\
The interconversion metric can also be used to construct entanglement measures. Indeed, for bipartite states $\ket{\Psi} \in \C^d \otimes \C^d$ the function $E_{\phi^+}(\Psi) \equiv - d_I(\phi^+,\Psi) = -\log[P(\Psi,\phi^+)]$, where $\ket{\phi^+} \propto \sum_{i=1}^{d}\ket{ii}$ is the maximally entangled state, is nonincreasing under LOCC and hence an entanglement measure (see Section \ref{Prel_ent}). Moreover, for a generic multipartite state $\ket{\chi} \in \mH_{n,d}$ the function $E_{\chi}(\Psi) = -d_I(\chi,\Psi)$ is an entanglement measure within the SLOCC class of $\ket{\chi}$. This follows from the fact that any LU-invariant function is an entanglement measure within the SLOCC class of a state with trivial stabilizer \cite{Sauerwein2018}.\\

Due to its clear operational meaning we expect that the interconversion metric (and its generalizations) will be useful in investigations of resource theories in general. Here, we show how it can be used to establish a connection between optimal SLOCC paths and geometry.

\subsection{Optimal paths and geodesics}
\label{sec:geodesic}
If a state $\ket{\Psi}$ can be transformed into $\ket{\Phi}$ via an optimal path $\{\ket{\psi_1(t)}\}$ and back to $\ket{\Psi}$ via another optimal path $\{\ket{\psi_2(t)}\}$ it is easy to see that $d_I(\Psi,\Phi)$ can be expressed in terms of the cumulative hazard rate (see Eq. (\ref{eq:hazard1})) as,
\bea
d_I(\Psi,\Phi) = \Lambda[\psi_1] + \Lambda[\psi_2].
\eea
That is, $d_I(\Psi,\Phi)$ can be interpreted as the minimal cumulative hazard rate (i.e. the minimal risk) one has to take if one wants to transform $\ket{\Psi}$ via an optimal path into $\ket{\Phi}$ and back to $\ket{\Psi}$ again. In the following we work out the connections between optimal paths and the interconversion metric in more detail.\\

Let us denote by $\{\ket{\psi_R(t)} = \ket{\psi(1-t)}\}_{0\leq t \leq1}$ the path which we obtain by going along the path $\{\ket{\psi(t)}\}$ in the reverse direction. We show in Appendix \ref{app:geodesic} that the length of an SLOCC path $\{\ket{\psi(t)}\}$ in the interconversion metric, $d_I$, can then be defined as
\begin{align}
 L[\psi] \equiv \int_{0}^1 ds(t), \label{eq:length1}
\end{align}
where $ds(t) = d_I[\psi(t),\psi(t+dt)] = [h_{\psi}(t) + h_{\psi_R}(1-t)]dt$ is the distance between the states $\ket{\psi(t)}$ and $\ket{\psi(t+dt)}$. Using the definition of the cumulative hazard rate in Eq. (\ref{eq:hazard1}), we can express Eq. (\ref{eq:length1}) as,
\begin{align}
 L[\psi] \equiv \Lambda[\psi] + \Lambda[\psi_R]. \label{eq:lengthfunct}
\end{align}
Note that the length of a path is, of course, independent of the direction in which it is traversed, i.e. $L[\psi] = L[\psi_R]$ holds (see also Appendix \ref{app:geodesic}).\\

The length functional in Eq. (\ref{eq:lengthfunct}) can now be used to characterize geodesics in the interconversion metric. Here, we are interested in so-called \emph{minimal geodesics}, which we define below, and their connection to optimal SLOCC paths.

Let us denote, as before, the set of all SLOCC paths connecting $\ket{\Psi}$ with $\ket{\Phi}$ by $\mathcal{Q}(\Psi,\Phi)$. A path $\{\ket{\psi(t)}\} \in \mathcal{Q}(\Psi,\Phi)$ is called a \emph{minimal geodesic} if it has the minimal length among all paths in $\mathcal{Q}(\Psi,\Phi)$ \footnote{Note that there can be geodesics that do not have minimal length (see e.g. Ref. \cite{burago2001}).}, i.e. if
\begin{align}
\label{eq:functional2}
L[\psi] = \min_{\gamma \in \mathcal{Q}(\Psi,\Phi)} (\Lambda[\psi] + \Lambda[\psi_R]).
\end{align}
Note that this definition is very similar to the variational characterization of optimal SLOCC paths in Observation \ref{obs:var}. Indeed, there are strong connections between these two variational problems, as we show in the following.\\

Clearly, the length of a minimal geodesic $\{\ket{\psi(t)}\}_{0\leq t \leq 1} \in \mathcal{Q}(\Psi,\Phi)$ is at least as large as the distance between $\ket{\Psi}$ and $\ket{\Phi}$, i.e.
\begin{align}
L[\psi] \geq d_I(\Psi,\Phi). \label{eq:ineqlength}
\end{align}
If equality holds in Eq. (\ref{eq:ineqlength}), the path $\{\ket{\psi(t)}\}$ is called a \emph{distance minimizing geodesic}. Using Eq. (\ref{eq:functional2}) and the definition $d_I(\Psi,\Phi) = -\log[P(\Psi,\Phi)P(\Phi,\Psi)]$ of the interconversion metric, we find that $\{\ket{\psi(t)}\} \in \mathcal{Q}(\Psi,\Phi)$ is a distance minimizing geodesic iff the following holds,
\begin{align}
\label{eq:var3}
-\log[P(\Psi,\Phi)] - \log[P(\Phi,\Psi)] = \min_{\gamma \in \mathcal{Q}(\Psi,\Phi)} (\Lambda[\psi] + \Lambda[\psi_R]).
\end{align}
It follows from the variational characterization of optimal SLOCC paths given in Observation \ref{obs:var} that Eq. (\ref{eq:var3}) is fulfilled iff $\{\ket{\psi(t)}\}$ and its reverse path, $\{\ket{\psi_R(t)}\}$, are optimal paths. We call a path with this property a \emph{twofold optimal path}. We have just proved the following observation.
\begin{observation}
\label{obs:geodesic}
 Let $\ket{\Psi},\ket{\Phi} \in \mH_n$ be $n$-partite states.  An SLOCC path $\{\ket{\psi(t)}\} \in \mathcal{Q}(\Psi,\Phi)$ is a distance minimizing geodesic iff it is twofold optimal, i.e. iff it is optimal and its reverse path, $\{\ket{\psi_R(t)}\} \in \mathcal{Q}(\Phi,\Psi)$, is optimal too.
\end{observation}
Clearly, if there exists one distance minimizing geodesic in $\mathcal{Q}(\Psi,\Phi)$, then any minimal geodesic has to be distance minimizing. Because of Observation \ref{obs:geodesic} this then implies that every minimal geodesic is twofold optimal.\\

Note that it is, \emph{a priori}, not clear whether a distance minimizing geodesic between two states exists \footnote{It is, in general, not even clear if a minimal geodesic exists.}. However, we show in this work that  bipartite states (see Section \ref{sec:bipartitepaths}) and generic multipartite states (see Section \ref{sec:geodgeneric}) can always be connected by a distance minimizing geodesic \footnote{Let us note that this shows that the metric space of LU classes in a bipartite and generic multipartite SLOCC class endowed with the interconversion metric is a so-called \emph{length space} (see e.g. Ref. \cite{burago2001} for the precise definition).}. That is, there always exists an optimal path for which the reverse path is also optimal. In Section \ref{sec:geodgeneric} we show that an even stronger result holds for generic $(n>4)$-qubit states, for which the sets of optimal paths, of twofold optimal paths and of minimal geodesics are identical.\\

These results reveal a strong connection between geometry and optimal (S)LOCC transformations. They provide a new perspective on SLOCC transformations and show that tools from geometry can be used to investigate them. Furthermore, they show the operational significance of the interconversion metric.

\section{Bipartite case}
\label{sec:bipex}
In this section we consider deterministic as well as probabilistic transformations among bipartite states. We characterize all optimal intermediate states and show that for deterministic transformations via LOCC this characterization can be done with the help of the \textit{majorization lattice}. Moreover, we show that the known optimal intermediate state introduced in Ref. \cite{Vidal1999}, which optimizes the fidelity with the final state, is not unique and can be easily generalized. Furthermore, we give a simple example showing that the optimal intermediate states do not necessarily optimize the fidelity with the final state.
We apply methods from survival analysis to characterize all optimal paths connecting two bipartite states $\ket{\Psi}, \ket{\Phi} \in \C^d \otimes \C^d$. Furthermore, we identify two paths along states with prominent entanglement properties. These are the \textit{most entangled} and the \textit{least entangled} path. The former consists of an OSBP path from $\ket{\Psi}$ to the most entangled state (in terms of all entanglement monotones $E_l$) that can still be  obtained with probability $P(\Psi, \Phi)$ and a subsequent LOCC path from this most entangled state to the final state $\ket{\Phi}$. The latter starts with an LOCC path from $\ket{\Psi}$ to the least entangled state, from which one can still obtain the final state $\ket{\Phi}$ optimally with the probability $P(\Psi, \Phi)$ on an OSBP path. Finally, we show that any two bipartite states can be connected via a twofold optimal SLOCC path. This shows that the sets of minimal geodesics in the interconversion metric and of twofold optimal paths are identical in the bipartite case. However, we also show that, for bipartite $(d>2)$-level systems, there are optimal paths that are not twofold optimal.

 \subsection{Optimal intermediate states}
 \label{sec:IntBip}
We present here a simple characterization of the set of all optimal intermediate states of two bipartite states $\ket{\Psi}, \ket{\Phi}$, i.e. of $\mathcal{I}(\Psi, \Phi)$. \par
 Recall, that $P (\Psi, \Phi) = P (\Psi, \chi) P (\chi, \Phi)$  has to hold for all optimal intermediate states $\ket{\chi}$. Lemma \ref{thm:monotones1} simplifies in the bipartite case to the following corollary.

\begin{corollary}
\label{cor:optIntBip}
Let $\ket{\Psi}, \ket{\Phi} \in \C^d \otimes \C^d$ with $P(\Psi,  \Phi) = \min_i \frac{E_i(\Psi)}{E_i(\Phi)} = \frac{E_l(\Psi)}{E_l(\Phi)}$. A state $\ket{\chi} \in \C^d \otimes \C^d$  is an optimal intermediate state, i.e. $\ket{\chi} \in \mathcal{I}(\Psi, \Phi)$, iff
\begin{equation}
\label{optIntBip}
P (\Psi, \chi) = \frac{E_l(\Psi)}{E_l(\chi)} \ \text{and} \ P (\chi, \Phi) = \frac{E_l(\chi)}{E_l(\Phi)}.
\end{equation}
\end{corollary}
Hence, by knowing $l$ such that $P(\Psi,\Phi)=\frac{E_l(\Psi)}{E_l(\Phi)}$ one can readily check whether a state is an optimal intermediate state, or not. Moreover, this characterization of all optimal intermediate states enables us to restrict the area of possible optimal paths, as any state on the path needs to be an optimal intermediate state of $\ket{\Psi}$ and $\ket{\Phi}$. Using Corollary \ref{cor:optIntBip} for $\ket{\Psi}, \ket{\Phi}$ such that $P(\Psi,  \Phi) = \frac{E_l(\Psi)}{E_l(\Phi)}$ the set of optimal intermediate states $\mathcal{I}(\Psi, \Phi)$ can be characterized via the following inequalities in terms of the entanglement monotones; $\ket{\chi} \in \mathcal{I}(\Psi, \Phi)$ iff
\begin{eqnarray}
\frac{E_l(\Psi)}{E_l(\chi)} \leq \frac{E_k(\Psi)}{E_k(\chi)}, \ \
\frac{E_l(\chi)}{E_l(\Phi)} \leq \frac{E_k(\chi)}{E_k(\Phi)}  \ \ \forall k.
\end{eqnarray}
This can be easily verified using that $P(\Psi, \chi) = \min_k  \frac{E_k(\Psi)}{E_k(\chi)}$ and similarly for $P( \chi, \Phi)$. \par

For deterministic transformations the optimal intermediate states can also be characterized with the help of the \textit{majorization lattice} $\mathcal{L}_M$, that can be defined due to the lattice structure of majorization (see. e.g. Ref. \cite{Cicalese2002} and Section\ \ref{Prel_ent} for the definition of the majorization lattice.). More precisely, for two states $\ket{\Psi}$ and $\ket{\Phi}$ fulfilling the majorization condition $\lambda(\Psi)\prec  \lambda(\Phi)$, all states $\ket{\chi}$ defined by the interval $[\Psi, \Phi] = \{ \lambda(\chi) \in \mathcal{L}_M : \lambda(\Psi)\prec \lambda(\chi) \prec \lambda(\Phi) \}$ are optimal intermediate states of $\ket{\Psi}$ and $\ket{\Phi}$. It can be easily seen that the interval is nonempty iff $\lambda(\Psi) \prec \lambda(\Phi)$.  Moreover, this nonempty interval is a sublattice of the majorization lattice $\mathcal{L}_M$. For any $\chi_1, \chi_2 \in [\Psi, \Phi]$ the supremum and infinum are also elements of this sublattice, i.e. $\chi_1 \vee \chi_2, \chi_1 \wedge \chi_2 \in [\Psi, \Phi]$. This follows simply from the fact that for $\chi_1, \chi_2 \in [\Psi, \Phi]$, $\lambda(\Psi) \prec \lambda(\chi_1), \lambda(\chi_2) \prec \lambda(\Phi)$. Thus, the supremum fulfills $\lambda(\Psi) \prec \lambda(\chi_1), \lambda(\chi_2)  \prec \lambda(\chi_1 \vee \chi_2) \prec \lambda(\Phi)$, as it is the "most entangled" state (measured in terms of all entanglement monotones $E_l$) that can be obtained from both states $\ket{\chi_1}, \ket{\chi_2}$. Similarily, for the infinum $\lambda(\Psi) \prec \lambda(\chi_1 \wedge \chi_2) \prec \lambda(\chi_1), \lambda(\chi_2) \prec \lambda(\Phi)$ holds. Therefore, the sublattice defined by the interval $[\Psi, \Phi]$ includes all possible intermediate states and hence, all possible LOCC paths. \par

Let us consider now optimal intermediate states for probabilistic transformations. We begin with the known intermediate state introduced in Ref. \cite{Vidal1999}. We show later that all of the optimal protocols discussed below can indeed be written as continuous SLOCC paths.
In Ref. \cite{Vidal1999}  the following optimal protocol for two bipartites states $\ket{\Psi}, \ket{\Phi}$ with $\lambda(\Psi) \not\prec \lambda(\Phi)$ was presented. This protocol achieving $P(\Psi, \Phi)$  consists of a deterministic transformations of $\ket{\Psi}$ into a certain intermediate state $\ket{\xi}$ via LOCC and a probabilistic transformation from $\ket{\xi}$ to the final state $\ket{\Phi}$ via an OSBP.  It was shown in Ref. \cite{Vidal2000} that, among all possible LOCC transformations of $\ket{\Psi} $ into an ensemble of mixed states, i.e. $\{p_k, \rho_k\}$, the maximal average fidelity with respect to the final state $\ket{\Phi}$, i.e. $\sum p_k \bra{\Phi} \rho_k \ket{\Phi}$, is obtained by the pure state $\ket{\xi}$ (for a review of the definition of the state $\ket{\xi}$ obtained in Ref. \cite{Vidal1999} see also Appendix \ref{app:bipartiteoptFi}). In particular, the pure state fidelity is then given by $F(\xi, \Phi) = \abs{\braket{\xi| \Phi}}^2 = \sqrt{1-d_{tr}(\xi, \Phi) ^2}$, with $d_{tr}$ the trace distance.\\

\begin{figure}[H]
   \centering
   \includegraphics[width=0.4\textwidth]{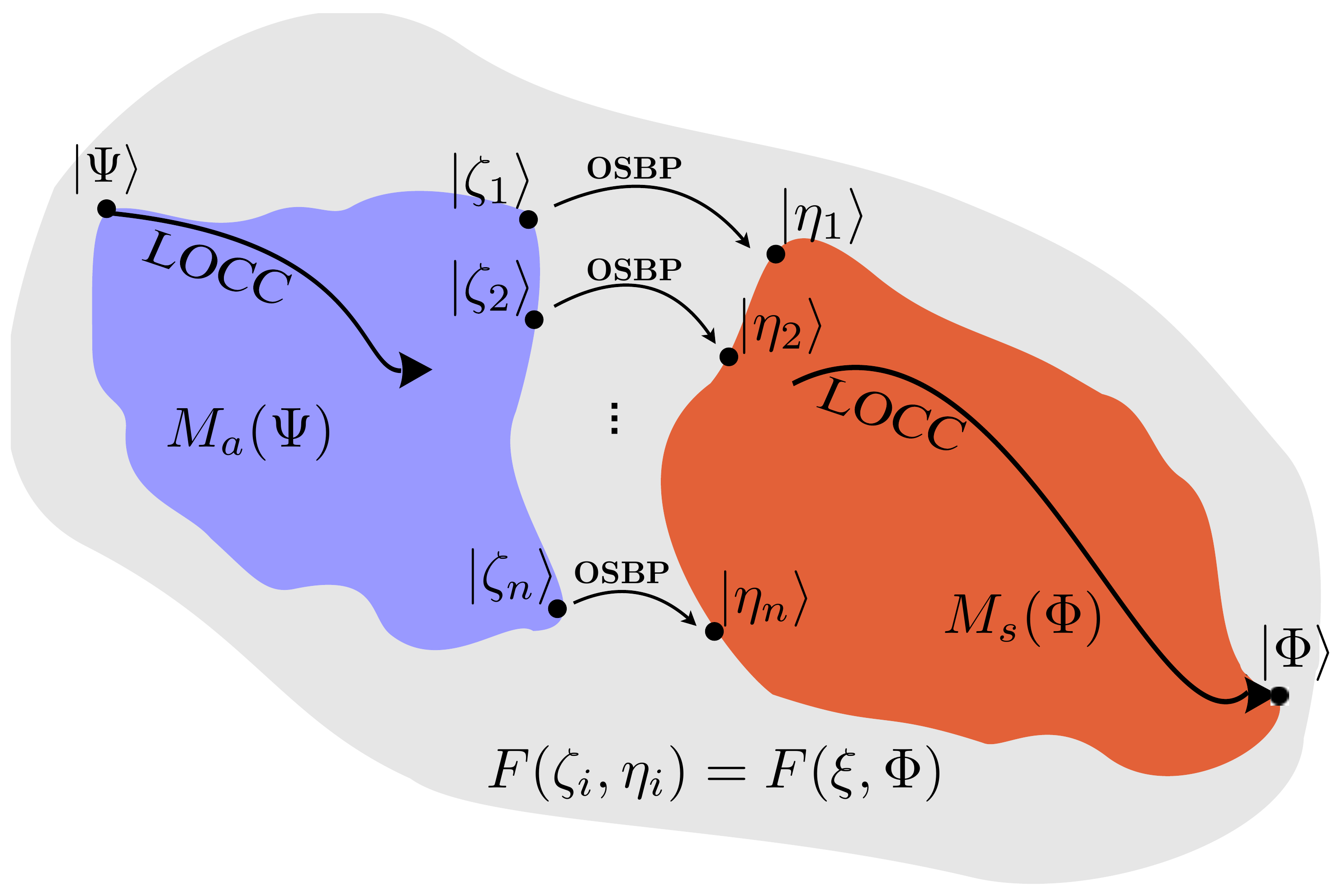}
   \caption{(color online). Schematic figure of the accessible set (blue) and the source set (red) of two states $\ket{\Psi}$ and $\ket{\Phi}$ in $\C^d \otimes \C^d$. Each state $\ket{\zeta_i}$ in the accessible set of $\ket{\Psi}$ optimally reaches the corresponding state $\ket{\eta_i}$ in the source set of $\ket{\Phi}$ via an OSBP for all $i$. Moreover, the fidelity $F(\zeta_i, \eta_i)$ is equal to the optimal fidelity $F(\xi, \Phi)$ for all $i$ and thus, these protocols are a natural generalization of the one introduced in Ref. \cite{Vidal1999}.   }
   \label{VaVsFi}
   \end{figure}

Considering the accessible set of $\ket{\Psi}$, $M_a(\Psi)$, and the source set of $\ket{\Phi}$, $M_s(\Phi)$, (see Fig.\ \ref{VaVsFi}) it seems natural that an optimal protocol starts with a transformations of $\ket{\Psi}$ to a state which is "as close as possible" to the source set, such that the probabilistic step is between close states. In fact, we show here that the optimal protocol from Ref. \cite{Vidal1999} can be easily generalized by using this idea. We explicitly construct certain states $\ket{\zeta} \in M_a(\Psi)$ and $\ket{\eta} \in M_s(\Phi)$, such that again the fidelity between them is optimized, i.e. 
\begin{equation}
F(\zeta, \eta) = \max_{\zeta' \in M_a(\Psi), \eta' \in M_s(\Phi)} F(\zeta', \eta'). 
\label{fidelity}
\end{equation}
For these states it can be easily seen that the protocol (see also Fig.\ \ref{VaVsFi})
\begin{equation}
\ket{\Psi} \xrightarrow{LOCC} \ket{\zeta} \xrightarrow{OSBP} \ket{\eta} \xrightarrow{LOCC} \ket{\Phi}
\label{optBipSLOCC}
\end{equation}
is optimal (see Appendix \ref{app:bipartiteoptFi} for details).

Note that there exist, generically, infinitely many pairs of intermediate states $\ket{\zeta}, \ket{\eta}$ fulfilling the relations stated above and $P (\zeta, \eta) = P (\Psi, \Phi)$, see also Fig.\ \ref{VaVsFi}. That these states allow for an optimal transformation is shown by explicit construction. We refer the reader to Appendix \ref{app:bipartiteoptFi} for the details. Thus, there exist infinitely many probabilistic protocols that achieve the optimal success probability for a transformation from $\ket{\Psi}$ to $\ket{\Phi}$ and for which the optimal intermediate states $\ket{\zeta}, \ket{\eta}$ optimize the fidelity, see Eq.\ \eqref{fidelity}.
    \par

The results stated above might suggest that there is a strong connection between the optimal success probability and the fidelity of the optimal intermediate states. However, we give now a simple counterexample of a different optimal transformation. The protocol includes a transformation from $\ket{\Psi}$ to the supremum state $\ket{\xi}_{sup} = \Psi \vee \Phi$ and is given by
\begin{equation}
\ket{\Psi} \xrightarrow{LOCC} \ket{\xi}_{sup} \xrightarrow{OSBP} \ket{\Phi}.
\end{equation} 
As shown in Ref. \cite{Bosyk2017}, the supremum state optimizes the distance on the majorization lattice in terms of the Shannon entropies with the target states, i.e. $d(\xi_{sup}, \Phi) = \min_{\xi' \in M_a(\Psi)} d(\xi', \Phi)$ (where $d(\Psi,\Phi)$ was defined in Section \ref{Prel_ent}), but it does not optimize the fidelity with the target state. Furthermore, it was shown in Ref.  \cite{Bosyk2017} that the relations
\begin{equation}
\lambda(\Phi) \prec \lambda(\xi_{sup}) \prec \lambda(\xi)
\end{equation}
hold. Hence, it is clear that the success probability $P(\xi_{sup}, \Phi)$ is equal to $P(\Psi, \Phi)$, as one could simply deterministically go from $\ket{\xi}_{sup}$ to $\ket{\xi}$. However, we can also easily find examples of a optimal protocol consisting of a direct OSBP from $\ket{\xi}_{sup}$ to $\ket{\Phi}$.
For this purpose let us first state the following observation.

\begin{observation}
A bipartite state $\ket{\Psi} \in \C^d \otimes \C^d$ with Schmidt vector $\lambda(\Psi) = (\Psi_1,\ldots,\Psi_d)$ can be optimally transformed via an OSBP into another bipartite state $\ket{\Phi} \in \C^d \otimes \C^d$  with Schmidt vector $\lambda(\Phi) = (\Phi_1, \ldots, \Phi_d)$ and with $\frac{\Psi_d}{\Phi_d}  \leq \frac{\Psi_l}{\Phi_l}$, for all $l \leq d$ iff the success probability is given by $P(\Psi, \Phi) = \frac{E_d(\Psi)}{E_d(\Phi)} = \frac{\Psi_d}{\Phi_d} < 1 $.
\label{OSBPbip}
\end{observation}
Recall that the entries of the Schmidt vector are nonincreasing (see Section \ref{Prel_ent}). For the proof of this observation we refer to Appendix \ref{app:bipartiteOSBP}.

Using the above observation it follows immediately that all states $\ket{\xi_{sup}}$ fulfilling $\frac{E_d(\xi_{sup})}{E_d(\Phi)} \leq  \frac{(\xi_{sup})_k}{\Phi_k} \ \forall k \neq d$ with $P(\xi_{sup}, \Phi) = \frac{E_d(\xi_{sup})}{E_d(\Phi)}$ can be optimally transformed via an OSBP into the final state $\ket{\Phi}$.
 Thus, also a direct OSBP obtains the same optimal success probability and this is the first example of an optimal protocol, where the OSBP part is not between states optimizing the fidelity. Let us now show that all of the protocols introduced above (among others) can be implemented as optimal SLOCC paths.

\subsection{Optimal SLOCC paths}
\label{sec:bipartitepaths}
Using the results on the optimal success probability for bipartite states, (see Eq. (\ref{eq:fewmonotones})), we show now that there always exists an optimal SLOCC path for bipartite states. \par
As $\ket{\Psi}$ can be written up to LUs as $\ket{\Psi} = D \otimes \mathbbm{1} \ket{\phi^+}$, any path connecting two bipartite states can be parametrized  (up to LUs) via
\begin{equation}
\{ \ket{\psi(t)} \} = \{ D(t) \otimes \mathbbm{1} \ket{\phi^+} \}_{0\leq t \leq1},
\end{equation}
with $\ket{\psi(0)} = D(0) \otimes \mathbbm{1} \ket{\phi^+} = \ket{\Psi}$ and $\ket{\psi(1)} = D(1) \otimes \mathbbm{1} \ket{\phi^+} = \ket{\Phi}$. As shown in Section\ \ref{sec:contLOCC}, Theorem \ref{thm:pathcar}, a differentiable path $\{ \ket{\psi(t)} \} $ is an optimal SLOCC path iff $\ket{\psi(t_1)} \in \mathcal{I}[\psi(0),\psi(t_2)]$ holds for all $0 \leq t_1 < t_2 \leq 1$. Using this together with Corollary \ref{cor:optIntBip} we immediately see that for $P(\Psi, \Phi) = \frac{E_l(\Psi)	}{E_l(\Phi)}$ the path $\{ \ket{\psi(t)} \}$ is optimal iff 
\begin{equation}
P[\psi(0), \psi(t_1)] = \frac{E_l[\psi(0)]}{E_l[\psi(t_1)]}, \ P[\psi(t_1), \psi(t_2)] = \frac{E_l[\psi(t_1)]}{E_l[\psi(t_2)]},
\end{equation}
for all $0 \leq t_1 < t_2 \leq 1$. Because of Theorem \ref{thm:pathcar}, the characterization of optimal paths can also be given as follows.

\begin{corollary}
\label{cor:optbippath}
Let $\ket{\Psi},\ \ket{\Phi} \ \in \C^d \otimes \C^d$ be two bipartite states with $P(\Psi, \Phi) = \frac{E_l(\Psi)}{E_l(\Phi)}$. A differentiable path $\{ \ket{\psi(t)} \} = \{ D(t) \otimes \mathbbm{1} \ket{\phi^+} \}$ from $\ket{\Psi} = \ket{\psi(0)}$ to $\ket{\Phi} = \ket{\psi(1)}$ is an optimal SLOCC path iff the hazard rate is given by
\begin{equation}
h_{\psi}(t) = \frac{1}{E_l[\psi(t)]} \frac{d}{dt} E_l[\psi(t)].
\label{hazardbiparite}
\end{equation}
\end{corollary}

The success probability of transforming $\ket{\Psi}$ into $\ket{\Phi}$ along the path $\{\ket{\psi(t)}\} $ is then given by
\begin{equation*}
P[\psi] = \exp\left(-\int_0^1 h_{\psi}(s) ds\right) =  \frac{E_l [\psi(0)] }{E_l [\psi(1)]}.
\end{equation*}

Using Eq.\ \eqref{hazardbiparite} we can now show that the particular bipartite SLOCC path $\{\ket{\psi_{str}(t)}\}_{0 \leq t \leq 1}$, given below, is always optimal. For this purpose let $\ket{\Psi}, \ket{\Phi} \in \C^d \otimes \C^d$ be bipartite states with Schmidt vectors $\lambda(\Psi), \lambda(\Phi)$, respectively and $P(\Psi, \Phi) = \frac{E_l(\Psi)}{E_l(\Phi)}$. We define the SLOCC path $\{\ket{\psi_{str}(t)}\}_{0 \leq t \leq 1}$ that connects $\ket{\Psi}$ with $\ket{\Phi}$, via the Schmidt vetor of $\ket{\psi_{str}(t)}$ as,
\begin{align}
\lambda[\psi_{str}(t)] = t \lambda(\Phi) + (1-t) \lambda(\Psi) \ \text{for} \ t \in [0,1]. \label{eq:strline}
\end{align}
Note that, in the space of Schmidt vectors, the path $\{\lambda[\psi_{str}(t)]\}$ is a straight path connecting $\lambda(\Psi)$ to $\lambda(\Phi)$. Because of this, we refer to $\{\ket{\psi_{str}(t)}\}$ as the straight path 
from $\ket{\Psi}$ to $\ket{\Phi}$. We now show that $\{\ket{\psi_{str}(t)}\}$ is optimal. To this end, it is sufficient to show that the following holds for all $t \in (0,1)$ (see Theorem \ref{thm:pathcar}),
\begin{align}
 \frac{1}{E_l[\psi_{str}(t)]} \frac{d}{dt} E_l[\psi_{str}(t)] = \max_k \frac{1}{E_k[\psi_{str}(t)]} \frac{d}{dt} E_k[\psi_{str}(t)]. \label{eq:ratio2}
\end{align}
Using that $P(\Psi, \Phi) = \frac{E_l(\Psi)}{E_l(\Phi)}$ and the fact that $E_k[\psi_{str}(t)] = E_k(\Psi) + t[E_k(\Phi) - E_k(\Psi)]$ is an affine function for all $k \in \{1,\ldots,d\}$, it is straightforward to show that Eq. (\ref{eq:ratio2}) holds.\\
This shows that the straight path of the form Eq. (\ref{eq:strline}) is an optimal path and thus, all bipartite states  $\ket{\Psi}$ can be optimally transformed into any other bipartite state $\ket{\Phi}$ along a path.  
 \par
The transformation of $\ket{\Psi}$ to $\ket{\Phi}$ along the straight path $\{\ket{\psi_{str}(t)}\}_{0 \leq t \leq 1}$ can be implemented via the transformation $\ket{\psi_{str}(t)} \rightarrow \{p_i, \ket{\psi_{str}(t+ dt)} \}$, with $\sum_i p_i = P(\Psi, \Phi)$. In order to illustrate this formalism let us consider the simple example of two states $\ket{\Psi} = D_{\Psi} \otimes \mathbbm{1} \ket{\phi^+}, \ket{\Phi}= D_{\Phi} \otimes \mathbbm{1} \ket{\phi^+} \in \C^3 \otimes \C^3$. Using the results in Ref. \cite{Torun2015} it can be easily shown that the three measurement operators $M_1 = p_1 D_{\Phi} D_{\Psi}^{-1}$, $M_2 = p_2 D_{\Phi} P_{1 \leftrightarrow 3} D_{\Psi}^{-1}$ and $M_3 = p_3 D_{\Phi} P_{2 \leftrightarrow 3} D_{\Psi}^{-1}$, with $P_{i \leftrightarrow j}$ permutation matrices (permuting $\ket{i}$ with $\ket{j}$), which are applied by Alice, allow for such a transformation. More precisely, the state $\ket{\Psi}$ is transformed with probability $P(\Psi, \Phi) = \sum_{i=1}^3 p_i$ into the ensemble $\{p_1, M_1 \otimes \mathbbm{1} \ket{\Psi}; p_2, M_2 \otimes P_{1 \leftrightarrow 3}  \ket{\Psi}; p_3, M_3 \otimes P_{2 \leftrightarrow 3} \ket{\Psi} \}$. All the states in the ensemble obviously correspond to $\ket{\Phi}$ and the transformation is achieved with probability $P(\Psi, \Phi)$. Any transformation from $\ket{\Psi} $ to $\ket{\Phi}$ along the straight path $\{\ket{\psi_{str}(t)}\}_{0 \leq t \leq 1}$ can thus, be implemented via the above described transformation.  \par
We consider now several other examples of bipartite optimal SLOCC paths. First, we show that all the protocols considered in the previous subsection can be implemented along a path. Then, we introduce two specific optimal paths, the \textit{most entangled path} and the \textit{least entangled path}, which exist for all bipartite states $\ket{\Psi}$, $\ket{\Phi}$ (with $\lambda(\Psi) \not\prec \lambda(\Phi)$). The most entangled path consists of an OSBP  from $\ket{\Psi}$ to the most entangled intermediate state $\ket{\chi^{max}}$, which can then be transformed deterministically into the final state $\ket{\Phi}$. Along the least entangled path $\ket{\Psi}$ is first transformed deterministically into the least entangled optimal intermediate state $\ket{\chi^{min}}$, which is then transformed into the final state $\ket{\Phi}$ via an OSBP. 

 \begin{figure}
   \centering
   \includegraphics[width=0.35\textwidth]{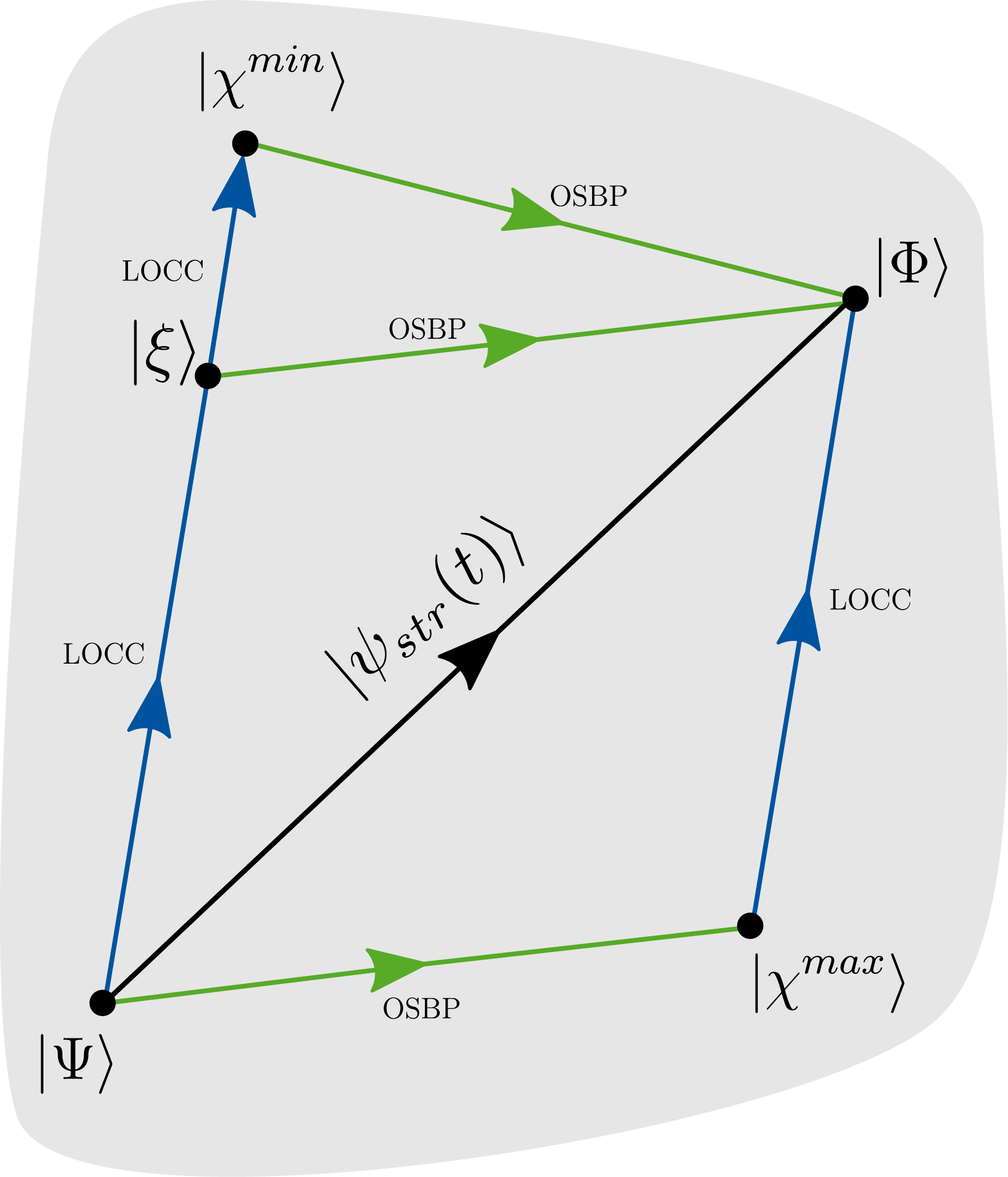}
   \caption{(color online). Transformations from a bipartite state $\ket{\Psi}$ to another bipartite state $\ket{\Phi}$ can be achieved via infinitely many optimal paths. These paths can be concatenations of LOCC (blue) and OSBP (green) transformations. Moreover, $\ket{\Psi}$ can always be transformed into $\ket{\Phi}$ along the straight path, $\{\ket{\psi_{str}(t)}\}$ (black). There are also two extremal transformations. In the first extremal transformation, the initial state is first transformed via an OSBP into the most entangled state $\ket{\chi_{max}}$ and then into the final state $\ket{\Phi}$ (see Section \ref{sec:mostent}). In the other extremal transformation, the initial state $\ket{\Psi}$ is transformed via LOCC into the least entangled state $\ket{\chi_{min}}$ and then via an OSBP into the final state $\ket{\Phi}$ (see Section \ref{sec:leastent}). The paths corresponding to these transformations are piecewise continuously differentiable paths.}
   \label{fig:BipPaths}
   \end{figure}

\paragraph{SLOCC paths along intermediate states optimizing the fidelity\\}

As we have shown above, there exist always optimal bipartite SLOCC paths, namely the straight paths, see Eq.\ \eqref{eq:strline}. Thus, the protocols introduced in Section\ \ref{sec:IntBip}, which are defined by the optimal intermediate states $\ket{\zeta}$ and $\ket{\eta}$ given in Eq.\ \eqref{nec+suffXiChi}, can be implemented along a concatenation of three straight paths. More precisely, along the LOCC path $\{\ket{{\psi_{str}^{LOCC}}_1(t)}\}$, which is the deterministic path from $\ket{\Psi}$ to $\ket{\zeta}$, the OSBP path $\{\ket{\psi_{str}^{OSBP}(t)}\}$, which is the probabilistic path from $\ket{\zeta}$ to $\ket{\eta}$ and the LOCC path $\{\ket{{\psi_{str}^{LOCC}}_2(t)}\}$, which is a deterministic path from $\ket{\eta}$ to $\ket{\Phi}$, see Fig.\ \ref{VaVsFi}. With $P(\Psi, \Phi) = \frac{E_l(\Psi)}{E_l(\Phi)}$ the deterministic paths fulfill $P[\psi] = \frac{E_l (\psi(0)) }{E_l (\psi(1))} = 1$ and the monotone with the largest relative growth ($E_l$) also has to stay the same, as the value of the entanglement monotones can only decrease via LOCC. For the probabilistic path $P[\psi] = \frac{E_l (\psi(0)) }{E_l (\psi(1))} = P(\Psi, \Phi)$. Note that the path corresponding to this transformation is a piecewise continuously differentiable path. \par
Clearly, also the protocol introduced in Ref. \cite{Vidal1999} (see Section \ref{sec:IntBip}) can be implemented along a path, which consists of an LOCC path from $\ket{\Psi}$ to $\ket{\xi}$ and an OSBP path from $\ket{\xi}$ to the final state $\ket{\Phi}$.

\paragraph{ Most entangled path\\}
\label{sec:mostent}
Let us consider now an optimal SLOCC path among bipartite states defined by a probabilistic transformation to the, what we call, \textit{most entangled} intermediate state $\ket{\chi^{max}}$, followed by a deterministic transformation of this state to the final state $\ket{\Phi}$. More precisely, this path is uniquely defined by the initial state $\ket{\Psi}$ and by the success probability $P$, which determines the state $\ket{\chi^{max}}$. Among all the states to which $\ket{\Psi}$ can be transformed with a given success probablity $P$, the state $\ket{\chi^{max}}$ maximizes all entanglement monotones $E_k$, i.e. $\text{argmax}_{\psi': P = P(\Psi,  \psi')} (E_2(\psi'), E_3(\psi'),...,E_d(\psi')) = \ket{\chi^{max}(P)}$, where the maximization is done with respect to all components, i.e. all entanglement monotones. The Schmidt coefficients of the most entangled state are given by
\begin{eqnarray}
\chi^{max}_k & = & \frac{\psi_k}{P}  \ , k \in \{2,...,d\} \nonumber \\
\chi^{max}_1 & = &  1- \sum_{k=2}^d \chi^{max}_k.
\label{maxentPconst}
\end{eqnarray}
Note, however, that the Schmidt coefficients in Eq.\ \eqref{maxentPconst} are  in general not automatically ordered decreasingly. In case they are not ordered, we show in Appendix \ref{app:bipartitemost} that the state $\ket{\chi^{max}}$ can be easily reordered and still fulfills all the properties of $\ket{\chi^{max}}$. \par
 As can be easily shown, it holds that $\frac{E_k(\Psi)}{E_k(\chi^{max})} = \frac{E_d(\Psi)}{E_d(\chi^{max})} = P(\Psi, \chi^{max}) = P $, for all $k \in \{2,...,d-1\}$ and $\frac{\Psi_d}{\chi^{max}_d} \leq \frac{\Psi_l}{\chi^{max}_l} \forall l$. Thus, as explained above (see Observation\ \ref{OSBPbip}), $\ket{\chi^{max}}$ can always be obtained from $\ket{\Psi}$ via a direct OSBP. We show now, that the states given in Eq.\ \eqref{maxentPconst} maximize all entanglement monotones for a given constant success probability $P$. This can be easily seen as for all states $\ket{\psi'}$ that can be obtained from $\ket{\Psi}$ with success probability $P$ it holds that $P = P(\Psi,  \psi') = \min_l \frac{E_l (\Psi)}{E_l(\psi')} = \frac{E_{l_1} (\Psi)}{E_{l_1}(\psi')} \leq \frac{E_k (\Psi)}{E_k(\psi')}$  for some $l_1 \in \{1,...,d\}$ (with different $l_1$ for different states $\ket{\psi'}$) and for all $k \neq l_1$. For the most entangled state it holds that  $P= P(\Psi, \chi^{max}) = \frac{E_k (\Psi)}{E_k(\chi^{max})} \ \forall k \in \{2,...,d\}$ and, thus, $E_k(\psi') \leq E_k(\chi^{max})$ for all $k \in \{1,...,d\}$. \par

Let us finally show that the state $\ket{\chi^{max}}$ with $P = P(\Psi, \Phi)$ can be transformed into the target state $\ket{\Phi}$ deterministically. From the definition of $\ket{\chi^{max}}$, see Eq.\ \eqref{maxentPconst} it can be easily shown that the majorization condition is fulfilled as $\forall l \in \{1,...,d\} $ it holds that $P(\Psi, \Phi) \leq \frac{E_l(\Psi)}{E_l(\Phi)}$ and thus
\begin{equation}
E_l (\chi^{max}) = \frac{1}{P(\Psi, \Phi)} \sum_{i=l}^d \Psi_i \geq E_l(\Phi) = \sum_{i=l}^d \Phi_i.
\end{equation}
Summarizing, the most entangled path (see Fig.\ \ref{fig:BipPaths}) is an optimal path, consisting of an OSBP path from the initial state $\ket{\Psi}$ to the state $\ket{\chi^{max}}$ (or the resorted state, see Appendix \ref{app:bipartitemost}) with $\chi_d^{max} = \frac{\Psi_d}{P(\Psi, \Phi)}$ and a deterministic path from $\ket{\chi^{max}}$ to the final state $\ket{\Phi}$.\\

\paragraph{Least entangled path\\}
\label{sec:leastent}
Let us define the \emph{least entangled} optimal intermediate state $\ket{\chi^{min}}$ of $\ket{\Psi}, \ket{\Phi}$ that can be reached from $\ket{\Psi}$ deterministically. For the definition of this least entangled state we set again all the ratios of the entanglement monotones equal to the optimal success probability. The intermediate state necessarily has to fulfill the condition $P(\Psi, \Phi) = \frac{E_l(\Psi)}{E_l(\Phi)} = P(\chi^{min}, \Phi) =\frac{E_l(\chi^{min})}{E_l(\Phi)}$ for $l \in \{2,...,d\}$. Moreover, $\ket{\chi^{min}}$ fulfills $E_k(\chi^{min}) \leq E_k(\Psi) \ \forall k$, as $\ket{\chi^{min}}$ can be obtained from $\ket{\Psi}$ via LOCC. Thus, we can set all ratios of the entanglement monotones equal to each other, i.e. $\frac{E_l(\chi^{min})}{E_l(\Phi)} = \frac{E_k(\chi^{min})}{E_k(\Phi)} \ \forall k \neq l$ and obtain the least entangled state $\ket{\chi^{min}}$. The Schmidt coefficients of this state are then given by
\begin{eqnarray}
\chi^{min}_k &=& P (\Psi, \Phi) \Phi_k \ \text{for all} \ k \in \{2,..,d\} \\
\chi^{min}_1 &=& 1-  P (\Psi, \Phi)  \sum_{i=2}^d \Phi_i,
\end{eqnarray}
which are sorted in decreasing order as the Schmidt coefficients of the final state $\ket{\Phi}$ are sorted.  Moreover, the state $\ket{\Phi}$ can be obtained from $\ket{\chi^{min}}$ via an OSBP on an SLOCC path given by the most entangled path defined above, which, in contrast to before, begins at $\ket{\chi^{min}}$. More precisely, in this case the reference state is $\ket{\chi^{min}}$ and the path is given by the Schmidt vector $\lambda(\chi^{max}) = (1- \sum_{k=2}^d \chi^{max}_k,  \frac{\chi^{min}_2}{\chi^{min}_d} \chi^{max}_d,...,\chi^{max}_d)$ for $\chi^{min}_d \leq \chi^{max}_d \leq \Phi_d$ (see also Fig. \ref{fig:BipPaths}).\\

\paragraph{Twofold optimal paths and minimal geodesics\\}
\label{sec:biplength}

In this section we show that two bipartite states can always be connected by a twofold optimal path. Combined with Observation \ref{obs:geodesic}, this shows that the set of minimal geodesics in the interconversion metric and the set of twofold optimal paths coincide in the bipartite case. However, we also show that, for bipartite systems of local dimension $d>2$, there are optimal paths that are not twofold optimal and thus do not correspond to minimal geodesics.\\

Let consider for two states $\ket{\Psi}, \ket{\Phi} \in \C^d \otimes \C^d$ again the straight path $\{\ket{\psi_{str}(t)}\}_{0 \leq t \leq 1}$ defined in Eq.\ \eqref{eq:strline}. 
We showed above that this straight path is an optimal path. Since its reverse path, $\{\ket{\psi_R(t)} = \ket{\psi(1-t)}\}_{0 \leq t \leq 1}$, is again a straight path, this shows that, indeed, the straight path connecting two bipartite states is twofold optimal. As mentioned above, this further proves that all minimal geodesics are twofold optimal in the bipartite case.\\

For 2-qubit states, any optimal path is twofold optimal. The reason for this is that there is (up to an irrelevant reparametrization \footnote{Note that two paths $\{\ket{\psi_1(t)}\}$ and $\{\ket{\psi_2(t)}\}$ are said to be equal up to reparametrization if there is a monotonously increasing function $f(t)$ such that $\ket{\psi_2(t)} = \ket{\psi_1[f(t)]}$ holds for all $t$. In this case, the two paths are said to describe the same \emph{curve} (see e.g. Ref. \cite{burago2001}).}) only one optimal path to transform an initial state $\ket{\Psi} \in \C^2 \otimes \C^2$ to a final state $\ket{\Phi} \in \C^2 \otimes \C^2$; namely the path along which the smallest Schmidt coefficient of $\ket{\Psi}$ is changed into the smallest Schmidt coefficient of $\ket{\Phi}$ in a strictly monotonous way (i.e. by increasing or decreasing it). However, for bipartite states of local dimension $d>2$ there are optimal paths that are not twofold optimal. We show this via an explicit example. For $d>2$, we consider the transformation from the maximally entangled state $\ket{\phi^+} \in \C^d \otimes \C^d$ with Schmidt vector $\lambda(\phi^+) = \frac{1}{d}(1,\ldots,1)$ to the state $\ket{\Phi}$ with Schmidt vector $\lambda(\Phi) = (\frac{6}{10}, \frac{3}{10}, \frac{1}{10 (d-2)}, \ldots, \frac{1}{10 (d-2)})$. This transformation can be achieved deterministically via an LOCC path $\{\ket{\psi_{opt}(t)}\}_{0\leq t \leq 1}$ that goes through the intermeditate state $\ket{\chi}$ with Schmidt vector $\lambda(\chi) = (\frac{5}{10}, \frac{4}{10}, \frac{1}{10 (d-2)}, \ldots, \frac{1}{10 (d-2)})$. In order for the reverse path, $\{\ket{\psi_{opt,R}(t)} = \ket{\psi_{opt}(1-t)}\}_{0\leq t \leq 1}$, to be optimal, the state $\ket{\chi}$ has to be an intermediate state of the transformation from $\ket{\Phi}$ to $\ket{\phi^+}$ (see Theorem \ref{thm:pathcar}). However, using Eq. (\ref{eq:fewmonotones}) for the optimal success probability of bipartite pure state transformations, it is easy to see that $P(\Phi, \chi) = \frac{4}{5}, P(\Phi,\phi^+) = P(\chi,\phi^+) = \frac{1}{10}\frac{d}{d-2}$ and hence $P(\Phi,\chi)P(\chi,\phi^+) < P(\Phi,\phi^+)$ holds. That is, $\ket{\chi}$ is not an intermediate state of the transformation from $\ket{\Phi}$ to $\ket{\phi^+}$ and thus the reverse path, $\{\ket{\psi_{opt,R}(t)}\}$, is not optimal.

\section{Multipartite case}
\label{sec:multiex}

In this section we consider SLOCC transformations of multipartite pure states. In Section \ref{sec:genmulti} we characterize optimal intermediate states and optimal SLOCC paths for almost all multiqudit SLOCC transformations. We show that there are infinitely many optimal SLOCC paths for these transformations. For the case of generic $(n\geq 5)$-qubit states we give an explicit decomposition for all optimal intermediate states and optimal paths. In contrast to the bipartite case and generic multipartite case we show in Section \ref{sec:noint} that there are multipartite LOCC transformations for which no intermediate state exists. 

\subsection{Generic multiqudit states}
\label{sec:genmulti}
In this section we determine optimal intermediate states and optimal SLOCC paths for transformations among generic multiqudit states; that is, for almost all multiqudit SLOCC transformations. As before, we call a state generic if it is an element of the full-measure set of states in $\mH_{n,d}$ that have trivial stabilizer (where here and in the following $n = 4, d>2$ or $n \geq 5, d \geq 2$ or $n = 3, d = 4,5,6$). We consider  transformations among states in the SLOCC class of a generic seed state $\ket{\Psi_s}$. It was shown in \cite{Gour2017,Sauerwein2018} that the optimal protocol to transform a state $\ket{\Psi} = g\ket{\Psi_s}$ into a state $\ket{\Phi} = h\ket{\Psi_s}$, where here and in the following $g,h \in \widetilde{G}$, is an OSBP. The optimal success probability reads,
\begin{align}
P(\Psi,\Phi) = \frac{\|\Phi\|^2}{\|\Psi\|^2} \frac{1}{\lambda_{max}(G^{-1}H)}, \label{eq:optgen}
\end{align}
where here and in the following $G = g^\dagger g, H = h^\dagger h$. In this section we often set $\ket{\Psi} = \ket{\Psi_s}$, as many results (e.g. the characterization of intermediate states) can then be stated in a particularly concise form. It is important to note that this in not a restriction since the seed state of a SLOCC class can be chosen at will. Moreover, all results in the following section can be straightforwardly reexpressed for transformations from $\ket{\Psi} = g \ket{\Psi_s}$ to $\ket{\Phi} = h \ket{\Psi_s}$ with $g \neq \one$.

\subsubsection{Optimal intermediate states}
\label{sec:genint}
Using Eq. (\ref{eq:optgen}) we find the following characterization of optimal intermediate states of SLOCC transformations among generic states.

\begin{theorem}
 \label{thm:neccsuff0}
 Let $\ket{\Psi} \in \mH_n$ be a generic state and let $\ket{\Phi} = h\ket{\Psi}$ be a state in the SLOCC class of $\ket{\Psi}$. A state $\ket{\chi} = g\ket{\psi}$ is an intermediate state of the SLOCC transformation from $\ket{\Psi}$ to $\ket{\Phi}$, i.e. $\ket{\chi} \in \mathcal{I}(\Psi,\Phi)$, iff
 \begin{align}
  \lambda_{max}(G^{-1} H) = \frac{\lambda_{max}(H)}{\lambda_{max}(G)}.
 \end{align}
\end{theorem}

\proof{ First, note that the following holds (see Eq. (\ref{eq:optgen})),
\begin{align}
 &P(\Psi,\chi) = \frac{\|\chi\|^2}{\|\Psi\|^2} \frac{1}{\lambda_{max}(G)}, \label{eq:prob1}\\
 &P(\chi,\Phi) =\frac{\|\Phi\|^2}{\|\chi\|^2} \frac{1}{\lambda_{max}(G^{-1} H)}.\label{eq:prob2}
\end{align}
Let us now show the ``only if''-part.
It is easy to see that for any $G,H > 0$ the following holds,
\begin{align}
\frac{1}{\lambda_{max}(G^{-1} H)}  \leq \frac{\lambda_{max}(G)}{\lambda_{max}(H)}. \label{ineq:1}
\end{align}
 Now suppose that the inequality in (\ref{ineq:1}) is strict for the intermediate state $\ket{\chi} = g\ket{\Psi}$.  Using Eqs. (\ref{eq:prob1}-\ref{eq:prob2}) we then get,
\begin{align*}
 &P(\Psi,\chi)P(\chi,\Phi) =\\
 & \frac{\|\chi\|^2}{\|\Psi\|^2} \frac{1}{\lambda_{max}(G)} \frac{\|\Phi\|^2}{\|\chi\|^2} \frac{1}{\lambda_{max}(G^{-1} H)}\\
 &< \frac{\|\Phi\|^2}{\|\Psi\|^2} \frac{1}{\lambda_{max}(H)} = P(\Psi,\Phi).
\end{align*}
This contradicts the fact that $\ket{\chi}$ is an intermediate state. Hence, the inequality in (\ref{ineq:1}) is in fact an equality.\\
Next, we show the ``if'''-part. In this case we have
\begin{align*}
 P(\chi,\Phi) &= \frac{\|\Phi\|^2}{\|\chi\|^2} \frac{1}{\lambda_{max}(G^{-1} H)}\\
  &= \frac{\|\Phi\|^2}{\|\chi\|^2} \frac{\lambda_{max}(G)}{\lambda_{max}(H)}.
\end{align*}
Using the formula for $P(\Psi,\chi)$ given in Eq. (\ref{eq:prob1}) it is then easy to see that this implies that $\ket{\chi} = g\ket{\Psi} \in \mathcal{I}(\Psi,\Phi)$. \qed\\
}

Note that Theorem \ref{thm:neccsuff0} provides an easy way to check if some fixed state is an intermediate state of a given transformation. However, we can also use Theorem \ref{thm:neccsuff0} to explicitly construct a large subset, and in the case of generic $(n>4)$-qubit states even the whole set, of optimal intermediate states.

\begin{theorem}
\label{thm:subintgen}
Let $\ket{\Psi},\ket{\Phi} = h\ket{\Psi} \in \mH_{n,d}$ be SLOCC equivalent generic states. A state $\ket{\chi} = g\ket{\Psi}$ is an optimal intermediate state of the transformation from $\ket{\Psi}$ to $\ket{\Phi}$, i.e. $\ket{\chi} \in \mathcal{I}(\Psi,\Phi)$,  if the following holds for all $i \in \{1,\ldots,n\}$,
\begin{itemize}
\item[(i)] $[G_i,H_i] = 0$,
\item[(ii)] $\frac{H_i}{\lambda_{max}(H_i)} \leq \frac{G_i}{\lambda_{max}(G_i)}$.
\end{itemize}
For transformations among generic $(n>4)$-qubit states all optimal intermediate states are of this form.
\end{theorem}
This shows, in particular, that there are always uncountably many different intermediate states for any given transformation among generic states.
We provide the proof of Theorem \ref{thm:subintgen} in Appendix \ref{app:genqb}.

\subsubsection{Optimal SLOCC paths}
\label{sec:genintcont}

In this section we consider optimal SLOCC paths for transformations of generic states in $\mH_{n,d}$.\\
Theorem \ref{thm:pathcar} states that an SLOCC path, $\{\ket{\psi(t)} = g(t)\ket{\Psi}\}$, from $\ket{\Psi} = \ket{\psi(0)}$ to $\ket{\Phi} = \ket{\psi(1)}$ is optimal iff $\ket{\psi(t_1)} \in \mathcal{I}[\psi(0),\psi(t_2)]$ holds for all $0 \leq t_1 < t_2 \leq 1$, i.e. iff
\begin{align}
\lambda_{max}[G(t_1)^{-1}G(t_2)] = \frac{\lambda_{max}[G(t_2)]}{\lambda_{max}[G(t_1)]}. \label{eq:charpathsgen}
\end{align}
Here, we used the characterization in Theorem \ref{thm:neccsuff0} of intermediate states. Using Eq. (\ref{eq:charpathsgen}) it is easy to see that a particular optimal path from  $\ket{\Psi}$ to $g\ket{\Psi}$ is $\{g_1(t) \otimes \ldots \otimes g_n(t)\ket{\Psi}\}$ with $g_i(t) = \sqrt{(1-t)\one + t G_i}$.\\
On the one hand, this shows that any intermediate state is an element of an optimal SLOCC path. Indeed, one can always connect the initial state with the intermediate state and then the intermediate state with the final state with an optimal path of the form described above.\\

Note that the characterization of optimal paths given in Eq. (\ref{eq:charpathsgen}) requires the comparison of properties of the path at two different values of the parameter $t$. In the following we use the variational approach introduced in Observation \ref{obs:var} to find a criterion that involves only the parameter $t$.\\

Without loss of generality we normalize $\ket{\psi(t)} = g(t)\ket{\Psi}$ again such that $\lambda_{max}[G(t)] = 1$.
Observation \ref{obs:var} states that $\{\ket{\psi(t)}\}$ is an optimal path iff it minimizes the cumulative hazard rate (see Eq. (\ref{eq:hazard1})), i.e. the functional
\begin{align}
 \Lambda[\psi] = \int_{0}^1 h_\psi(s) ds,
\end{align}
Hence, we first have to calculate the hazard rate $h_\psi(t)$ for the path $\{\ket{\psi(t)}\}$ described above, which is defined as (see Eq. (\ref{eq:hazard})),
\bea
h_{\psi}(t) = \lim_{\Delta t \rightarrow 0} \frac{1 - P[\psi(t),\psi(t+\Delta t)]}{\Delta t}. \label{eq:hazard2}
\eea
Using Eq. (\ref{eq:optgen}) one can show that the following holds,
\begin{align}
&P[\psi(t),\psi(t+\Delta t)) = \nonumber\\
&\frac{n(t+\Delta t)^2}{n(t)^2} \frac{1}{\lambda_{max}({G(t)^{-1}} G(t+\Delta t))}, \label{eq:smallstep}
\end{align}
where we used the notation $n(t) = \|g(t)\ket{\Psi}\|$. Since $g(t)$ is differentiable we can expand the right-hand side of Eq. (\ref{eq:smallstep}) as,
\begin{align*}
 &P[\psi(t),\psi(t+\Delta t)]\\
 &= 1 + \left[2\frac{n'(t)}{n(t)} - \lambda_{max}\left[G(t)^{-1} G'(t)\right]\right] \Delta t\\
 & + O(\Delta t^2). \label{eq:tailor}
\end{align*}
Here, $G'(t) = \frac{d}{dt}G(t)$ denotes the derivative of $G(t)$. Note that this derivative exists since  $\{\ket{\psi(t)} = g(t) \ket{\Psi}\}$ is an SLOCC path and, thus, $g(t)$ (and therefore $G(t)$) are differentiable (see Section \ref{sec:pathdef}).
We insert the expression in Eq. (\ref{eq:tailor}) into Eq. (\ref{eq:hazard2}) and see that the hazard rate is given by,
\begin{align}
 h_{\psi}(t) = - 2\frac{n'(t)}{n(t)} + \lambda_{max}\left[G(t)^{-1} G'(t)\right].
\end{align}
The cumulative hazard can then be written as 
\begin{align}
 \Lambda[\psi] = \Lambda_o[\psi]  + \Lambda_s[\psi],
\end{align}
where 
\begin{align*}
 &\Lambda_o[\psi] = -2 \int_0^1 \frac{n'(s)}{n(s)} ds,\\
 &\Lambda_s[\psi] = \int_0^1 \lambda_{max}\left[G(s)^{-1} G'(s)\right] ds.
\end{align*}

Using the relation between the cumulative hazard rate and the probability, $P[\psi]$, to continuously transform $\ket{\Psi}$ into $\ket{\Phi}$ along the path $\{\ket{\psi(t)}\}$ given in Eq. (\ref{eq:probfunc}), we obtain 
\begin{align}
P[\psi] = e^{-\Lambda[\psi]} = e^{-\Lambda_o[\psi]} e^{-\Lambda_s[\psi]} = P(\Psi,\Phi) e^{-\Lambda_s[\psi]}. \label{eq:Pfunc3}
\end{align}
Here, we used that $P(\Psi,\Phi) = \frac{n(1)^2}{n(0)^2}$ if $\lambda_{max}(G) = 1$ (see Eq. (\ref{eq:optgen})).\\

Equation (\ref{eq:Pfunc3}) shows that $\Lambda_o[\psi]$ can be interpreted as the  rate of the optimal part of the SLOCC transformation along $\{\ket{\psi(t)}\}$, while $\Lambda_s[\psi]$ corresponds to the nonoptimal part of the transformation. Furthermore, it is clear from Eq. (\ref{eq:Pfunc3}) that the path $\{\ket{\psi(t)}\}$ is an optimal SLOCC path iff $\Lambda_s[\psi] = 0$, which is satisfied iff $\lambda_{max}\left[G(t)^{-1} G'(t)\right] = 0$ for all $t \in (0,1)$. In summary, we proved the following characterization of all optimal SLOCC paths.
\begin{theorem}
\label{thm:charpath1}
Let $\ket{\Psi}, \ket{\Phi} = g\ket{\Psi}$ be generic states. An SLOCC path $\{\ket{\psi(t)} = g(t)\ket{\Psi}\}$, with $\lambda_{max}[G(t)] = 1$, is optimal iff
 \begin{align}
  \lambda_{max}\left[G(t)^{-1} G'(t)\right] = 0 \ \text{for all} \ t \in (0,1). \label{eq:localbnd}
 \end{align}
\end{theorem}
Let us mention again that the condition $\lambda_{max}[G(t)] = 1$ in Theorem \ref{thm:charpath1} is not a constraint, as we can always renormalize the states on the path to fulfill this condition.
Theorem \ref{thm:charpath1} shows that all optimal entanglement transformations among generic states can be obtained as solutions of a "differential equation". Note further that the path given after Eq. (\ref{eq:charpathsgen}) fulfills this condition (after proper normalization) since it is optimal.\\

It is interesting to see how the characterization in Theorem \ref{thm:charpath1} relates to the examples
(and for generic qubit states, the characterization) of optimal intermediate states given in Theorem  \ref{thm:subintgen}. We know from Theorem \ref{thm:pathcar} that a path $\{\ket{\psi(t)} = g(t)\ket{\Psi}\}$ from $\ket{\Psi} = \ket{\psi(0)}$ to $\ket{\Phi} = \ket{\psi(1)}$ is an optimal SLOCC path iff $\ket{\psi(t_1)} \in \mathcal{I}[\psi(0),\psi(t_2)]$ holds for all $0 \leq t_1 < t_2 \leq 1$. Let us construct the path $\{\ket{\psi(t)}\}$ in such a way that $\ket{\psi(t_1)}$ is an optimal intermediate state of the transformation from $\ket{\psi(0)}$ to  $\ket{\psi(t_2)}$ as given in Theorem \ref{thm:subintgen}. Renormalizing $\ket{\Phi}$ such that $\lambda_{max}(G) = 1$ and expressing $G = g^\dagger g$ in spectral decomposition as,
\bea
G = \bigotimes_{i=1}^n U^{(i)} \text{diag}[1,r_2^{(i)},\ldots,r_d^{(i)}]{U^{(i)}}^\dagger,
\eea
where $U^{(i)}$ are local unitaries and $0 < r_k^{(i)} \leq 1$ for $k \in \{2,\ldots,d\}$ and $i \in \{1,\ldots,n\}$, we find that this is the case iff the following holds for all $t \in [0,1]$,
\bea
G(t) = \bigotimes_{i=1}^n U^{(i)} \text{diag}[1,r_2(t)^{(i)},\ldots,r_d(t)^{(i)}]{U^{(i)}}^\dagger, \label{eq:examplepaths}
\eea
where $r_k(0)^{(i)} = 1, r_k(1)^{(i)} = r_k^{(i)}$ and $\frac{d}{dt}[r_k(t)^{(i)}] \leq 0$. It is straightforward to see that these paths, as expected, fulfill Eq. (\ref{eq:localbnd}). Moreover, Eq. (\ref{eq:examplepaths}) shows that there are always infinitely many different optimal SLOCC paths along which one can continuously transform a generic state $\ket{\Psi}$ into an SLOCC equivalent state $\ket{\Phi} = g\ket{\Psi}$. In the case of generic $(n>4)$-qubit states, all optimal paths are of the form given in Eq. (\ref{eq:examplepaths}). That is, we get the following characterization of all optimal paths of generic qubit states.
\begin{lemma}
\label{lem:charqubit}
Let $\ket{\Psi},\ket{\Phi} = g\ket{\Psi} \in \mH_{n,2}$ be generic $(n>4)$-qubit states, with
\begin{align}
 G = \bigotimes_{i=1}^n U^{(i)} \text{diag}(1,r^{(i)}){U^{(i)}}^\dagger,
\end{align}
where $U^{(i)}$ are local unitaries and $0 < r^{(i)} \leq 1$.\\
 A path $\{\ket{\psi(t)} = g(t)\ket{\Psi}\}$, with $\lambda_{max}[G(t)] = 1$, from $\ket{\Psi} = \ket{\psi(0)}$ to $\ket{\Phi} = \ket{\psi(1)}$ is an optimal SLOCC path of the transformation from $\ket{\Psi}$ to $\ket{\Phi}$ iff
 \begin{align}
  G(t) = \bigotimes_{i=1}^n U^{(i)} \text{diag}[1,r(t)^{(i)}]{U^{(i)}}^\dagger,
 \end{align}
 with $\frac{d}{dt}[r^{(i)}(t)] \leq 0$, for all $t \in (0,1)$.
\end{lemma}

\subsubsection{Twofold optimal paths and minimal geodesics}
\label{sec:geodgeneric}
In this section we first show that the sets of optimal paths, of twofold optimal paths and of minimal geodesics in the interconversion metric are all identical for generic $(n>4)$-qubit states. For generic qudit states with local dimension $d>2$, any minimal geodesic is a twofold optimal SLOCC path. However, there are also optimal (but not twofold optimal) SLOCC paths that are not minimal geodesics.\\

Recall from Section \ref{sec:geodesic} that an optimal SLOCC path $\{\ket{\psi(t)}\}$ is called twofold optimal if the reverse path, $\{\ket{\psi_R(t)} = \ket{\psi(1-t)}\}$, is an optimal path aswell. Using Lemma \ref{lem:charqubit} it is straightforward to see that indeed any optimal path of generic $(n>4)$-qubit states fulfills this property. Moreover, this shows that these paths, and hence all minimal geodesics of generic $(n>4)$-qubit states, are distance minimizing (see Observation \ref{obs:geodesic}). This shows that, for generic $(n>4)$-qubit states, the sets of optimal paths, of twofold optimal paths and of minimal geodesics are identical. Stated differently, in this case any optimal path is also optimal for the reverse transformation.\\
However, for higherdimensional generic states there exist optimal SLOCC paths that are not twofold optimal. To give an example, consider the $(n>3)$-qutrit path
$\{\ket{\psi(t)} = g(t)\ket{\Psi}\}_{0\leq t \leq 1}$, where $\ket{\Psi}$ has trivial stabilizer (examples of such states are given in Ref. \cite{Sauerwein2018}) and $g(t) = u(t)\Delta(t)u(t)^\dagger \otimes \one^{\otimes {n-1}}$, with,
\begin{align*}
&u(t) =
\begin{pmatrix}
1 & 0 & 0\\
0 & \sin(\frac{t}{2}) & \cos(\frac{t}{2})\\
0 & -\cos(\frac{t}{2}) & \sin(\frac{t}{2})
\end{pmatrix}, \\
&\Delta(t) = \text{diag}\left(1, 1 - \frac{t}{4}, 1 - \frac{t}{2}\right).
\end{align*}
It is straightforward to show that this path fulfills Eq. (\ref{eq:localbnd}) and is, hence, an optimal SLOCC path, but the reverse path $\{\ket{\psi_R(t)} = \ket{\psi(1-t)}\}$ is not. Notice that the eigenbasis of $G(t)$ changes with $t$. This cannot happen for generic $(n>4)$-qubit states (see Lemma \ref{lem:charqubit}).\\
However, for any  generic $n$-partite state $\ket{\Psi}$ with trivial stabilizer and any $g = g_1 \otimes \ldots \otimes g_n \in \widetilde{G}$ we can construct a twofold optimal path from $\ket{\Psi}$ to $\ket{\Phi} = g\ket{\Psi}$. To this end, we first write $G_i = g_i^\dagger g_i$ as,
\begin{align*}
G_i = u_i \text{diag}(\lambda_1^{(i)},\ldots,\lambda_d^{(i)})u_i^\dagger,
\end{align*}
where $\lambda_1^{(i)} \geq \ldots \geq \lambda_d^{(i)} > 0$ and $u_i u_i^\dagger = \one$. We then define the path $\{\ket{\psi(t)} = g_1(t) \otimes \ldots \otimes g_n(t)\ket{\Psi}\}_{0 \leq t \leq 1}$ with
\begin{align*}
g_i(t) = \begin{cases}
      \one & 0 \leq t < \frac{i-1}{n} \\
       u_i \text{diag}[\lambda_1^{(i)}(t),\ldots,\lambda_d^{(i)}(t)]u_i^\dagger& \frac{i-1}{n} \leq t  < \frac{i}{n} \\
      g_i & \frac{i}{n} \leq t \leq 1
   \end{cases},
\end{align*}
where $\lambda_k^{(i)}(t) = \left(\lambda_k^{(i)}\right)^{nt - i + 1}$. Hence, the local operators $g_i$ acting on $\ket{\Psi}$ are, one at a time, differentiably changed from $\one$ to $g_i$. After an appropriate normalization it is easy to show that both, $\{\ket{\psi(t)}\}$ and $\{\ket{\psi_R(t)} = \ket{\psi(1-t)}\}$, fulfill Eq. (\ref{eq:localbnd}). That is, $\{\ket{\psi(t)}\}$ is a twofold optimal path and therefore a distance minimizing path (see Observation \ref{obs:geodesic}). Hence, for generic multiqudit states any minimial geodesic is a distance minimizing path and therefore twofold optimal.

\subsection{Multipartite LOCC transformations without intermediate states}
\label{sec:noint}

In this section we show that there are LOCC transformations of multipartite qubit states for which there exists no intermediate state. More precisely, we show that such transformations exist for any $n$-qubit system with $n = 2^k$ and $k \geq 2$. Moreover, in these cases intermediate states do not even exist if transformations via the (in general more powerful) separable operations are considered. This is in stark contrast to the bipartite (see Section \ref{sec:bipex})  and the generic multipartite case (see Section \ref{sec:multiex}), where always infinitely many intermediate states exist.\\

We first provide an example of a 4-qubit LOCC transformation without intermediate states. More precisely, we consider states in the SLOCC class of a generic 4-qubit seed state $\ket{\psi_s}$ \cite{DeVicente2013}. These states have nontrivial local symmetries and are therefore not generic multiqudit states as defined in Section \ref{sec:mainresults}. In Ref. \cite{Sauerwein2015} all LOCC and separable (SEP) transformations among states in these SLOCC classes were characterized. Here, we consider LOCC transformations among a subset of these states, namely the set,
\bea
\{g \otimes \one^{\otimes 3}\ket{\psi_s} \ | \ g \in GL(\C,2)\}. \label{eq:setS}
\eea
In Ref. \cite{Sauerwein2015} it was shown that, for any state $\ket{\psi}$ in this set, there exists a unique vector
$\vec{\gamma} = (\gamma_1, \gamma_2, \gamma_3) \in \R_{\geq 0}^2 \times \R$, with $\|\vec{\gamma}\| < \frac{1}{2}$, such that $\ket{\psi}$ is LU equivalent to the state,
\bea
\ket{\psi(\vec{\gamma})} = g(\vec{\gamma}) \otimes \one^{\otimes 3}\ket{\psi_s}, \label{eq:4qbstate1}
\eea
where $g(\vec{\gamma}) = \sqrt{G(\vec{\gamma})}$, with $G(\vec{\gamma}) = \frac{1}{2} \one + \sum_{k=1}^3 \gamma_k \sigma_k$. Here, $\{\sigma_i\}$ denote the Pauli matrices. In particular, this implies that for $\vec{\gamma},\vec{\xi} \in \R_{\geq 0}^2 \times \R$, $\vec{\gamma} \neq \vec{\xi}$ (with $\|\vec{\gamma}\|, \|\vec{\xi}\|  < \frac{1}{2}$) there exists no local unitary $u = u_1 \otimes u_2 \otimes u_3 \otimes u_4$ such that $\ket{\psi(\vec{\gamma})} = u\ket{\psi(\vec{\xi})}$, i.e. the LU classes of $\ket{\psi(\vec{\gamma})}$ and $\ket{\psi(\vec{\xi})}$ are at finite distance to each other. We use this fact below.  In Appendix \ref{app:noint} we prove the following theorem.

\begin{theorem}
\label{thm:noint}
There exist vectors $\vec{\gamma} \in \R_{\geq 0}^2 \times \R_{>0}$ and $\vec{\xi} \in \R_{\geq 0}^2 \times \R_{<0}$ (in particular $\vec{\gamma} \neq \vec{\xi}$) with $\|\vec{\gamma}\|, \|\vec{\xi}\| < \frac{1}{2}$ such that
$\ket{\psi(\vec{\gamma})}$ can be deterministically transformed via LOCC into $\ket{\psi(\vec{\xi})}$, but this transformation does not have an optimal intermediate state, i.e. $\mathcal{I}[\psi(\vec{\gamma}),\psi(\vec{\xi})] = \{\}$.
\end{theorem}

Since $\vec{\gamma} \neq \vec{\xi}$ holds for the states in Theorem \ref{thm:noint}, the LU classes of $\ket{\psi(\vec{\gamma})}$ and $\ket{\psi(\vec{\xi})}$ are at finite distance to each other. As there is no optimal intermediate state for the transformation from $\ket{\psi(\vec{\gamma})}$ and $\ket{\psi(\vec{\xi})}$, this shows that one really has to ``jump'' in the Hilbert space from the initial to the final state in order to complete it optimally.\\

Finally, note that in Refs. \cite{Spee2017,DeVicente2017} SLOCC classes of $2^k$-qubit states (with $k \geq 2$) were constructed for which LOCC can be characterized using the results on LOCC transformations of generic 4-qubit states presented in Ref. \cite{Sauerwein2015}. Theorem \ref{thm:noint} can be straightforwardly generalized to these cases. That is, for any of these classes of $2^k$-qubit states there are LOCC transformations that can only be implemented via a ``jump'' in Hilbert space as no optimal intermediate state exists. For these classes it moreover holds that LOCC = SEP \cite{Spee2017} and thus these transformations do not even have intermediate states if the (in general more powerful) separable operations are considered.

\section{Entanglement monotones}
\label{sec:monotones}

In this section we introduce an infinite set of entanglement monotones for generic pure multipartite states. Moreover, we show that a specific finite set of them can be regarded as the generalization of the bipartite entanglement monotones. In particular, we show that this finite set, together with the description of the SLOCC class a state belongs to, completely characterizes the entanglement contained in the multiparite state.

As mentioned before, it was shown that the optimal success probability
 to transform a state $\ket{\Psi}=g\ket{\Psi_s} \in
\mH_{n,d}$ in the SLOCC class of a generic state into a state $\ket{\Phi} = h\ket{\Psi_s}$, where $g,h
\in \widetilde{G}$, is given by \cite{Gour2017}
\bea
P(\Psi,\Phi) = \frac{\|\Phi\|^2}{\|\Psi\|^2}
\frac{1}{\lambda_{max}(G^{-1} H)}. \label{eq:optprobgenEnMon}
\eea

As before, $G=g^\dagger g$ and $H=h^\dagger h$ denote local operators. We included here the norms of the states. However, in the following the states are considered to be normalized. Recall that we call an SLOCC class generic if it is comprised of states that do not have any nontrivial local symmetry. As mentioned before, for the Hilbert spaces, ${\cal H}_{n,d}$, with $n>4$ and $d$ arbitrary, it has been shown in Refs.  \cite{Gour2017,Sauerwein2018} that the set of states without any nontrivial local symmetries is of full measure.

Scrutinizing how the expression in Eq. (\ref{eq:optprobgenEnMon}) for the maximal success probability is related to the general expression,
\bea \label{eq_PsuEnMOns} P(\Psi,\Phi) = \min_\mu \frac{\mu(\Psi)}{\mu(\Phi)},\eea where the minimization is over all entanglement monotones, we now identify infinitely many entanglement monotones for pure states. Moreover, we show that it suffices to consider only those entanglement monotones in the optimization above.
More precisly, we show now that the success probability (for generic states) can be written as
\bea \label{Eq:auxP}
P(\Psi,\Phi) = \min_{\vec{x}}
\frac{E^{\Psi_s}_{\vec{x}}(\ket{\Psi})}{E^{\Psi_s}_{\vec{x}}(\ket{\Phi})},
\eea
where
\bea \label{eq:EnMonsSecEnMOns} E^{\Psi_s}_{\vec{x}}(\ket{\Psi})=\bra{\vec{x}}G\ket{\vec{x}}\eea are entanglement monotones for pure states (within SLOCC classes of generic states). Here, $\ket{\vec{x}}$ denotes a product state, i.e. $\ket{\vec{x}}=\ket{x_1}\otimes
\ldots \otimes \ket{x_n}$, with $\ket{x_i}\in \C^d$.

This can be easily seen by noting that
\begin{align*} \lambda_{max}(G^{-1} H) &=\max_{\vec{x}} \frac{\bra{\vec{x}} (g^\dagger)^{-1} H
g^{-1}\ket{\vec{x}}}{\bra{\vec{x}}\vec{x}\rangle}\\
&= \max_{\vec{x}} \frac{\bra{\vec{x}}  H \ket{\vec{x}}}{\bra{\vec{x}}G |\vec{x}\rangle}.
\end{align*}

Here, the optimization is only over product states, as $h$ and $g$ are local operators.

Let us now show that the function $E^{\Psi_s}_{\vec{x}}$ are, for any generic state $\Psi_s$ and any product state $\vec{x}$, entanglement monotones for pure states. In order to do so, we use the necessary and sufficient conditions for a function to be an entanglement monotone \cite{Vidal2000mono} and show that $E^{\Psi_s}_{\vec{x}}$ is
(i) invariant under LUs and (ii) is nonincreasing on average under unilocal operations.

Regarding (i); note that for any LU operator $\otimes_i U_i$, it holds that
\begin{align*}
  E^{\Psi_s}_{\vec{x}}[(\otimes_i U_i) g \ket{\Psi_s}] &= \bra{\vec{x}}(\otimes_i U_i g_i)^\dagger (\otimes_i U_i g_i) \ket{\vec{x}}\\
  &=\bra{\vec{x}}G\ket{\vec{x}}=E^{\Psi_s}_{\vec{x}}[ g \ket{\Psi_s}].
\end{align*}

Hence, the entanglement monotones are invariant under local unitaries.

Regarding (ii); consider a local operation on the first party, which transforms the state $g \ket{\Psi_s}$ into the ensemble $\{p_i, \frac{A_i g \ket{\Psi_s}}{||A_i g \ket{\Psi_s}||}\}$. Using the completeness relation $\sum_i A_i^\dagger A_i = \one$, it is easy to see that
\begin{align*}
&\sum_i p_i E^{\Psi_s}_{\vec{x}}\left[ \frac{A_i g \ket{\Psi_s}}{||A_i g \ket{\Psi_s}||} \right] =\sum_i p_i \frac{\bra{\vec{x}} (A_i g)^\dagger A_i g\ket{\vec{x}}}{p_i}\\
&=\bra{\vec{x}}G\ket{\vec{x}} = E^{\Psi_s}_{\vec{x}}\left[ g \ket{\Psi_s}\right].
\end{align*}

Hence, $E^{\Psi_s}_{\vec{x}}$ is, in fact, invariant under unilocal operations. As the same argument applies to all unilocal operations, that is for all parties, this shows that $E^{\Psi_s}_{\vec{x}}$ is (for any generic state $\Psi_s$  and any product state $\vec{x}$) an entanglement monotone.

Note that this proof shows that the entanglement monotone, $E^{\Psi_s}_{\vec{x}}$,  evaluated for a state $\ket{\Psi}$ coincides with the average amount of entanglement of any ensemble which is reached from $\ket{\Psi}$. This equivalence holds even for separable operations (SEP), as we needed nowhere in the proof that the ensemble is reached from LOCC (but only that it is reached from SEP). Hence, we have that
\begin{align*}
E^{\Psi_s}_{\vec{x}}\left[ g \ket{\Psi_s}\right] = \sum_i p_{i_1,\ldots,i_n} E^{\Psi_s}_{\vec{x}}\left[ \frac{A_{i_1}\otimes \ldots \otimes A_{i_n} g \ket{\Psi_s}}{||A_{i_1}\otimes \ldots \otimes A_{i_n} g \ket{\Psi_s}||} \right],
\end{align*}
for any SEP map with Kraus operators $\{A_{i_1}\otimes \ldots \otimes A_{i_n}\}$ such that $\sum_{i_1,\ldots,i_n} (A_{i_1}\otimes \ldots \otimes A_{i_n})^\dagger (A_{i_1}\otimes \ldots \otimes A_{i_n}) =\one$.

Interestingly, these entanglement monotones are the analog of the functions $E_l$ in the bipartite case (see Eq. (\ref{Eq:auxP})). Moreover, as $G$ is a local operator they can be easily computed. Let us also note here that an entanglement measure is simply not increasing under LOCC, in contrast to an entanglement monotone, which does not increase on average. In the case of pure states, the former is indeed a weaker condition than the latter. In fact, in the case considered here, there exists no LOCC transformation among pure states within the same Hilbert space. Hence, to check that a function is an entanglement measure, one needs to consider local unitaries and transformations from a higher dimensional Hilbert space to a lower dimensional one \cite{Sauerwein2018}. Here, we consider entanglement monotones within a specific SLOCC class and therefore do not consider transformations to lower dimensional Hilbert spaces.

Let us now discuss some properties of the entanglement monotones.
It is important to note that, in contrast to most of the entanglement monotones introduced for multipartite states, the entanglement monotones introduced here are not SLOCC invariant \footnote{Let us remark here, that the monotones (for different seed states) can lead to the same values for two states in different SLOCC classes. That is,
$ E^{\Psi_{s_1}}_{\vec{x}}(g\ket{\Psi_{s_1}})=E^{\Psi_{s_2}}_{\vec{x}}(g\ket{\Psi_{s_2}}), \forall \vec{x} ,$ where $\ket{\Psi_{s_i}}$ for $i=1,2$ denote two different seed states, which possibly belong to different SLOCC classes.
}. Hence, they allow to compare the entanglement contained in two states within the same SLOCC class. Actually, as we show now, these measures completely characterize the entanglement contained in generic states (as long as the SLOCC class is known).

Whereas there are infinitely many monotones of the form $E^{\Psi_{s}}_{\vec{x}}$, a finite set suffices to uniquely define the state, given that the SLOCC class, e.g. $\ket{\Psi_s}$, is known. This can be easily seen by noting that there exists a finite set of product states, $\{\ket{\vec{x}_i}\}_{i \in I}$, where $I$ is a finite index set, such that $\{\proj{\vec{x}_i}\}_{i \in I}$ forms a basis in the space $\mathcal{B}(\mH_n)$ of all linear operators that map $\mH_n$ to $\mH_n$. Hence, a generic multipartite state, $\ket{\Psi}$, is completely determined by the quantities $\{E_{\vec{x}_i}(\Psi)\}_{i}$ and the seed state.
This resembles very much the case of bipartite states, whose entanglement is completely determined by the monotones $E_l$. In contrast to the bipartite case, where all fully entangled states correspond to a single SLOCC class and therefore the seed states can always be chosen to be the maximally entangled $\ket{\phi^+}$ state, here, the SLOCC class has to be specified by the state $\ket{\Psi_s}$. Note, however, that the algorithm presented in \cite{Verstraete2003} can be used to determine a uniquely defined representative (a so-called critical state) of the SLOCC class to which $\ket{\Psi} \in \mH_n$ belongs to.

Summarizing, we have shown the following theorem.
\begin{theorem} \label{Th:EnMons}
  Let $\ket{\Psi}=g\ket{\Psi_s} \in \mH_n$ be generic. Then, the (infinitely many) functions $E^{\Psi_s}_{\vec{x}}$ are entanglement monotones (within a SLOCC class). Let furthermore the SLOCC class to which $\ket{\Psi}$ belongs to, i.e. $\ket{\Psi_s}$, be known. Then, the finite set of entanglement monotones, $\{E^{\Psi_s}_{\vec{x}_i}\}_{i}$, where $\{\proj{\vec{x}_i}\}_{i}$ denotes a finite product state basis of $\mathcal{B}(\mH_n)$, uniquely characterizes the entanglement of the state, i.e. characterizes the state up to local unitaries.
\end{theorem}

Let us finally remark here that it was already shown in Ref. \cite{Gour2011} that the expression on the right hand side of Eq. (\ref{eq:optprobgenEnMon}) leads to a lower bound on the success probability for arbitrary (also nongeneric ones) states. Note however, that the inequality can be strict for nongeneric states, as it is for instance the case for bipartite systems, where no direct OSBP is optimal. In this case, the local symmetry of the seed state can be used to achieve a better bound \cite{Gour2011}. First examples of (nongeneric) multipartite cases, where the inequality is strict are derived in Ref. \cite{Sw2018}, where we show that a OSBP does not succeed to reach the optimal success probability.

\section{Conclusion}
\label{sec:conclusion}

We have studied all possible optimal probabilistic and deterministic transformations between bipartite as well as multipartite states. We have shown that, generically, optimal transformations can be split into finitely many steps and the optimal intermediate states were characterized. The total success probability of transforming the initial to the final state remains maximal. Moreover, we have demonstrated that, generically, any pair of states can be connected via an optimal path. We determined the success probability of a transformation along this path using methods from survival analysis. The necessary and sufficient conditions for a path to be optimal are then given by a differential equation. We introduced a physically motivated distance measure, the interconversion metric, which measures the difficulty of interconverting two states. For the bipartite and the generic multipartite case we showed that all minimal geodesics in the interconversion metric correspond to twofold optimal paths, i.e. optimal paths for which also the reverse path is optimal. Moreover, we introduced an infinite set of easily computable entanglement monotones for generic multipartite pure states. A given finite set of these monotones, together with the knowledge of the SLOCC class to which a state belongs to, completely determines the entanglement contained in the state. As we show elsewhere, these monotones can also be used to relate the optimality of the transformation along a path with the entanglement that is lost during this process \cite{Sw2018}.\\
The methods introduced in this work can also be used to derive new optimal transformations for nongeneric multipartite states aswell as for mixed states \cite{Sw2018}. Moreover, the ideas and mathematical tools presented here have applications beyond entanglement theory. They can, for example, be applied to other quantum resource theories \cite{Chitambar2018}, such as generalized LOCC operations \cite{Chitambar2017,Contreras-Tejada2018} and quantum thermodynamics \cite{Goold2016}. It would be appealing to see whether new features of the corresponding resources can be revealed using this approach, as it was the case for the theory of entanglement investigated here.\\

\noindent\textit{Acknowledgments:---} This research was supported by the Austrian Science Fund (FWF) through Grants No. Y535-N16 and No. DK-ALM:W1259-N27.

\appendix

\section{Alternative derivation of the probability functional}
\label{sec:altder}

In this appendix we provide an alternative derivation for the formula given in Eq. (\ref{eq:PfuncInt}) of the probability $P[\Psi,\psi(t)] \equiv S_{\psi}(t)$ to transform $\ket{\Psi}$ via the SLOCC path $\{\ket{\psi(t)}\}$ into $\ket{\psi(t)}$. We show here how the expression in Eq. (\ref{eq:PfuncInt}) follows as the limit of $N \rightarrow \infty$ optimal transformations via finer sequences of $N$ intermediate states. To this end, we use the theory of product integration. We refer the reader to,  e.g., Refs. \cite{Gill, slavik2007product} for a more thorough exposition of this theory.\\

For $\ket{\Psi},\ket{\Phi} \in \mH_n$ and a sequence of $N$ (not necessarily optimal) intermediate states $\chi^{(N)} = (\ket{\chi_i})_{i=0}^{N+1}$ (where $\ket{\chi_0} = \ket{\Psi}$ and $\ket{\chi_{N+1}} = \ket{\Phi}$) of the transformation from $\ket{\Psi}$ to $\ket{\Phi}$ we define the \emph{mesh} of $\chi^{(N)}$ as
\bea
 |\chi^{(N)}| = \max_{i} \|\ket{\chi_i} - \ket{\chi_{i-1}}\|.
\eea
That is, the mesh of $\chi^{(N)}$ quantifies how large the steps in the decomposition of the transformation from $\ket{\Psi}$ to $\ket{\Phi}$ via the intermediate states in $\chi^{(N)}$ are. We can further define a partial order on sequences of intermediate states: We say that a $N$-sequence $\chi^{(N)} = (\ket{\chi_i})_{i=0}^{N+1}$ is a refinement of an $M$-sequence $\xi^{(M)} = (\ket{\xi_i})_{i=0}^{M+1}$ if $\{\ket{\xi_i}\}_{i=1}^M \subset \{\ket{\chi_i}\}_{i=1}^N$; that is, if $\chi^{(N)}$ contains all elements of $\xi^{(M)}$ (and possibly more). With these definitions we can consider sequences of finer divisions via intermediate states of the transformation from $\ket{\Psi}$ to $\ket{\Phi}$ and provide an alternative derivation of Eq. (\ref{eq:PfuncInt}).\\

Let $\{\ket{\psi(t)}\}_{0\leq t \leq 1}$ be a (not necessarily optimal) SLOCC path connecting $\ket{\Psi} = \ket{\psi(0)}$ and $\ket{\Phi} = \ket{\psi(1)}$. Let $(\chi^{(N)})_{N \in \mathbb{N}}$ be a sequence of finer $N$-sequences of intermediates states that lie on the path $\{\ket{\psi(t)}\}$. That is, for any $N \in \mathbb{N}$ there are parameters  $\{t_k^{(N)}\}_{k=0}^{N+1} \in [0,1]$ (with $t_0^{(N)} = 0$ and $t_{N+1}^{(N)} = 1$) such that
\begin{align}
&\chi^{(N)} = \left(\ket{\psi(t_k^{(N)})}\right)_{k=0}^{N+1},\\
&\lim_{N \rightarrow \infty} |\chi^{(N)}| = 0. \label{eq:mesh2zero}
\end{align}
Clearly, for any $N \in \mathbb{N}$ the probability to transform $\ket{\Psi}$ into $\ket{\Phi}$ via the intermediate states in $\chi^{(N)}$ can then be expressed as,
\begin{align}
P(\Psi,\Phi) &= \prod_{k=1}^{N+1} P[\psi(t_{k-1}^{(N)}),\psi(t_k^{(N)})]. \label{eq:prod1}
\end{align}
Due to Eq. (\ref{eq:mesh2zero}), $P[\psi(t_{k-1}^{(N)}),\psi(t_k^{(N)})]$ is arbitrarily close to one for large enough $N$. We can now use that, by definition of an SLOCC path (see Section \ref{sec:pathdef}), $P[\psi(t),\psi(t+dt)] = 1-r_{\psi} dt$ holds, with $r_{\psi}(t) = h_{\psi}(t)$. For sufficiently large $N$, the expression in Eq. (\ref{eq:prod1}) is then arbitrarily close to
\bea
P_N = \prod_{k=1}^{N+1} (1 - h_{\psi}(t_k) \Delta t_k). \label{eq:prod2}
\eea
The limit of $N \rightarrow \infty$ of Eq. (\ref{eq:prod2}) is called the right product integral of the function $-h_{\psi}(t)$ and is
denoted by
\bea
\tilde{P}[\psi] = (1-h_{\psi}(s)ds)\prod_{s=0}^t.
\eea

Since $h_{\psi}(t)$ is an integrable function, the following holds (see e.g. Example 2.5.6 of \cite{slavik2007product}),
\bea
\tilde{P}[\psi] = P[\psi].
\eea
This concludes the proof of Eq. (\ref{eq:PfuncInt}).

\section{Length of curves}
\label{app:geodesic}
We use the same notation as in Section \ref{sec:geodesic}. The length of an SLOCC path $\{\ket{\psi(t)} = g(t)\ket{\Psi}\}_{0\leq t \leq 1}$ in the interconversion metric, $d_I$, can be defined as
the supremum over the length of all polygonal chains that can be inscribed in $\{\ket{\psi(t)}\}$ (see e.g. Refs. \cite{rudin1976, burago2001} for details), i.e. as
\begin{align}
 &L[\psi] \equiv \nonumber \\
 &\sup\left\{\sum_{i=1}^{n-1} d_I[\psi(t_i),\psi(t_{i+1})] \ | \ n \in \mathbb{N}, 0 \leq t_0 \leq \ldots \leq t_n \leq 1\right\}. \label{eq:sup}
\end{align}
Note that it is, \emph{a priori}, not clear if $L[\psi]$ is well-defined for every SLOCC path $\{\ket{\psi(t)}\}_{0\leq t \leq 1}$, i.e. it is not clear if any SLOCC path is \emph{rectifiable} (see e.g. Ref. \cite{burago2001}). However, in the following we show  that this is indeed the case, by deriving an explicit formula for $L[\psi]$ for any SLOCC path $\{\ket{\psi(t)}\}_{0\leq t \leq 1}$. \\
To this end, let $n \in \mathbb{N}, 0 \leq t_0 \leq \ldots \leq t_n \leq 1$. Let further $\{\ket{\psi_R(t)} = \ket{\psi(1-t)}\}$ denote the reverse path of $\{\ket{\psi(t)}\}$. It follows from the definition Eq. (\ref{eq:hazard}) of the hazard rate as a limit that there are functions $f_i(t)$ and $f_i^R(t)$ for $i \in \{1,\ldots,n-1\}$ with $\lim_{t\rightarrow 0} \frac{f_i(t)}{t} = 0$ and $\lim_{t\rightarrow 0} \frac{f_i^R(t)}{t} = 0$ such that the following holds,
\begin{align*}
 &1-P[\psi(t_i),\psi(t_{i+1})] = h_{\psi}(t_i) \Delta t_i+ f_i(\Delta t_i),\\
 &1-P[\psi(t_{i+1}),\psi(t_i)] = h_{\psi_R}(1-t_{i}) \Delta t_i + f_i^R(\Delta t_i),
\end{align*}
where $\Delta t_i = t_{i+1} - t_i$. Inserting this into $d_I[\psi(t_i),\psi(t_{i+1})] = -\log\{P[\psi(t_i),\psi(t_{i+1})]P[\psi(t_{i+1}),\psi(t_{i})]\}$ and using that $-\log(1-x) = x + O(x^2)$ we find the following equation,
\begin{align}
 d_I[\psi(t_i),\psi(t_{i+1})] = [h_{\psi}(t_i)+h_{\psi_R}(1-t_{i})] \Delta t_i + F_i(\Delta t_i), \label{eq:approx1}
\end{align}
where $F_i(t)$ is a function that satisfies $\lim_{t \rightarrow 0} \frac{F_i(t)}{t} = 0$. Since $d_I$ fulfills the triangle inequality, it is clear that the supremum in Eq. (\ref{eq:sup}) can only be attained for $n \rightarrow \infty$ and $\max_i\{\Delta t_i\} \rightarrow 0$. Using Eq. (\ref{eq:approx1}) we see that for all $\epsilon > 0$ there exists an $n_0 \in \mathbb{N}$ and a $\delta > 0$ such that for all $n \geq n_0$ and all $0 \leq t_0 \leq \ldots \leq t_n \leq 1$ with $\max_i \{\Delta t_i\} < \delta$ the following holds,
\begin{align}
 \left |\sum_{i=1}^{n-1} d_I[\psi(t_i),\psi(t_{i+1})] - \sum_{i=1}^{n-1} [h_{\psi}(t_i)+h_{\psi_R}(1-t_{i})] \Delta t_i \right| < \epsilon. \label{eq:approx2}
\end{align}
Note that the right sum on the left-hand side of Eq. (\ref{eq:approx2}) is simply a Riemann-sum approximation of the following integral,
\begin{align}
 \tilde{L}[\psi] = \int_{0}^1 ds(t),
\end{align}
 where $ds(t) = d_I[\psi(t),\psi(t+dt)] = [h_{\psi}(t)+h_{\psi_R}(1-t)] dt$. Hence, inequality \ref{eq:approx2} shows that the length of $\{\ket{\psi(t)}\}$ is indeed given by $L[\psi] = \tilde{L}[\psi]$.

\section{Bipartite optimal intermediate states and paths}

\subsection{Optimal intermediate states $\ket{\zeta}$ and $\ket{\eta}$}
 \label{app:bipartiteoptFi}
Let us review here the definition of the optimal intermediate state $\ket{\xi}$ of the protocol introduced in Ref. \cite{Vidal1999}. There, it was shown that a bipartite state $\ket{\Psi}$ can always be transformed deterministically into a state $\ket{\xi}$, which can then be transformed optimally into the final state $\ket{\Phi}$, that is $P(\xi, \Phi) = P(\Psi, \Phi)$. Moreover, $\ket{\xi}$ optimizes the fidelity with the final state (see Section \ref{sec:IntBip} of the main text for more details). With the help of the definition of $\ket{\xi}$ we can then show how one can generalize this type of protocol by introducing the two optimal intermediate states $\ket{\zeta}$ and $\ket{\eta}$, for which the fidelity is equal to the optimal one, i.e. $F(\zeta, \eta) = F(\xi, \Phi)$. Moreover, the transformation $\ket{\Psi} \xrightarrow{LOCC} \ket{\zeta} \xrightarrow{OSBP} \ket{\eta} \xrightarrow{LOCC} \ket{\Phi}$ is also optimal (see also Fig.\ \ref{VaVsFi}), that is $P(\zeta, \eta) = P(\Psi, \Phi)$. \par
For the definition of $\ket{\xi}$ (see Ref. \cite{Vidal1999}) let $l_1$ be the smallest integer in $\{1,...,d\}$ such that
\begin{equation}
\frac{E_{l_1}(\Psi)}{E_{l_1}(\Phi)} = \min_{l \in \{1,...,d\}} \frac{E_{l}(\Psi)}{E_{l}(\Phi)} \equiv r_1.
\end{equation}
Thus, $r_1$ is equal to the optimal success probability $P(\Psi, \Phi)$. In case $r_1 \neq 1$, let $l_2$ be the smallest integer in $\{1,...,l_1-1\}$ such that
\begin{equation}
\frac{E_{l_2}(\Psi)-E_{l_1}(\Psi)}{E_{l_2}(\Phi)-E_{l_1}(\Phi)} = \min_{l \in \{1,...,l_1-1\}} \frac{E_{l}(\Psi)-E_{l_1}(\Psi)}{E_{l}(\Phi)-E_{l_1}(\Phi)} \equiv r_2.
\end{equation}
If one repeats this process until $l_k =1$ for some $k$, a series of $k+1$ positive integers $l_0 \equiv d+1 > l_1 >\ldots> l_k$ and a series of $k$ positive real numbers $0<r_1<...<r_k$ is obtained, which defines the state $\ket{\xi}$ via
\begin{equation}
\ket{\xi} = \sum_{i=1}^d \sqrt{ \xi_i} \ket{ii}, \ \xi_i = r_j \Phi_i \ \text{if} \ i \in [l_j, l_{j-1} -1].
\end{equation}
It can be easily shown that, indeed, $\lambda(\Psi) \prec \lambda(\xi)$. \par

Using now the above defined $\{l_j\}$ we can introduce the states $\ket{\zeta}, \ket{\eta}$. Let us denote by $\hat{e}_{l_j}(\zeta),\hat{e}_{l_j}(\eta) \in \R^d$, for $j \in \{1,\ldots,k\}$, nonnegative unit vectors, i.e. $\Vert  \hat{e}_{l_j}(\zeta) \Vert_1 = \Vert  \hat{e}_{l_j}(\eta) \Vert_1 =1$, with nonzero components only for indices in $\{l_{j},...,l_{j-1}-1\}$ (again with $l_0 = d+1$) . Moreover, the unit vectors have to fulfill the conditions
\begin{eqnarray}
\hat{e}_{l_j}(\zeta) = \hat{e}_{l_j}(\eta)\nonumber, \\
\hat{e}_{l_j}(\Psi) \prec \hat{e}_{l_j}(\zeta) \prec \hat{e}_{l_j}(\Phi), \label{nec+suffXiChi}
\end{eqnarray}
where $\hat{e}_{l_j}(\Psi)  = \frac{1}{E_{l_j}(\Psi)-E_{l_{j-1}}(\Psi)} \lambda_{[l_j, l_{j-1}-1]}(\Psi)$. Here, $\lambda_{[l_j, l_{j-1}-1]}(\Psi)$ denotes the vector whose components with indices in $\{l_j,...,l_{j-1}-1\}$ are equal to the ones of the Schmidt vector $\lambda(\Psi)$, and similar for $\hat{e}_{l_j}(\Phi)$. Then, the Schmidt vectors of $\ket{\zeta}$ and $\ket{\eta}$ are given by (with $E_{l_0} =0$)
\begin{eqnarray}
\lambda(\zeta) &= \sum_{j=1}^k [E_{l_j}(\Psi) - E_{l_{j-1}}(\Psi)] \hat{e}_{l_j}(\zeta), \\
\lambda(\eta) &= \sum_{j=1}^k [E_{l_j}(\Phi) - E_{l_{j-1}}(\Phi)] \hat{e}_{l_j}(\eta).
\end{eqnarray}
 Using the results from Ref. \cite{Vidal1999} it can be easily shown that the success probability fulfills $P (\zeta, \eta) = P (\Psi, \Phi)$. Moreover, it can be easily seen that the fidelity $F(\zeta, \eta)$ is equal to the fidelity $F(\xi, \Phi)$ computed in Ref. \cite{Vidal2000}. As the conditions in Eq.\ \eqref{nec+suffXiChi} are  fulfilled by infinitely many states $\ket{\eta}, \ket{\zeta}$, there exist infinitely many probabilistic protocols, that achieve the optimal success probability for a transformation from $\ket{\Psi}$ to $\ket{\Phi}$ and where the optimal intermediate states are chosen to optimize the fidelity.

\subsection{Proof of Observation \ref{OSBPbip}}
\label{app:bipartiteOSBP}
Let us provide here the proof of Observation \ref{OSBPbip}, which we restate here for the sake of readability.\\

\noindent \textit{{\bf Observation \ref{OSBPbip}.}
A bipartite state $\ket{\Psi} \in \C^d \otimes \C^d$ with Schmidt vector $\lambda(\Psi) = (\Psi_1,\ldots,\Psi_d)$ can be optimally transformed via an OSBP into another bipartite state $\ket{\Phi} \in \C^d \otimes \C^d$  with Schmidt vector $\lambda(\Phi) = (\Phi_1, \ldots, \Phi_d)$ and with $\frac{\Psi_d}{\Phi_d}  \leq \frac{\Psi_l}{\Phi_l}$, for all $l \leq d$ iff the success probability is given by $P(\Psi, \Phi) = \frac{E_d(\Psi)}{E_d(\Phi)} = \frac{\Psi_d}{\Phi_d} < 1 $.
}

\begin{proof}
Let us, w.l.o.g., write the bipartite states in the decomposition $\ket{\Psi} = D_{\Psi} \otimes \mathbbm{1} \ket{\phi^+}, \ \ket{\Phi} = D_{\Phi} \otimes \mathbbm{1} \ket{\phi^+}$, where $\ket{\phi^+} = \sum_{i=1}^d \ket{i,i}$, $D_{\Psi} = \text{diag}(\sqrt{\Psi_1},\ldots, \sqrt{\Psi_d})$ and $D_{\Phi} = \text{diag}(\sqrt{\Phi_1},\ldots, \sqrt{\Phi_d})$. Then, the most general local operator, transforming $\ket{\Psi}$ to $\ket{\Phi}$, that can be applied by party A is of the form
\begin{equation}
M = \sqrt{p} D_{\Phi} U D^{-1}_{\Psi},
\label{OSBPbipartite}
\end{equation}
where $U \in \mathcal{U}(d)$ is an arbitrary unitary matrix. Note that the unitary is allowed due to the symmetries of the maximally entangled state $\ket{\phi^+}$, i.e. $U \otimes U^* \ket{\phi^+} = \ket{\phi^+}$ (see also \cite{Gour2011}). Moreover, only unitary symmetries have to be considered as it was shown in \cite{Lo2001}, that any local transformation among bipartite pure states can be done by party A performing the quantum measurement and party B applying the corresponding unitary. Then, $M$ and the corresponding unitary applied to $\ket{\Psi}$ yields the final state with probability $p$, i.e. $M \otimes U^* \ket{\Psi} = \sqrt{p} \ket{\Phi}$. Moreover, the square of the maximum singular value of the operator $M$ in Eq.\ \eqref{OSBPbipartite}, i.e. the maximum eigenvalue of $M^{\dagger} M$ can be lower bounded by
\begin{eqnarray}
\lambda_{max}(p (D^{-1}_{\Psi})^2 U^{\dagger} D_{\Phi}^2 U) &\geq & p \lambda_{max}( (D^{-1}_{\Psi})^2) \lambda_{min} (U^{\dagger} D_{\Phi}^2 U) \nonumber \\ &=& p \frac{\Psi_d}{\Phi_d}.
\end{eqnarray}
This lower bound can always be obtained for $U = \mathbbm{1}$, as the Schmidt coefficients of the two states fulfill $\frac{\Psi_d}{\Phi_d}  \leq \frac{\Psi_l}{\Phi_l}$, for all $l \leq d$. 
Note that for an OSBP the maximum singular value of the measurement operator corresponding to the successful branch has to be equal to one, i.e. $\Vert M \Vert =1$, such that the state of the failure branch is no longer fully entangled in $ \C^d \otimes \C^d$ and can, thus, no longer be transformed into the final state with a finite probability of success. As $\lambda_{max} (M^{\dagger} M) = p \frac{\Psi_d}{\Phi_d}$, this can only be fulfilled iff $p = P(\Psi, \Phi) = \frac{E_d(\Psi)}{E_d(\Phi)}$.
\end{proof}

\subsection{Reordered states on the most entangled path}
 \label{app:bipartitemost}
Here, we construct the state $\ket{\chi^{max'}}$, which has the same properties as the state $\ket{\chi^{max}}$ defined as
 \begin{eqnarray}
\chi^{max}_k & = & \frac{\Psi_k}{P}  \ , k \in \{2,...,d\} \nonumber \\
\chi^{max}_1 & = &  1- \sum_{k=2}^d \chi^{max}_k,
\end{eqnarray}
if the Schmidt coefficients above are not ordered decreasingly, i.e. $\chi^{max}_1 \not\geq \chi^{max}_{k}$ for some $k$. We denote the reordered state by $\ket{\chi^{max'}}$, which we obtain in the following way.
For $\chi^{max}_d > \frac{\Psi_d}{2 \Psi_2+\Psi_3}$ the first Schmidt coefficient is smaller than the second, i.e. $\chi^{max}_1 < \chi^{max}_2$ and we define them to be equal to each other for the new state $\ket{\chi^{max'}}$, i.e. $\chi^{max'}_1 = \chi^{max'}_2 = (1- \chi^{max}_d \sum_{i=3}^d \frac{\Psi_i}{\Psi_d} )/2$, and leave all other Schmidt coefficients as before. If the first two Schmidt coefficients of this new state are now smaller than the third Schmidt coefficient we proceed in the same way. In general we set the first $m$ Schmidt coefficients equal to each other, i.e. $\chi^{max'}_1 = \chi^{max'}_2 = ...=\chi^{max'}_m =  (1- \chi^{max}_d \sum_{i=m+1}^d \frac{\Psi_i}{\Psi_d} )/m $ for some $m \in \{2,...,d-1\}$ if $\chi^{max'}_d >  \frac{\Psi_d}{m \Psi_m+\sum_{i=m+1}^d \Psi_i}$. The last $d-m$ Schmidt coefficients are still given by the relation in Eq.\ \eqref{maxentPconst} in Section \ref{sec:mostent}. The Schmidt coefficients of this redefined state $\ket{\chi^{max'}}$ are then sorted in decreasing order. Let us now show that this state still maximizes all entanglement monotones for a constant success probability $P$. As the smallest $m+1$ monotones are still equal to the one of $\ket{\chi^{max}}$, i.e. $E_k (\chi^{max'}) = E_k (\chi^{max}) \ \forall k \in \{m+1,...,d\}$, they still fulfill the condition $E_k (\chi^{max'}) \geq E_k (\psi')$ for all states $\ket{\psi'}$ with $P(\Psi, \psi') = P(\Psi, \chi^{max'})$. Let $m$ be the smallest number such that this is not fulfilled for the $m^{th}$ monotone, i.e. $E_m (\chi^{max'}) = E_{m+1}(\chi^{max'}) + \chi^{max'}_1 < E_m (\psi') = E_{m+1}(\psi') + \psi'_{m}$. Then, $ \chi^{max'}_1 < \psi'_{m}$ and, moreover, $E_m (\chi^{max'}) = 1-(m-1) \chi^{max'}_1 < E_m (\psi') = 1- \sum_{i=1}^{m-1} \psi'_i$. From the first inequality we get that $1- (m-1) \chi^{max'}_1 > 1-(m-1) \psi'_{m} $ and as the Schmidt coefficients of $\ket{\psi'}$ are sorted, we have also that $1-(m-1) \psi'_{m} > 1- \sum_{i=1}^{m-1} \psi'_i$, which  then contradicts $1-(m-1) \chi^{max'}_1 < 1- \sum_{i=1}^{m-1} \psi'_i$. Thus, also $E_m (\chi^{max'}) \geq E_m (\psi')$ and the same holds for all $E_k$ with $k \in \{1,...,m-1\}$. \par
Let us also show that $\ket{\chi^{max'}}$ can obtain the final state $\ket{\Phi}$ deterministically. Recall that for the state $\ket{\chi^{max}}$ the following holds
\begin{equation}
E_l (\chi^{max}) = \frac{1}{P(\Psi \rightarrow \Phi)} \sum_{i=l}^d \Psi_i \geq E_l(\Phi) = \sum_{i=l}^d \Phi_i.
\end{equation}
Using this we can also show that $\ket{\chi^{max'}}$ is majorized by $\ket{\Phi}$. As the last $d-m$ Schmidt coefficients of $\ket{\chi^{max'}}$ are equivalent to the Schmidt coefficients of $\ket{\chi^{max}}$ it follows immediately that $E_l(\chi^{max'}) = E_l(\chi^{max}) \geq E_l (\Phi)$ $\forall l \in \{m+1,...,d\}$. Let $m$ be the smallest number such that this is not fulfilled for the $m^{th}$ entanglement monotone, i.e.
\begin{align}
E_m (\chi^{max'}) &= E_{m+1} (\chi^{max'}) + \chi^{max'}_1 \nonumber \\ 
&< E_m(\Phi) = E_{m+1}(\Phi) + \Phi_{m} \nonumber \\
&\Rightarrow  \chi^{max'}_1 < \Phi_{m}.
\end{align}
Thus, we also have that
\begin{align}
&E_m(\chi^{max'}) = 1- (m-1) \chi^{max'}_1 \nonumber \\
&> 1- (m-1) \Phi_m >  1 - \sum_{i =1}^{m-1} \Phi_i = E_m (\Phi),
\end{align}
contradicting the initial assumption that $E_m(\chi^{max'})  < E_m (\Phi)$. This shows that also the resorted state fulfills the majorization condition $E_l(\chi^{max'})  \geq E_l(\Phi)$ $\forall l \in \{1,...,d\}$.

\section{Proof of Theorem \ref{thm:subintgen}}
\label{app:genqb}
In this appendix we provide the proof of Theorem \ref{thm:subintgen}, which we restate here for the sake of readability.\\

\noindent \textit{{\bf Theorem \ref{thm:subintgen}.}
Let $\ket{\Psi},\ket{\Phi} = h\ket{\Psi} \in \mH_{n,d}$ be SLOCC equivalent generic states. A state $\ket{\chi} = g\ket{\Psi}$ is an optimal intermediate state of the transformation from $\ket{\Psi}$ to $\ket{\Phi}$, i.e. $\ket{\chi} \in \mathcal{I}(\Psi,\Phi)$,  if the following holds for all $i \in \{1,\ldots,n\}$,
\begin{itemize}
\item[(i)] $[G_i,H_i] = 0$,
\item[(ii)] $\frac{H_i}{\lambda_{max}(H_i)} \leq \frac{G_i}{\lambda_{max}(G_i)}$.
\end{itemize}
For transformations among generic $(n>4)$-qubit states all optimal intermediate states are of this form.
}

\proof{
Using Theorem \ref{thm:neccsuff0} it is easy to see that the conditions in Theorem \ref{thm:subintgen} are sufficient for $g\ket{\chi}$ to be an optimal intermediate state. Let us show that they are also necessary for  $(n>4)$-qubit states.\\
Using Theorem \ref{thm:neccsuff0} we see that $\ket{\chi} \in \mathcal{I}(\Psi,\Phi)$ iff
\begin{align}
 \frac{\lambda_{max}(G)}{\lambda_{max}(H)}\lambda_{max}(G^{-1}H) = 1.
\end{align}
We express this equation as,
\begin{align}
\prod_{i=1}^n \frac{\lambda_{max}(G_i)}{\lambda_{max}(H_i)}\lambda_{max}(G_i^{-1}H_i) = 1. \label{eq:prodsqb1}
\end{align}
Note that each factor on the left-hand side of Eq. (\ref{eq:prodsqb1}) is smaller equal to one. Hence, Eq. (\ref{eq:prodsqb1}) is fulfilled iff
\begin{align}
 \frac{\lambda_{max}(G_i)}{\lambda_{max}(H_i)}\lambda_{max}(G_i^{-1}H_i) = 1, \ \forall \ i \in \{1,\ldots,n\}.\label{eq:prodsqb2}
\end{align}
Let us write,
\begin{align}
 &\frac{H_i}{\lambda_{max}(H_i)} = U_i\text{diag}(1,d_i)U_i^\dagger,\\
 &\frac{G_i}{\lambda_{max}(G_i)} = V_i\text{diag}(1,d_i')V_i^\dagger,
\end{align}
where $U_i, V_i$ are unitaries and $ 0 < d_i, d_i' \leq 1$.
Using the notation $\tilde{V}_i = V_i^\dagger U_i$, Eq. (\ref{eq:prodsqb2}) is equivalent to,
\begin{align}
 \lambda_{max}\left[\text{diag}\left(1,\frac{1}{d_i'}\right)\tilde{V}_i\text{diag}\left(1,d_i\right)\tilde{V}_i^\dagger\right]=1. \label{eq:prodsqb3}
\end{align}
Expressing $\tilde{V}_i = e^{i\phi_i} e^{i\alpha_i \sigma_z}e^{i\beta_i \sigma_x}e^{i\gamma_i \sigma_z}$ in Euler decomposition, where $\phi,\alpha_i,\beta_i,\gamma_i \in [0,2\pi)$, we see that
Eq. (\ref{eq:prodsqb3}) is fulfilled iff $\alpha_i,\gamma_i = 0$ and $\beta_i \in \{0,\pi\}$, i.e. only if $\tilde{V}_i$ is diagonal, and $d_i \leq d_i' \leq 1$. It is easy to see that these conditions are equivalent to the conditions given in Theorem \ref{thm:neccsuff0}.
\qed\\
}

\section{Proof of Theorem  \ref{thm:noint}}
\label{app:noint}
In this appendix we provide the proof of Theorem \ref{thm:noint}. Before that, we summarize some results shown in Refs. \cite{DeVicente2013, Sauerwein2015}, which we use for the proof. The notation is the same as in Section \ref{sec:noint} of the main text.
\begin{lemma}
\label{lem:facts}
Let $\vec{\gamma} = (\gamma_1,\gamma_2,\gamma_3), \vec{\xi} = (\xi_1, \xi_2, \xi_3) \in \R_{\geq 0}^2 \times \R$. Then the following holds.
\begin{itemize}
\item[(i)]\cite{Sauerwein2015} A state $\ket{\psi(\vec{\gamma})}$ with  $\gamma_i \neq 0$ for all $i \in \{1,2,3\}$ can (up to LUs) only be transformed to/ reached from a 4-qubit state that is of the form  $\ket{\psi(\vec{\xi})}$ with $\xi_i \neq 0$ for all $i \in \{1,2,3\}$.
\item[(ii)] \cite{Sauerwein2015} If $\gamma_i, \xi_i \neq 0$ for all $i \in \{1,2,3\}$ then $\ket{\psi(\vec{\gamma})}$ can be transformed via LOCC into $\ket{\psi(\vec{\xi})}$ iff the following conditions are fulfilled,
\begin{align}
& 1 + r_1 + r_2 + r_3 \geq 0 \label{eq:r0}\\
& 1 + r_1 - r_2 - r_3 \geq 0 \\
& 1 + r_2 - r_1 - r_3 \geq 0 \\
& 1 + r_3 - r_1 - r_2 \geq 0 \label{eq:r3},
\end{align}
where $r_i = \frac{\gamma_i}{\xi_i}$ for $i \in \{1,2,3\}$.
\item[(iii)] \cite{DeVicente2013} $\ket{\Psi(\vec{\gamma})}$ can only be transformed to $\ket{\Psi(\vec{\xi})}$ via LOCC if $\|\vec{\gamma}\| \leq \|\vec{\xi}\|$. That is, $\frac{1}{2} - \|\vec{\gamma}\|$ is an entanglement measure.
\end{itemize}
\end{lemma}
Using (i) and (ii) of Lemma \ref{lem:facts} it is easy to show the following observation.
\begin{observation}
\label{obs:norm}
Let $\vec{\gamma}, \vec{\xi} \in \R_{\geq 0}^2 \times \R^\times$ with $\|\vec{\gamma}\| = \|\vec{\xi}\|$ and $\vec{\gamma} \neq \vec{\xi}$.
Then $\ket{\Psi(\vec{\gamma})}$ cannot be transformed into $\ket{\Psi(\vec{\xi})}$ and $\ket{\Psi(\vec{\xi})}$ cannot be transformed into $\ket{\Psi(\vec{\gamma})}$ via LOCC.
\end{observation}

Using these results we now prove Theorem \ref{thm:noint}, which we restate here for the sake of readability.\\

\noindent \textit{{\bf Theorem \ref{thm:noint}.}
There exist vectors $\vec{\gamma} \in \R_{\geq 0}^2 \times \R_{>0}$ and $\vec{\xi} \in \R_{\geq 0}^2 \times \R_{<0}$ (in particular $\vec{\gamma} \neq \vec{\xi}$) with $\|\vec{\gamma}\|, \|\vec{\xi}\| < \frac{1}{2}$ such that
$\ket{\psi(\vec{\gamma})}$ can be deterministically transformed via LOCC into $\ket{\psi(\vec{\xi})}$, but this transformation does not have an optimal intermediate state, i.e. $\mathcal{I}[\psi(\vec{\gamma}),\psi(\vec{\xi})] = \{\}$.
}\\

\begin{figure}
   \centering
   \includegraphics[width=0.35\textwidth]{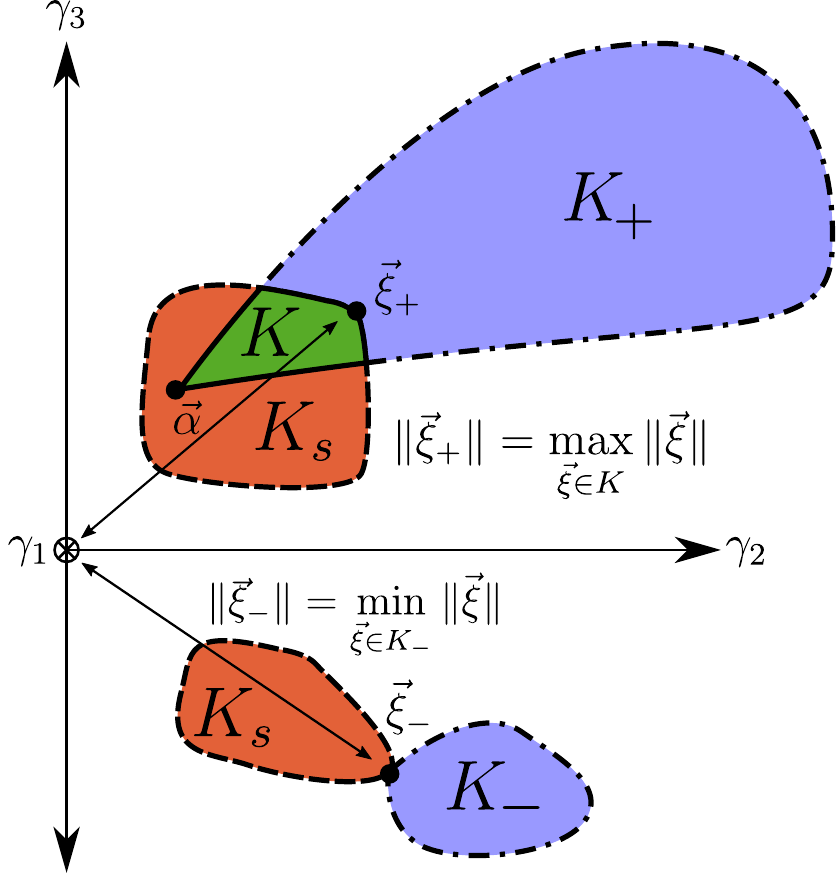}
   \caption{(color online). Schematic figure of the sets defined in the proof of Theorem \ref{thm:noint}. The coordinate axis of $\gamma_1$ is pointing out of the paper plane. Note that $K_s$ consists of two components. }
   \label{fig:noint}
   \end{figure}

\proof{
We consider the state $\ket{\psi(\vec{\alpha})}$ with $\vec{\alpha} = (0.09,0.1,0.08)$. Using (i) and (ii) of Lemma \ref{lem:facts} it is straightforward to show that the accessible set of the state $\ket{\psi(\vec{\alpha})}$ can be expressed (up to LUs) as,
\begin{align}
\mathcal{M}_a[\psi(\vec{\alpha})] = \left\{\ket{\psi(\vec{\beta})} \ | \ \vec{\beta} \in K_+ \cup K_- \right\},
\end{align}
where $K_+ \subset \R_{>0}^2 \times \R_{>0}$ and $K_- \subset \R_{>0}^2 \times \R_{<0}$ (note the different sign of the last component) are two compact sets in $\R^3$ that are at nonzero distance to each other, i.e.
$d(K_+,K_-) = \min_{\vec{\zeta}_+ \in K_+, \vec{\zeta}_- \in K_-} \| \vec{\zeta}_+ - \vec{\zeta}_-\| > 0$ (see Fig. \ref{fig:noint}).\\

Since $K_-$ is compact, there exists a $\vec{\xi}_- \in K_-$ such that $\|\vec{\xi}_-\| = \min_{\vec{\xi} \in K_-} \|\vec{\xi}\|$. Note that, due to (i) and (ii) of Lemma \ref{lem:facts} there exists a compact subset $K_s \subset \R^3$ such that the source set of $\ket{\psi(\vec{\xi}_-)}$ can (up to LUs) be expressed as,
\begin{align}
\mathcal{M}_s[\psi(\vec{\xi_-})] = \left\{\ket{\psi(\vec{\gamma'})} \ | \ \vec{\gamma'} \in K_s \right\}.
\end{align}
Let us denote the intersection of $K_s$ and $K_+$ by $K = K_s \cap K_+$. Note that $K$ is compact and not empty (as $\vec{\alpha} \in K$). Hence, there exists a $\vec{\xi}_+ \in K$ such that $\|\vec{\xi}_+\| = \max_{\xi \in K} \|\vec{\xi}\|$.

The different compact sets defined above are depicted in Fig. \ref{fig:noint}. It follows from the definition of the two states that $\ket{\psi(\vec{\xi}_+)}$ can be transformed  via LOCC into $\ket{\psi(\vec{\xi}_-)}$. However, this transformation does not have an intermediate state, as we now show by contradiction. Suppose that there is an intermediate state $\ket{\psi(\vec{\beta})} \in \mathcal{I}[\psi(\vec{\xi}_+),\psi(\vec{\xi})_-]$. Then it has to hold that $\vec{\beta} \neq \vec{\xi}_+,\vec{\xi}_-$ and $\vec{\beta} \in K \cup K_- $. However, due to (iii) of Lemma \ref{lem:facts} and Observation \ref{obs:norm} the state $\ket{\psi(\vec{\xi}_+)}$ cannot be transformed to any state $\ket{\psi(\vec{\beta})}$ with $\vec{\beta} \in K$ and therefore $\vec{\beta} \in K \cup K_-$ has to be an element of $K_-$.  Analogously, the state $\ket{\psi(\vec{\xi}_-)}$ cannot be reached from any state $\ket{\psi(\vec{\beta})}$ with $\vec{\beta} \in K_-$. Hence, $\vec{\beta} \not \in K$ and $\vec{\beta} \not \in K_-$ hold, which contradicts the fact that $\vec{\beta} \in K \cup K_-$ has to holds. Hence, such a $\vec{\beta}$ cannot exist and therefore $\mathcal{I}[\psi(\vec{\xi_+}),\psi(\vec{\xi}_-)] = \{\}$ holds.  Using $\ket{\psi(\vec{\gamma})} = \ket{\psi(\vec{\xi}_+)}$ and $\ket{\psi(\vec{\xi})} = \ket{\psi(\vec{\xi}_-)}$ this completes the proof of the theorem.\qed}

\bibliography{BIBSLOCC}

\end{document}